\documentclass[12pt]{iopart}

\usepackage{utf8math}
\usepackage{todonotes}
\usepackage{graphicx}
\usepackage{algpseudocode, algorithm}
\expandafter\let\csname equation*\endcsname\relax
\expandafter\let\csname endequation*\endcsname\relax
\usepackage{amsmath}
\usepackage{amssymb}
\usepackage{amsthm}
\usepackage{subcaption}
\usepackage{graphicx}
\usepackage{siunitx}
\usepackage{dirtytalk}
\usepackage[version=3]{mhchem} 
\usepackage{bm}
\usepackage{xr}
\usepackage{xcolor}
\usepackage{hyperref}
\usepackage{comment}
\usepackage[normalem]{ulem}
\usepackage{listings}
\usepackage{mathtools}
\renewcommand{\vec}[1]{\mathbf{#1}}
\renewcommand{\Re}{\mathbb R}
\graphicspath{{figures}}
\usepackage{cite}

\begin{document}
\title[ILP for Unsupervised Training Set Selection in Molecular Machine Learning]
{Integer Linear Programming for Unsupervised Training Set Selection in Molecular Machine Learning}
\author{
Matthieu Haeberle,\textsuperscript{1,2,$\ddag$}
Puck van Gerwen,\textsuperscript{1,3,$\ddag$}
Ruben Laplaza,\textsuperscript{1,3}
Ksenia R. Briling,\textsuperscript{1}
Jan Weinreich,\textsuperscript{1,3}
Friedrich Eisenbrand\textsuperscript{2} and
Cl\'emence Corminboeuf\textsuperscript{*,1,3}}
\address{
\textsuperscript{1}Laboratory for Computational Molecular Design, Institute of Chemical Sciences and Engineering, \'Ecole Polytechnique F\'ed\'erale de Lausanne, 1015 Lausanne, Switzerland\\
\textsuperscript{2}Chair of Discrete Optimization, \'Ecole Polytechnique F\'ed\'erale de Lausanne, 1015 Lausanne, Switzerland\\
\textsuperscript{3}National Center for Competence in Research-Catalysis (NCCR-Catalysis), Zurich, Switzerland \\
\textsuperscript{$\ddag$}These authors contributed equally to this work.
}
\ead{clemence.corminboeuf@epfl.ch}
\vspace{10pt}
\begin{indented}
\item[]\today
\end{indented}
\begin{abstract}
Integer linear programming (ILP) is an elegant approach to solve linear optimization problems, naturally described using integer decision variables. Within the context of physics-inspired machine learning applied to chemistry, we demonstrate the relevance of an ILP formulation to select molecular training sets for predictions of size-extensive properties.
We show that our algorithm outperforms existing unsupervised training set selection approaches, especially when predicting properties of molecules larger than those present in the training set.
We argue that the reason for the improved performance is due to the selection that is based on the notion of local similarity (\textit{i.e.}, per-atom) and a unique ILP approach that finds optimal solutions efficiently.
Altogether, this work provides a practical algorithm to improve the performance of physics-inspired machine learning models and offers insights into the conceptual differences with existing training set selection approaches.
\end{abstract}

\section{Introduction}
Linear programming (LP) is the method that optimizes an objective function subject to linear constraints on the variables. When framing an optimization problem, decision variables represent the choices or unknowns, and they are manipulated to optimize the objective function while satisfying the constraints. With Integer Linear Programming (ILP), these variables are constrained to integer values, which allow for a natural description of specific classes of problems in which discrete choices have to be made. However, this integer restriction also renders ILP problems NP-hard, meaning that no known algorithm can solve a general integer linear program in polynomial time. The art of integer programming is to find models that can be solved efficiently. We pursue this line of research here with the aim of finding a performing molecular training set for physics-inspired machine-learning tasks relevant to chemistry (\textit{vide infra}).\\

Molecular science presents several problems that are naturally formulated with ILP.
For example, Multiple Sequence Alignment (MSA) is a key challenge in biology, where three or more biological sequences (amino acids, nucleobases) are aligned based on their similarity (homology, superposability, or functionality).\cite{msa_2006} A number of gaps are introduced into the sequences until they conform to the same length. From an ILP perspective, the goal is to maximize the similarity between aligned sequences while minimizing the number of gaps introduced.\cite{book_ILP_bio, seq_alignment_ILP_1999, seq_alignment_ILP_2002, seq_alignment_ILP_2006, seq_alignment_ILP_2008} The decision variables represent the (integer) assignment of residues to alignment positions. Constraints may include prohibiting gaps within certain regions of the sequences or ensuring that every residue is aligned to at most one position. Various approaches\cite{book_ILP_bio, seq_alignment_ILP_1999, seq_alignment_ILP_2002, seq_alignment_ILP_2006, seq_alignment_ILP_2008} enable MSA of ten or more sequences to be solved in polynomial time.
This objective function can be formulated as a linear combination of similarity scores between aligned residues and penalties for introducing gaps.
Similarly, ILP can be used for sequence to structure alignment,\cite{rna_alignment_ILP_1998, rna_alignment_ILP_2002} protein-protein docking\cite{protein_protein_ILP_2000} and side-chain positioning,\cite{side_chain_positioning_ILP_2004} among other applications in structural biology.\cite{book_ILP_bio} \\

ILP-ready problems exist in chemistry as well. Atom mapping is, for instance, the identification of the one-to-one correspondence between atoms in reactants and products of a chemical reaction. This mapping effectively gives a description of the reaction mechanism, which helps in automated reaction planning,\cite{route_retro_2009, route_retro_2015, segler_retro_2017, coley_retro_2018, coley_retro_2020} reaction classification\cite{rxn_classification_2013} and search,\cite{rxn_searching_2002, rxn_searching_2016} and prediction of reaction properties.\cite{digdisc, vangerwen2023equireact} Automatic identification of atom mapping is a challenging task. ILP approaches to this vary primarily on the definition of the objective function: whether based on subgraphs,\cite{bahiense2012maximum, huang2006maximum} the number of bonds breaking and forming,\cite{first_stereochem_2012} or the so-called \textit{Imaginary Transition Structure}\cite{mann2014atom} (ITS,\cite{fujita1986description} or later referred to as the CGR\cite{nugmanov2019cgrtools, varnek2005substructural}), a superimposition of reactant and product structures.
The decision variables are the atom maps from reactant atoms to product atoms. Constraints include mapping atoms exactly once, ensuring balanced reactions, and enforcing stereochemistry. Other ILP-problems in chemistry include chemical equation balancing,\cite{eq_balancing_ILP} bond order assignment,\cite{bond_order_assignment_ILP} automated mass spectra interpretation,\cite{mass_spectra_ILP, mass_spec_ILP_2008} determination of reaction networks,\cite{rxn_networks_ILP_2012, rxn_networks_ILP_2016} and molecular design.\cite{mol_design_MILP_2020, mol_design_MILP_2_2020, mol_design_ILP_2022, mol_design_ILP_2_2022, mol_design_ILP_2021, mol_design_ILP_2002, mol_design_ILP_2007, vaccine_design_ILP_2008} In quantum chemistry, ILP has also been used in deterministic optimization solutions to the Hartree--Fock equations.\cite{Lavor_2007_deterministic, janes2013deterministic}\\

Here, we explore the relevance of ILP in the domain of physics-inspired machine learning (ML).
Physics-inspired machine learning aims at circumventing the resolution of the Schr\"odinger equation by using ML models to infer quantum-chemical molecular properties.\cite{huang2021ab, musil2021physics, deringer_gpr, langer_representations_2022, noe2020machine} Such models often rely upon kernel ridge regression (KRR), exploiting the concept of similarity:\cite{fabregat2022metric, yang_liu_distance_2006} two molecules that are similar in representation space result in a higher kernel similarity (up to 1) whereas dissimilar molecules result in a kernel similarity close to 0. There are several key ingredients in KRR: (i) the vectorial \textit{representation} used to represent/featurize molecules or atoms in molecules, (ii) the \textit{distance metric} used to compare the representations in the kernel and finally (iii) the \textit{data} used to train the model. \\

\textit{Molecular representations} should encode the fundamental physics: often in the form of symmetry constraints and/or physical laws as exemplified by the use of non-linear functions $\vec{X}(\vec{Z}, \vec{R})$ of nuclear charges and atomic coordinates.\cite{Dral2015, huang2021ab, Reddy2021, musil2021physics, deringer_gpr, langer_representations_2022, noe2020machine}
They might be \textit{local} (\textit{i.e.}, one feature vector per atomic environment)\cite{FCHL, bartok2013representing, behler_generalized_2007, drautz2019atomic, nigam2020recursive} or \textit{global} (\textit{i.e.}, a vector per-molecule).\cite{rupp2012fast, hansen2015machine, huo2017unified}
It has been shown that for size-extensive properties such as atomization energies\cite{rupp2012fast, fabregat2022local, huang2020quantum} free energies of solvation,\cite{lemm2023improved} dipole polarizability tensors,\cite{wilkins2019accurate} or electron density\cite{grisafi2018transferable} (or atomic properties such as \textsuperscript{13}C or \textsuperscript{1}H chemical shifts, core electron level shifts and atomic force components\cite{rupp2015machine, huang2020quantum}), local approaches allow for accurate property predictions of larger molecules, using a training set of smaller molecules (``extrapolation'').
Additional steps, such as representation normalization or the use of linear kernels,\cite{jung2020size} need to be employed to enable global variations to extrapolate.
Conversely, if the molecules in the training and test sets are of similar size (``interpolation''), global representations are sufficient.
KRR takes $\vec{X}$ of an unseen (\textit{i.e.}, target) system as input and predicts its properties based on the distance between its representation and those of the molecules in a training set.
It has been shown repeatedly that the choice of distance metric affects the quality of the regression.\cite{hansen2013assessment, bartok2017machine, unke2017toolkit, gallarati_reaction-based_2021}
Metric learning can also be used to learn a dedicated distance metric for a specific regression task.\cite{weinberger_metric_2007, fabregat2022metric} \\

The success of supervised ML methods also heavily relies on a high-quality (\textit{i.e.}, diverse, relevant and non-redundant) labelled \textit{training dataset}. To this end, there has been a growing effort in curating datasets in computational chemistry.\cite{Ullah_2024, bo2018role, nandy2022audacity, draxl2018nomad, smith2017ani, ramakrishnan2014quantum, jain2013commentary, nakata2017pubchemqc, hoja2021qm7, qm7_b, tddft_data, qmugs, electrolyte_genome_project, wahab2022compas}
Nevertheless, accurately predicting properties of unseen target molecules with a model trained
on existing labelled dataset remains a challenge.
Building customized datasets for each new problem
-- including new regions of chemical space and new properties --
is the standard in the field (which is usually more expensive than computing the properties of the target molecules in the first place).
More efficient solutions have included transfer learning and active learning.
Transfer learning\cite{yamada_transfer_2019, jha2019transfer, grambow_2019_transfer, bai_transfer_2020, jackson2021transfer, shim2022transfer, king2024transfer} involves freezing the weights of a model trained on an initial dataset, and continuing training (``fine-tuning'') on an additional dataset that is more relevant to the target problem. Note that training a model from scratch on the tailored dataset still sometimes results in better performance than fine-tuning.\cite{raffel_transfer_transformer_2020, taylor2009transfer, pesciullesi2020transfer}
Given that the model effectively transfers to the new task, this approach offers the possibility of reduced data generation needs.
Active learning\cite{smith2018less, gubaev2018machine, sivaraman2020machine, podryabinkin2017active, reker_active_2015, wen2023active, dodds2024active, huang2020quantum} involves iteratively labelling more data, according to the convergence of the model that is continuously fine-tuned on the added datapoints.
In principle, this approach minimizes the number of additional computations performed for labelling.
\\

\textit{Training set selection} instead aims at choosing a subset from an existing dataset, removing redundant or irrelevant datapoints.
While this also reduces the computational resources needed for labelling,
the emphasis is typically on improving model efficiency.\cite{heinen2023reducing} For instance, so-called \textit{coresets} are shown to guarantee equivalent performance to training on the full dataset.\cite{bachem2017coresets, lucic_coresets_2018, borsos_coresets_2020, mirzasoleiman2020coresets}
Alternatively, a breadth of methods have been shown to enable \textit{stable} machine-learned force fields for molecular dynamics (MD). The emphasis is on selecting maximally diverse training sets, such that a trained ML model is prepared for a range of possible conformations that may be encountered in an MD simulation.
Such methods include: Farthest Point Sampling (FPS),\cite{Dral_2017, imbalzano2018automatic, rossi_solvation_2020} which selects the datapoints with the largest distance between any two points; CUR decomposition,\cite{mahoney2009cur, imbalzano2018automatic, cersonsky2021improving} which is a low-rank matrix approximation (of the matrix of representations of the training set); and Orthogonal Matching Pursuit (OMP)\cite{fabregat2022local} as a greedy (supervised) optimization algorithm.
Browning \textit{et al.}\cite{browning2017genetic} were the first to illustrate that selecting a tailored training set from existing data can \textit{improve}, rather than just maintain, prediction accuracy.
Huang \textit{et al.}\cite{huang2020quantum} developed the Atoms in Molecules (AMONs) approach to construct dedicated, minimal training sets for specific target molecules as an enumeration of the substructures of the graph of the target.
They illustrated that a training set of small molecules (up to 7 heavy atoms) could allow for accurate prediction of size-extensive properties of biomolecules: namely, the atomization energy, and local properties such as atomic charges, NMR shifts, core level shifts and force components.
However, there is a significant computational cost associated with constructing the labelled training dataset for each target molecule. \\

Similarity Machine Learning (SML),\cite{lemm2023improved} introduced by the same group, explores how to select an optimal (w.r.t.\ a certain objective function) training subset from existing datasets (namely QM9\cite{ramakrishnan2014quantum} and Enamine REAL\cite{enamine}).
SML is based on the notion of \textit{similarity} that is critical to KRR and other local learning algorithms:\cite{vapnik_local} by selecting molecules the closest to the target (based on the distance between their representations), an improvement is found over random selection.
While SML circumvents the expense of labelling a new dataset, it focuses on the prediction of properties of molecules within the same datasets as used for training (with an appropriate train/test split). A more general use of SML would imply the availability of data that perfectly resembles the chemistry of the target molecule (\textit{i.e.}, the ML model only needs to interpolate), which is not compatible with out-of-sample applications. Another issue that limits the approach to the interpolation setting is the notion of global similarity on which the selection algorithm relies. Finally, the \textit{greedy} selection \textit{i.e.},\ sequential in terms of their distance ranking, does not necessarily guarantee an optimal solution.\\

This work addresses the limitations of other training set selection algorithms by introducing an ILP-based solution that rigorously selects an optimal training subset from an existing database of small molecules. Then, we use the selected training sets to build lightweight KRR models to predict the atomization energy of larger organic molecules, a representative application of ML in chemistry.\cite{rupp2012fast,rupp2015machine,huang2020quantum,ramakrishnan2014quantum} Conceptually, our approach builds on the SML work\cite{lemm2023improved} by searching for \textit{similar} environments in an unsupervised fashion, but uses \textit{local} representations to measure similarity between \textit{atomic environments} rather than global ones to look for similar molecules.
The proposed algorithm can tackle realistic \textit{extrapolation} scenarios, necessarily more challenging than interpolation, but also most valuable, as computations of quantum-chemical molecular properties scale unfavorably with system size.
\textit{ILP}\cite{schrijver} is used to find atomic environments similar to those in the target molecule, which allows for the identification of globally optimal (rather than greedy) solutions.
Our approach ultimately constructs an atom mapping between atoms in the target molecule and a subset of atoms in a molecular database.
Unlike previous uses of ILP, which identify atom maps,\cite{bahiense2012maximum, huang2006maximum, first_stereochem_2012, mann2014atom} to solve the traditional task of recognizing a pseudo-reaction mechanism, here, we leverage similar atomic environments for similarity-based machine learning.
Our integer programming model is based on a well-adapted variant of the \emph{bipartite matching problem} which has an efficient ILP-formulation, \cite{wolsey2014integer} in which we incorporate penalization of undesirable solutions as well as forcing new molecules into the solution.
We illustrate the advantages of our training set selection algorithm in several train/test scenarios: training on QM7\cite{rupp2012fast} molecules and testing on held-out QM7 molecules, larger QM9\cite{ramakrishnan2014quantum} molecules and finally larger drug-like molecules. We compare to existing unsupervised training set selection approaches, namely SML,\cite{lemm2023improved} FPS,\cite{Dral_2017, imbalzano2018automatic, rossi_solvation_2020} CUR\cite{mahoney2009cur, imbalzano2018automatic} and random selection.

\section{Theory}
The goal is to infer the properties of a large molecule (the \textit{target}) using a minimal training set of $N$ small molecules (\textit{fragments}), selected from a pre-existing, larger, molecular database likely containing redundant and/or irrelevant instances. In other words, the selected training set should be tailored to the specific chemistry of the target to achieve a good performance despite its reduced size.

\begin{figure}[h!]
    \centering
    \includegraphics[width=.92\textwidth]{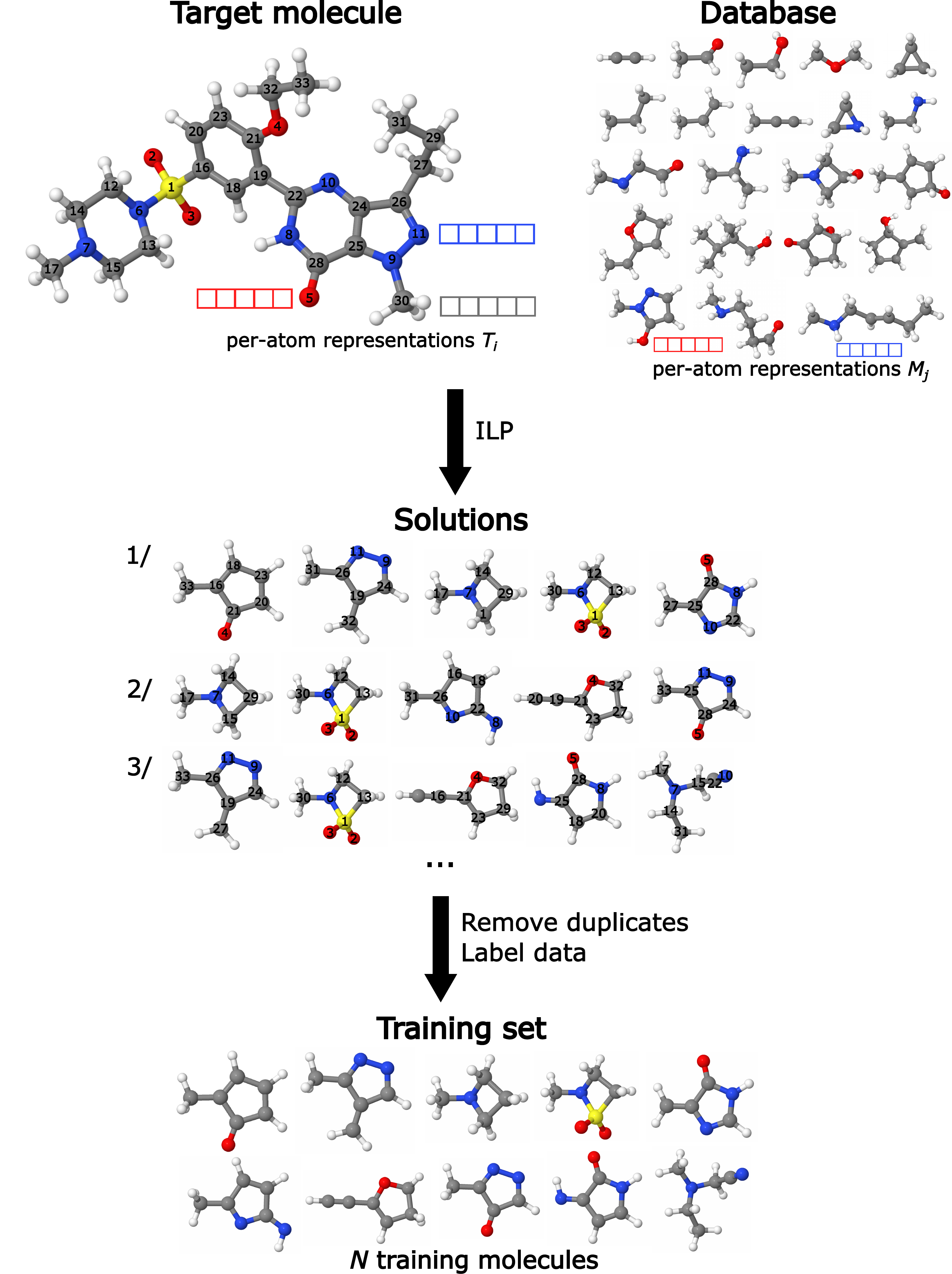}
    \caption{Outline of our approach to identify an optimal training set from an existing database (here, QM7\cite{rupp2012fast}), for a dedicated target molecule (here, sildenafil). Target and fragment molecules are represented with atom-centered feature vectors, $\vec{T}_i$ and $\vec{M}_j$ respectively, where $i$ and $j$ are the corresponding atom indices. With ILP, we find solutions as combinations of atoms in fragment molecules that best match the target atoms. Here, the top three solutions (smallest objective value) are shown. After removing duplicates, and labelling molecules with properties if necessary, an optimal training set is obtained.
    \cite{puck_thesis}}
    \label{fig:method-outline}
\end{figure}

Figure~\ref{fig:method-outline} summarizes the approach: a target molecule $T$ (here, the molecule sildenafil) is described by a molecular representation formed by atom-centered feature vectors (labelled $\vec{T}_i$ in the Figure, where $i$ indexes each atom in the target). Similarly, all of the atoms in a database of fragments $D$ are represented as $\vec{M}_j$ (with $j$ as the atom index).
\textit{Solutions} as labelled in the Figure (and $F$ in Algorithm~\ref{algo:1}) correspond to the output of the ILP algorithm: an \textit{atom mapping} $\pi$ between atoms of the target molecule and atoms in chosen fragment molecules. The map is selected based on the Euclidean distance between representations of atomic environments in the target and fragments. Algorithm~\ref{algo:1} presents the general ILP approach and its implementation is detailed in Section~\ref{sec:algo}. \\
\clearpage

\clearpage
\begin{algorithm}[h!]
\small{
    \hspace*{\algorithmicindent} \textbf{Input}: database $D$ of fragments, target $T$, size $N$. \\
    \hspace*{\algorithmicindent} \textbf{Output}: selected subset of $D$ of size $N$.
    \begin{algorithmic}[1]
        \State Initialize the set $S=\emptyset$. All elements of $S$ are unique.
        \While{$S$ has less than $N$ elements}
            \State Compute an atom mapping $\pi$ from $T$ to $D$ of near-minimal value.
            \State The solution $F$ is the set of fragments of $D$ mapped to by $\pi$.
            \State Add each fragment of $F$ to $S$ while recording the value of $\pi$.
            \State Query the ILP algorithm for new fragments.
        \EndWhile
        \State Order the elements of $S$ with increasing value.
        \State Return the first $N$ elements of $S$.
    \end{algorithmic}
    }
    \caption{Subset selection}
    \label{algo:1}
\end{algorithm}
Solutions resemble decomposition ``reactions'' of the target molecule, since we enforce that all atoms in the target molecule must be matched by fragment atoms, in each of the solutions (Section~\ref{sec:algo}). In Figure~\ref{fig:method-outline}, the top 3 solutions are shown (as measured by their objective value).
Two additional steps are undertaken to transform these solutions into a training set for supervised machine learning. First, since the same molecules can appear in several ILP solutions, the set of unique molecules is extracted.
Second, if the initial dataset was unlabeled (\textit{i.e.},\ consisted only of structures and no property labels), the unique set of molecules is labelled with properties at this stage. The algorithm is implemented (Section~\ref{sec:constraint_new_solns}) to search for solutions until $N$ molecules are selected. The labelled set of $N$ small molecules is then used as the training set for supervised ML. In Section~\ref{sec:results}, we construct learning curves where $N$ is increased systematically until $N_{\max}=1024$ (\textit{i.e.}, a sufficient amount of data to train an accurate model), to showcase the effectiveness of small-to-medium sized training sets selected via the different algorithms.\\

\begin{figure}[!h]
    \centering
    \includegraphics[width=0.45\textwidth]{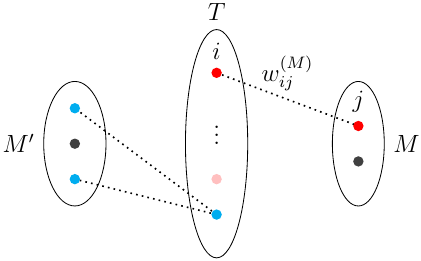}
    \caption{Weighted bipartite graph corresponding to a target $T$ and database with two elements $D = \{M, M'\}$.
    For each fragment $M \in D$ and for each atom $i \in T$ and $j \in M$ with same atom types,
    the edge between $i$ and $j$ has cost $w_{ij}^{(M)} = \Vert \vec{T}_i - \vec{M}_j \Vert_2^2$, where $\vec{T}_i$ and $\vec{M}_j$
    are the representations of each atom.
    }
    \label{fig:loc-vect}
\end{figure}

The problem of optimal mapping is represented in Figure~\ref{fig:loc-vect} as a so-called weighted bipartite graph.\cite{schrijver} Such a graph has its vertices divided into two disjoint sets: the atoms of the target molecule $T$, and the atoms of the fragments, corresponding here to the two database molecules $M$ and $M'$. Its edges connect vertices from the disjoint sets whenever their corresponding atoms match, with a weight equal to the distance of the corresponding endpoints in the representation space.
Finding a maximum matching of minimal weight is achievable polynomial in time (with respect to the number of vertices) using the \emph{Hungarian method} through augmenting paths, which does not require ILP.\cite{schrijver}
However, formulating the problem with ILP enables the incorporation of additional constraints and modification of the objective value. For example, we may force particular fragments to appear in the solutions, or penalize undesired ones. Our choices of constraints are detailed in Sections \ref{sec:algo}--\ref{sec:penalty}.
The results shown in Section~\ref{sec:results} demonstrate the effectiveness of this approach.

\subsection{ILP formulation}
\label{sec:algo}

\begin{figure}[!h]
    \centering
    \includegraphics[width=.99\textwidth]{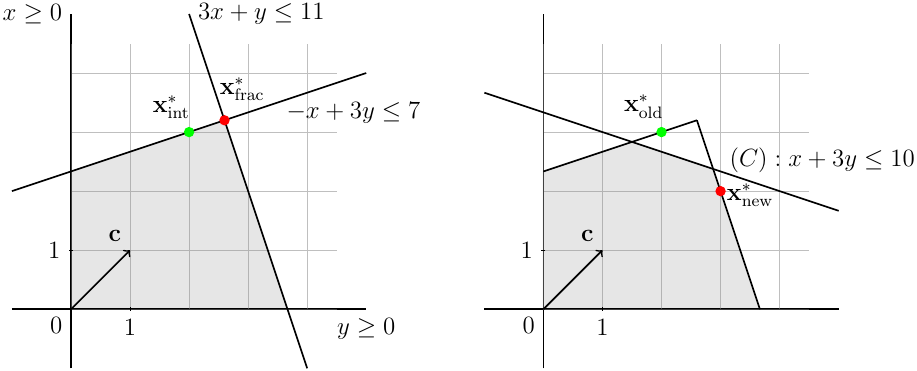}
    \caption{Example of linear programs in two dimensions. Left: The point $\vec{x}^*_\mathrm{frac}$ maximizes $\vec{c}^\top \vec{x}$ for points $\vec{x}\equiv(x,y)$ restricted to the grey area. The point $\vec{x}^*_\mathrm{int}$ is a maximum within the grey area intersected with the $\mathbb{Z}^2$ grid. Right: An additional constraint $(C)$ is added to the ILP algorithm from the left, making the old maximum $\vec{x}^*_\mathrm{old}$ infeasible, leaving $\vec{x}^*_\mathrm{new}$ as the only integer maximizer of $\vec{c}^\top \vec{x}$.}
    \label{fig:IP-vs-ILP}
\end{figure}

An ILP optimization problem has the form
\begin{align}
    \text{maximize} \qquad\qquad & \vec{c}^\top \vec{x} \label{eq:max} \\
    \text{subject to} \qquad\qquad & \vec{A}\vec{x} \leq \vec{b}, \label{eq:linear}\\
     & \vec{x} \in \mathbb{Z}^n, \label{eq:ILP_constraint}
\end{align}
where $\vec{A} \in \Re^{m \times n}$, $\vec{b} \in \Re^m$, and $\vec{c} \in \Re^n$. In other words, \eqref{eq:max} and \eqref{eq:linear} describe a linear optimization objective with linear constraints, and universal quantifier \eqref{eq:ILP_constraint} describes the ILP-specific constraint on the integer variables. Figure~\ref{fig:IP-vs-ILP} illustrates the difference between LP and ILP problems. In both cases, there is only one global minimum since for every point inside the grey polyhedron there is a direction that improves the objective.

Our problem setting is formulated using ILP as follows.
For each possible atom mapping, we construct a tuple $(i,j,M)$ with $i \in T$, $M \in D$, and $j \in M$ of the same atom type as $i$.
Let us denote $i\overset{M}{\mapsto}j$ if atom $i \in T$ is mapped to $j \in M$. We introduce the following decision variable $\vec{x}$ as
\[
    x_{ij}^{(M)} = \begin{cases} 1 \text{ if } i \overset{M}{\mapsto} j, \\ 0 \text { otherwise.} \end{cases}
\]
This expression gives the interpretation of $\vec{x}$ and not its definition: we deduce the mapping from the values of the output.
The cost of each mapping is defined by the $L^2$ distance between the mapped atoms in the representation space.
The problem is thus formulated as follows:
\begin{align}
\text{minimize }  \qquad \qquad & \sum_{\mathclap{\substack{i \in T, \\ M \in D, ~ j \in M}}} \Vert \vec{T}_i - \vec{M}_j \Vert_2^2\  x_{ij}^{(M)} && \label{expr:obj}  \\[10pt]
\text{subject to } \qquad \qquad & \sum_{\mathclap{M \in D, \ j \in M}} x_{ij}^{(M)} = 1 && \forall i \in T, \label{eq:bij} \\[10pt]
    			   & \sum_{i \in T} x_{ij}^{(M)} \leq 1 && \forall M \in D, ~ \forall j\in M, \label{eq:inj} \\[10pt]
			         & x_{ij}^{(M)} \in \{0,1\} && \forall i \in T, ~ \forall M \in D, ~ \forall j \in M. \label{var:xint}
\end{align}
Expression~\eqref{expr:obj} is equal to the sum of squares of the cost associated to each individual mapping. Conceptually, it implements the search for \textit{similar} atom-centered environments to those in the target molecule (using Euclidean similarity).
Technically, it corresponds directly to the cost of a choice of weighted edges in the bipartite graph illustrated in Figure~\ref{fig:loc-vect}.\\

Equation~\eqref{eq:bij} imposes that each atom of the target must be matched exactly once. This ensures that all atoms in the target molecule are well-represented by appropriate fragment environments.

Replacing the equal sign in \eqref{eq:bij} by a ``$\leq$'' would make the algorithm map singular atoms of the target. Such a \emph{closest-point search} over atoms would not provide solutions representing the target in its entirety, which is necessary to predict global (per-molecule) properties, as is our focus here. However, the closest-point search variation would be well-suited to select local environments to predict local (per-atom) properties of the target, such as atomic charges, force components, NMR shifts or core level shifts.\\

Inequality~\eqref{eq:inj} imposes that each atom of the database must be matched at most once. This effectively diversifies the selected fragment pool, increasing the efficiency of the search algorithm (since it runs until $N$ fragment molecules are found).
Equation~\eqref{var:xint} enforces our particular ILP setting: atoms can either be mapped (1) or not (0).

\subsection{Imposing new molecules in each solution}
\label{sec:constraint_new_solns}

Once a solution is found by the ILP algorithm, fragments are added to the generated subset $S \subset D$, and the procedure is repeated.
Without additional constraints, individual solutions might consist of the same sets of fragments with different atom maps.
Our goal is the creation of a training set of $N$ molecules for supervised ML models of global (per-molecule) properties, rather than directly making use of the atom-mapping solutions. Therefore, our algorithm is more effective (similarly to inequality \ref{eq:inj} in Section~\ref{sec:algo}) if each of the ILP solutions contain at least one new fragment.
This is imposed with a \emph{lazy constraint}.\\

We first introduce indicator variables $\{y_M\}_{M \in D}$  that attest whether a fragment is taken.
The original ILP formulation with forced new fragments may be rewritten as follows:
\begin{align}
\text{minimize } \qquad \qquad & \sum_{\mathclap{\substack{ i \in T, \\ M \in D, ~ j \in M}}} \Vert \vec{T}_i - \vec{M}_j \Vert_2^2 \ x_{ij}^{(M)} \nonumber  \\[10pt]
\text{subject to} \qquad\qquad & \eqref{eq:bij}, \eqref{eq:inj}, \text{ and } \eqref{var:xint}, \nonumber \\[10pt]
			& \left\{ \begin{aligned}
			y_M &\leq \sum_{\substack{i \in T, \\ j \in M}} x_{ij}^{(M)} && \forall j \in M,  \\[10pt]
			y_M &\geq x_{ij}^{(M)} && \forall i \in T, ~ \forall M \in D, ~ \forall j \in M.
			\end{aligned}\right.
            \label{eq:log_disj} \\[10pt]
			& \sum_{\mathclap{M \in D \setminus S}} y_M \geq 1, \label{eq:force} \\[10pt]
			& y_M \in \{0, 1\}  \quad\ \qquad \forall M \in D. \label{var:yint}
\end{align}

Inequalities \eqref{eq:log_disj} express the logical disjunction
$ y_M = \bigvee\limits_{\mathclap{\substack{i \in T, ~ j \in M}}} ~ x_{ij}^{(M)}$: $y_M = 1 $ if and only if at least one of the $x_{ij}^{(M)} = 1$.

Inequality~\eqref{eq:force} requires that at least one fragment not in the set $S$ be taken (effectively imposing new molecules in each of the solutions).
Using this formulation, only inequality~\eqref{eq:force} has to be updated each time a new solution is found.

\subsection{Penalizing undesirable solutions}
\label{sec:penalty}

In the present ILP formulation, solutions contain fragments that introduce additional atoms relative to the those in the target molecule. As discussed in Section~\ref{sec:algo}, selecting the entire molecule instead of only the best matching atoms is unavoidable since the ML model is trained on global (per-molecules) energies.
These spare atoms may reduce the efficiency of the chosen subsets given that only the mapped portions of the fragments are known to contain environments that are similar to the target atoms.
Therefore, we introduce a penalty for the size difference between the target and the fragments.
For this, denote by $|M|$ the number of non-hydrogen atoms of molecule $M$, and $|T|$ the number of non-hydrogen atoms of the target $T$, and let $p \geq 0$ be a constant of choice.
The final ILP algorithm is given then by the following,
\begin{align}
\text{minimize }  &&\sum_{\mathclap{\substack{i \in T, \\ M \in D, ~ j \in M}}} \Vert \vec{T}_i - \vec{M}_j \Vert_2^2 \ x_{ij}^{(M)} \ + \ &p \cdot \left[ \sum_{M} \left( |M| \cdot y_M \right) - |T| \right]\label{expr:objpen}  \\[10pt]
\text{subject to } & &\eqref{eq:bij} - \eqref{var:yint}.& \nonumber
\end{align}

In Section~\ref{sec:results}, we present results for penalty constants $p=0$ and $1$. Note that an alternative penalty could be written as a function of the number of non-hydrogen atoms added by the fragments.

\section{Computational details}
\subsection{Datasets}\label{sec:datasets}
\begin{figure}[h!]
    \centering
    \includegraphics[width=\textwidth]{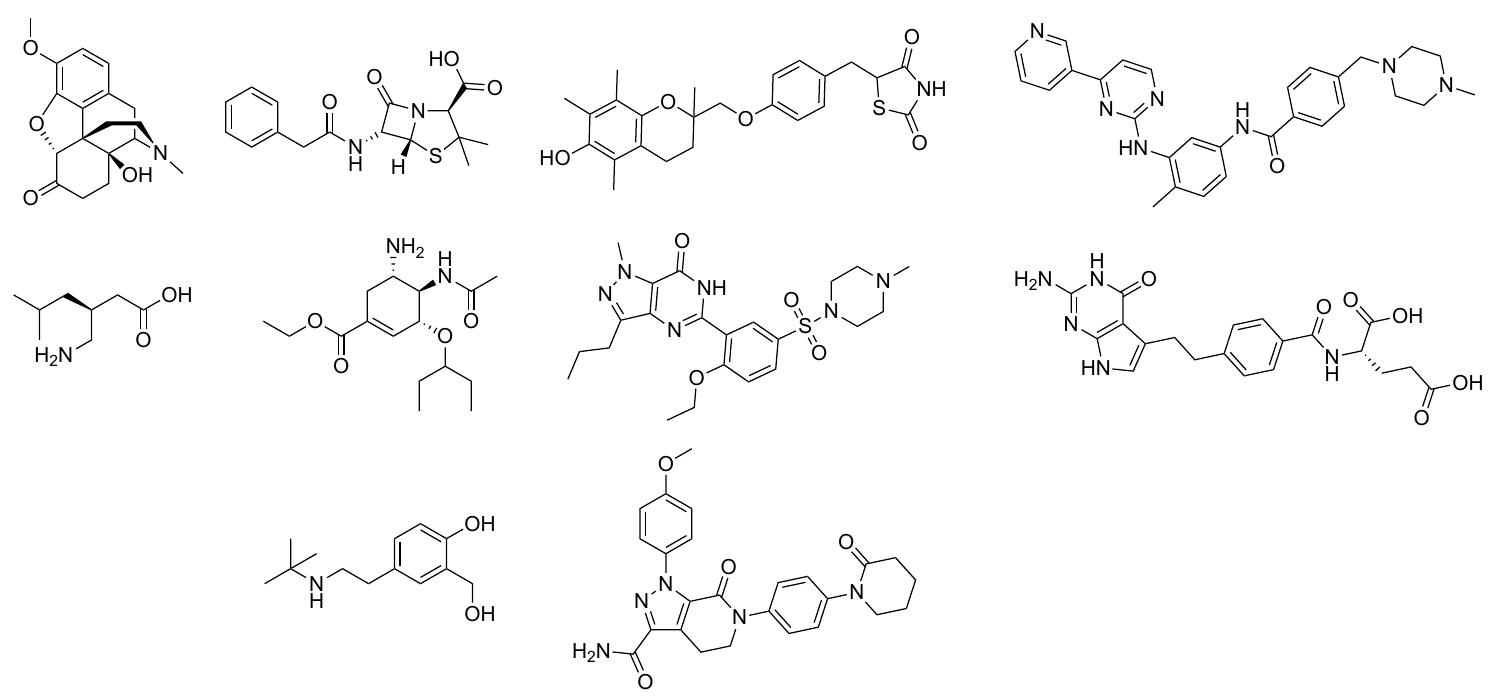}
    \caption{2D structures of the ten drug molecules used as targets. From left to right; 1st row: oxycodone, penicillin, troglitazone, imabinib; 2nd row: pregabalin, oseltamivir, sildenafil, pemetrexed; 3rd: salbutamol, apixaban.}
    \label{fig:drug_mols}
\end{figure}

Experiments are performed using three datasets: QM7,\cite{rupp2012fast} QM9\cite{ramakrishnan2014quantum} and ten drug molecules (illustrated in Figure~\ref{fig:drug_mols}). In all cases, models are trained on QM7, which contains small organic molecules (with seven or fewer atoms of elements C, N, O, S).
Ten molecules, containing eight or nine C, N, O atoms,
were selected from QM9\cite{ramakrishnan2014quantum}
to constitute the first extrapolation test QM9*.
Ten common drug molecules with heavy-atom types C, N, O, S were chosen from the literature and constitute the more challenging extrapolation scenario, with an average of $27 \pm 8$ heavy atoms.

\begin{figure}[h!]
    \centering
    \includegraphics[width=0.8\textwidth]{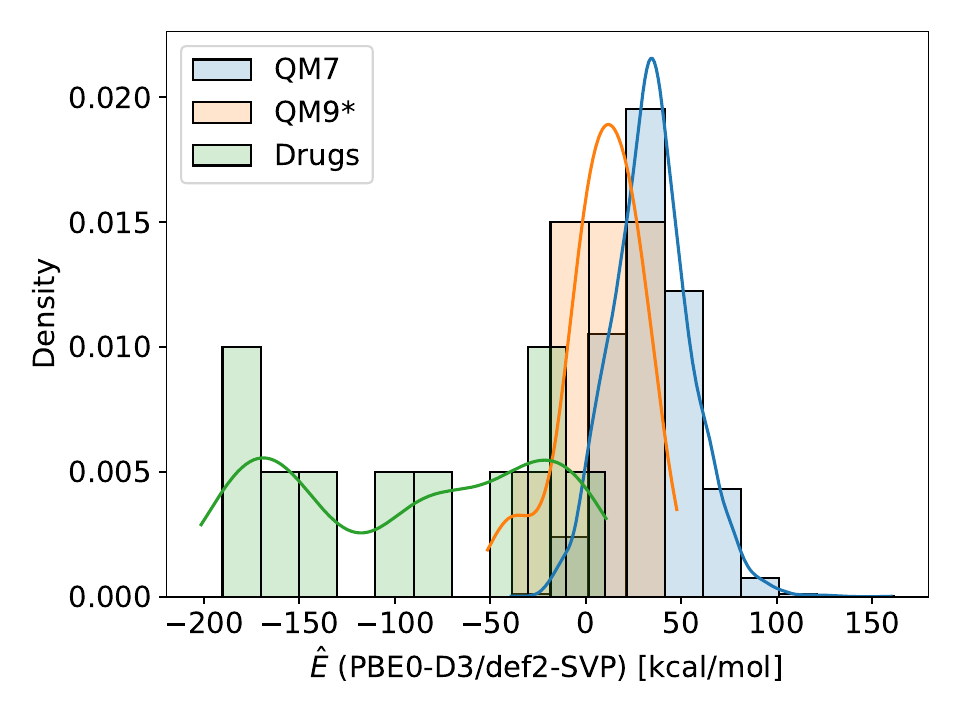}
    \caption{Density plot (\textit{i.e.}, $d = \frac{c}{\Delta b \sum{c}}$, where $d$ is density, $c$ is counts, and $\Delta b$ is the discrete difference between the number of bins, as implemented in \texttt{matplotlib.pyplot.hist} with \texttt{density=True}) of energy correction $\hat{E}$ of QM7 molecules, ten random QM9 molecules (QM9*), and ten drug molecules, computed using PBE0-D3/def2-SVP. The smooth curves are continuous fits to the histograms using kernel density estimates, as implemented in \texttt{sns.kdeplot}.}
    \label{fig:data_dists}
\end{figure}

QM7 molecules were re-optimized at the PBE0-D3/def2-SVP level\cite{Adamo1999,Weigend2005,Grimme2010} with ORCA version 5.0,\cite{Neese2022} starting from the original geometries.\cite{rupp2012fast} Analogously, the QM9* and drug molecules (\textit{vide infra}) were optimized at the PBE0-D3/def2-SVP level using the same version of ORCA for consistency.
Figure~\ref{fig:data_dists} illustrates a density plot of the total energy of molecules in QM7, the slightly larger molecules in QM9 (QM9*) and the drug molecules (Drugs).
These distributions illustrate that depending on the train/test scenario, the predictions may be interpolative (which is straightforward for ML models), or extrapolative (challenging for ML models).
Training on QM7 and predicting on QM7 is interpolative and the most straight-forward prediction scenario, similar to the experiments performed in the SML work\cite{lemm2023improved}.
The distribution of QM9* falls partially outside of that of QM7: training on QM7 and predicting on QM9* corresponds to what we call ``limited'' extrapolation.
The distribution of the drug molecules falls almost entirely outside of that of QM7: training on QM7 and predicting on the drug molecules is pure extrapolation. Each of these experiments will be tested in Section~\ref{sec:results}.

As a size-extensive quantity, the total energies $E$ are the key property of interest.
In order to facilitate the machine learning of a large range of absolute energies, the target energies are transformed as follows. A linear regression model is fitted on QM7 to learn average contributions for \ce{H}, \ce{C}, \ce{N}, \ce{O}, and \ce{S}:
\begin{equation}
    E_I \approx a N^{\ce{H}}_I + b N^{\ce{C}}_I + c N^{\ce{N}}_I + d N^{\ce{O}}_I + e N^{\ce{S}}_I,
\end{equation}
where $E_I$ is the single-point energy, $a, b, c, d, e$ are the coefficients learned by the model
(in other words, dressed atomic energies)
and $N^{\ce{X}}_I$ is the number of atoms of type $\ce{X}$ in each molecule $I$. With the learned coefficients, total energies are then transformed:
\begin{equation}
    \hat{E}_I = E_I - a N^{\ce{H}}_I - b N^{\ce{C}}_I - c N^{\ce{N}}_I - d N^{\ce{O}}_I - e N^{\ce{S}}_I.
\end{equation}
The energy ``correction'' terms $\hat{E}$ are the quantities learned by the ML models in the next sections. The total energies are recovered by adding the correction term to the sum of the dressed atomic energies back per-molecule.

\subsection{Molecular representations}
\label{sec:representations}

In this work, molecular representations are used for two distinct tasks. The representation first serves as a basis to evaluate the molecular similarity (or the similarity of the atomic environments) in the feature space, in order to select molecules to constitute the training set. The selected molecules are then used to train the ML model, which itself takes a representation as input for the prediction of the molecular properties.
For the first task, the training set is generated using either a global or local representation of molecules, depending on the selection algorithm. Our ILP algorithm uses local representations. The other methods tested (Section~\ref{sec:baseline_methods}) instead use global representations.
For the property prediction task, regardless of whether the selection method used local or global representations, we use local representations. This effectively partitions the global energy property into atomic contributions, and enables the prediction of energies of larger molecular systems (Section~\ref{sec:KRR}).\\

Global representations correspond to a single vector of size $L$ per molecule $\tilde{\vec{X}} \in \Re^L$, such that a dataset of representations of $N$ molecules is a matrix $\tilde{\vec{D}} \in \Re^{N \times L}$.
Local representations constitute a matrix of dimensions $n_\mathrm{at} \times L$ per molecule, \textit{i.e.},\ $\vec{X} \in \Re^{n_\mathrm{at} \times L}$ where $n_{\mathrm{at}}$ is the number of atoms in the molecule. A dataset of local representations of $N$ molecules is a tensor of rank 3, $\vec{D} \in \Re^{N \times \max(n_\mathrm{at}) \times L}$. The $\max(n_\mathrm{at})$ pads the first dimension with zeros such that molecules of different size can be accommodated.\\

For both the selection and training tasks, we use the FCHL19 representation.\cite{FCHL, christensen2020fchl}
computed with the \texttt{qml}\cite{qml} package.
FCHL19 is an atom-centered (local) representation, molecules are each represented as a matrix $\vec{X}$ by default. The global variant of the representation is constructed using a sum over the atoms:
\begin{equation}
    \tilde{\vec{X}} = \sum_{i}^{n_{\mathrm{at}}} \vec{X}_i \label{eq:sum_fchl}.
\end{equation}

Our choice of representation is motivated by the fact that the FCHL19 representation, constructed purely from molecular geometry, is comparatively less expensive to obtain than the property of interest. Nevertheless, note that our algorithm is compatible with any other local representation or featurization strategy.

\subsection{Machine learning models}
\label{sec:KRR}

KRR models are used to make property predictions.
Energy corrections $\hat{E}$ are predicted according to
\begin{align}
    \hat{E}(\vec X_I) &= \sum_{J=1}^{\mathclap{N}} \alpha_J K(\vec{X}_I, \vec{X}_J), \\
    K(\vec{X}_I, \vec{X}_J) &= \sum_{i \in I} \sum_{j \in J} \exp \big(- ||\vec{X}_{Ii} - \vec{X}_{Jj}||_2^2 / 2 \sigma^2 \big),
\end{align}
where the global kernel $K$ decomposes the global target property $\hat{E}$ into a sum of local contributions. Per-atom representations $\vec{X}$ are used directly to compare atoms $i$ and $j$ in molecules $I$ and $J$ with a Gaussian kernel.
The coefficients $\bm{\alpha}$ are learned from the training set,
\begin{equation}
    \bm{\alpha} = (\vec{K} + \lambda\vec{I})^{-1}{\hat{\vec E}}^{\mathrm{train}},
\end{equation}
where $\lambda$ is a regularization coefficient, which is a hyperparameter to be tuned alongside the Gaussian width $\sigma$.

\subsection{Learning curves and hyperparameters}
Learning curves are generated by taking seven increasing fractions of the training data. The number of molecules in the training set $N$ is increased up to a total of $N_{\max}=1024$ training points. Hyperparameters (the kernel width $\sigma$ and regularization coefficient $\lambda$) are then optimized by further splitting the training data into a train and validation set.
The hyperparameters are parametrized as
$\sigma = 10^{n_\sigma / 8}, n_\sigma \in \mathbb{N} $
and
$\lambda = 10^{-n_\lambda}, n_\lambda \in \mathbb{N}$,
and
optimized over grids
$0 \leq n_\sigma \leq 24 $
and $4 \leq n_\lambda \leq 9$,
for each training set size.
The test data is always the target molecule of interest. In other words, energy predictions are made for a single target molecule at a time, in contrast to typical ML workflows where predictions are made for several test molecules at once. The optimal hyperparameters are listed in Table~\ref{S-tab:hypers_KRR} of the supplementary information.

Learning curves correspond to energies averaged over ten target molecules (whether held-out QM7 molecules, QM9* molecules or drug molecules) to illustrate how different training set selection approaches perform on average. The individual learning curves for the ten drugs molecules are given in Section~\ref{S-sec:learning_curves} of the supplementary information.

For the random sampling baseline, learning curves were averaged over five random shuffles of the training data. The error bars show the standard deviation of the predictions across shuffles of the training data.

\subsection{ILP implementation and parameters}
The optimization algorithms are implemented using the Gurobi Optimization \cite{gurobi} software, which has state-of-the-art solvers for ILP. We make use of the python interface \texttt{gurobipy}, version 10.0.2, which wraps around Gurobi's optimization algorithms written in C.

A callback function is specified in \texttt{Model.optimize()} in order to count the total number of unique molecular fragments already found, and add the corresponding lazy constraint thought \texttt{Model.cbLazy()}.
The algorithm runs until $N_S$ unique fragments are stored, as described by Section~\ref{sec:algo}.

Hydrogen atoms were pruned as a pre-processing step. While this lowers the precision in the true objective value of the solutions, it significantly speeds up the solving process.

\subsection{Baseline methods}
\label{sec:baseline_methods}

We compare the proposed ILP-based selection to existing methods for unsupervised training set selection: random selection, CUR,\cite{mahoney2009cur, imbalzano2018automatic} SML\cite{lemm2023improved} and FPS.\cite{Dral_2017, imbalzano2018automatic,rossi_solvation_2020}

Random selection consists of randomly sampling the training data for $N$ molecules.
CUR decompositions are low-rank matrix decompositions, expressed in terms of a subset of actual columns and/or rows of the data matrix.\cite{mahoney2009cur} We use the implementation in the python package \texttt{scikit-matter},\cite{goscinski2023scikit} version 0.2.0. CUR makes use of global representations of the target and fragment molecules.
SML selects the $N$ datapoints with the smallest distance to the target molecule, according to global representations of datapoints and target. We implemented SML using the python package \texttt{numpy}.
Finally, FPS selects the $N$ molecules with the largest distance between any two points, according to the global representations of the molecules. FPS was implemented using \texttt{gurobipy} according to the following ILP approach:

Let $\{\vec{X}_1, \dots, \vec{X}_{n_D}\}$ be $n_D$ representations of the molecules in a dataset (\textit{i.e.}, a global representation per-molecule),
and $C$ the maximal Euclidean distance between any two points on the set.
The following mixed integer program extracts $N$ points (\textit{i.e.},\ the training subset) with maximum minimal distance between any pair:
\begin{align}
\text{maximize } \qquad & y && \\[10pt]
\text{subject to }  \qquad &\sum_{i=1}^{N_D} x_i = N, && \\[10pt]
            & y \leq ||(\vec{X}_i - \vec{X}_j)||_2^2 + C \cdot (2-x_i-x_j), && \forall i,j=1, \dots, N_D, \\[10pt]
            & x_i \in \{0, 1\} && \forall i=1,\dots,N_D, \\[10pt]
            & y \in \Re.  &&
\end{align}

While proving optimality of a subset takes a long time in general, the heuristic solutions provided by Gurobi are better than a naive greedy approach typically implemented for uses of FPS.\cite{Dral_2017, imbalzano2018automatic,rossi_solvation_2020}

\section{Results and Discussion}\label{sec:results}

\subsection{Prediction of molecular properties}
\label{sec:predictions}

This section tests the performance of the proposed ILP approaches: with and without penalty constants ($p=1$ or $p=0$). As introduced in Section~\ref{sec:penalty}, the penalty constant $p$ penalizes solutions that introduce spare atoms compared to the target molecule for which the property is predicted. For example, consider that the target molecule is an alkyl chain with a carboxylic acid unit (\ce{-COOH}), and that two fragment molecules in the dataset contains the same unit, but the first with few additional atoms (\textit{e.g.}, \ce{HCOOH}) and the second attached to ``unwanted'' atoms (\textit{e.g.}, \ce{PhCOOH}, where \ce{Ph} is the phenyl ring). Without the penalty constant, the algorithm arbitrarily selects either of the two molecules. However, the penalty constant prevents the introduction of spare atoms and thus penalizes the selection of \ce{PhCOOH}. \\

The ILP variants are compared to SML,\cite{lemm2023improved} FPS,\cite{Dral_2017, imbalzano2018automatic,rossi_solvation_2020} CUR,\cite{mahoney2009cur} and random subset selection on the prediction of the energy corrections $\hat{E}$ of the target molecules, which are increasingly large compared to the training set molecules (vide infra). FPS was devised to ensure \textit{diversity} by selecting the most dissimilar datapoints in a training set. It is often used to select a representative subset of a dataset for visualization,\cite{de_comparing_2016, fabregat2020hamiltonian} labelling,\cite{gallarati2024genetic, fabregat2020hamiltonian, imbalzano2021uncertainty, raimbault2019using} and/or training of molecular dynamics potentials.\cite{celerse2024organic, rossi_solvation_2020, mouvet2022recent} Diversity or dissimilarity in the representation space generally leads to dissimilarity in the property space. Therefore, maximizing dissimilarity of the training set is not necessarily best suited to improve the accuracy of an ML model for a dedicated target molecule.\cite{helfrecht2020structure, fabregat2022local} CUR\cite{mahoney2009cur} is not explicitly designed to maximize dissimilarity but rather to best approximate the full data matrix. This makes it conceptually more suitable than FPS for improving the property predictions of a target molecule, but at the same time not tailored for the task either. SML,\cite{lemm2023improved} on the other hand, was specifically devised to solve the problem tackled here: optimizing low-data regime predictions on dedicated target molecules. We will therefore focus in particular on the differences between the results with SML and our ILP approach. Random selection serves as a sanity check: for any of the training set selection methods to be useful, they should perform better than random. In all cases considered here, the training set is selected from the QM7 database which contains 7165 molecules in total.

\subsubsection{Interpolation and ``limited'' extrapolation tasks}
\begin{figure}[h!]
    \centering
    \includegraphics[width=\linewidth]{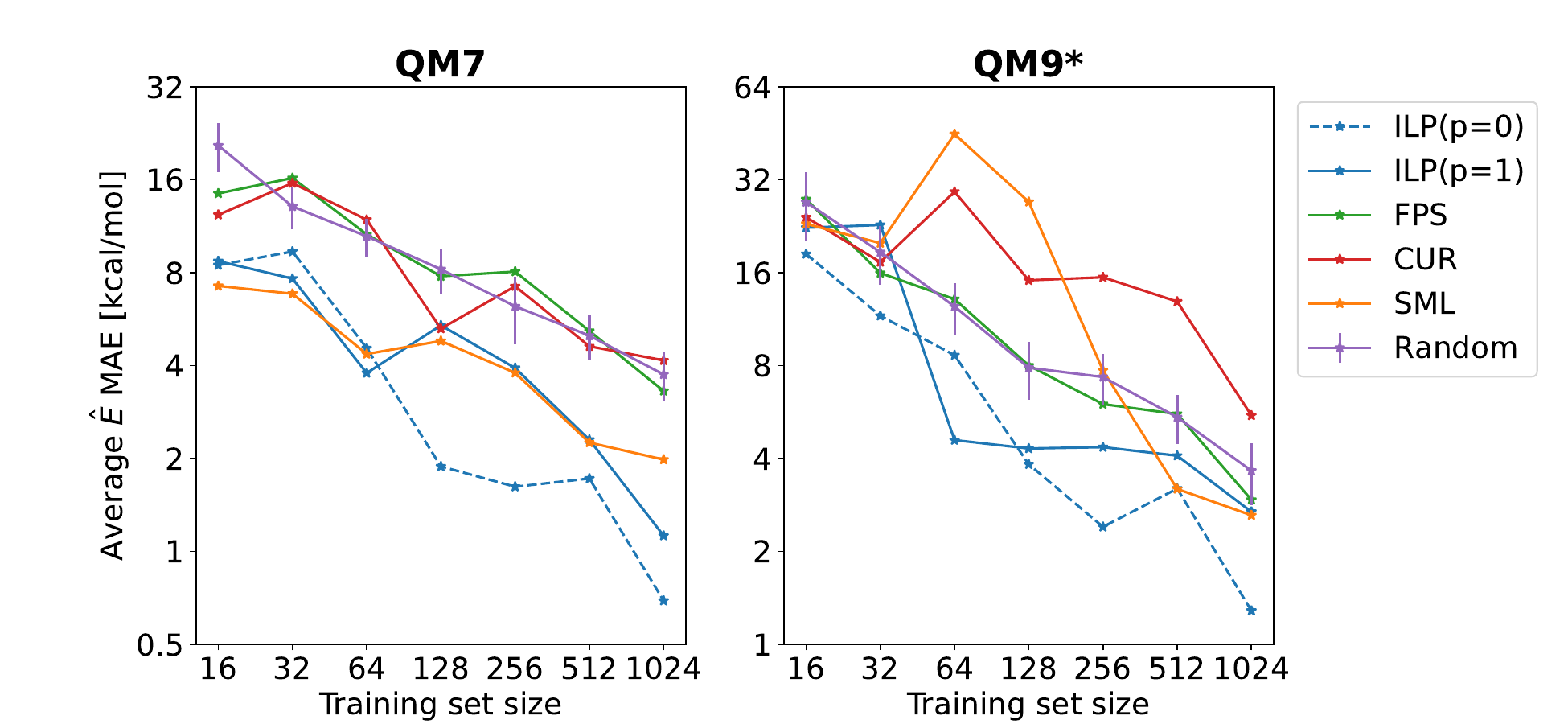}
    \caption{Learning curves demonstrating how the mean absolute error (MAE) of predictions of $\hat{E}$ varies as a function of training set size $N$ for two train/test setups. Left: training on QM7 molecules and testing on (ten held-out) QM7 molecules and right: training on QM7 molecules and testing on ten QM9* molecules. Models include the ILP algorithm (this work) with ($p=1$) and without ($p=0$) penalty term, SML, FPS, CUR, and random selection.}
\label{fig:combined}
\end{figure}

The first experiment is an \textit{interpolation} task: training on QM7 molecules and predicting $\hat{E}$ of ten (held-out) test molecules from the same dataset. The test molecules resemble the size (up to seven heavy atoms) and composition (C, N, O, S) of those in the training set. The second task is training on QM7 molecules and predicting $\hat{E}$ of QM9*. This constitutes an ``limited'' extrapolation task, since the distribution of $\hat{E}$ in the target overlaps in part with that of QM7 (Figure~\ref{fig:data_dists}).
Figure~\ref{fig:combined} shows the learning curves for the two tasks, for each training set selection methods. The mean absolute error (MAE) is obtained as an average over the energy target ($\hat{E}$) of the ten target molecules and shown as a function of training set size $N$.

Both models with and without a penalty constant (ILP($p=1$) and ILP($p=0$), respectively) perform well, with MAEs $\leq \SI{1.5}{kcal/mol}$ for the QM7 molecules and $\sim \SI{4}{kcal/mol}$ for the QM9* molecules, using 1024~training points. The ILP($p=0$) results in lower MAEs than ILP($p=1$) except for at $N = 32, 64$ for the QM7 targets and $N=64$ for the QM9* targets.
This illustrates that the penalty term is unnecessary for small target molecules, which is a consequence of the representation: FCHL19 uses a radial cut-off $r_{\mathrm{cut}} = \SI{8}{\AA}$. The description of each atomic environment thus extends \SI{8}{A} away from each atom's nucleus. This is in the range of 3--5 bonds away from the central atom. Therefore, the entire molecular environment of small molecules is well-described using local representations, and distant ``unwanted'' atoms will be detected in the representation, making the ILP correction unnecessary. \\

SML also results in strong performance, lagging slightly behind the ILP($p=1$) for the QM7 targets and oscillating around the ILP learning curves for the QM9*. Since the target molecules are mostly of comparable size to the training set (QM7 consists of mostly seven-heavy atom molecules), the use of global representations in SML allows for an intelligent training set selection strategy. SML does result in a more unstable (\textit{i.e.}, non-monotonic) learning curve for the QM9* targets, suggesting that already in the case of a minor size mismatch between fragments and target (for example, three \textit{vs.}\ nine heavy atoms), the global selection approach suffers. \\

The other training set selection methods, namely random, FPS and CUR, lead to worse performance for QM7, but FPS and random perform reasonably well for QM9* targets, approaching the MAEs of ILP($p=1$) at 1024 training points. This suggests that using random sampling or FPS to get a training set that resembles the distribution of QM7 is a good strategy to predict $\hat{E}$ of held-out QM7 targets or related QM9* targets, owing to the transferability of organic chemistry motifs.

\subsubsection{Extrapolation tasks}
\begin{figure}[h!]
    \centering
    \includegraphics[width=0.63\linewidth]{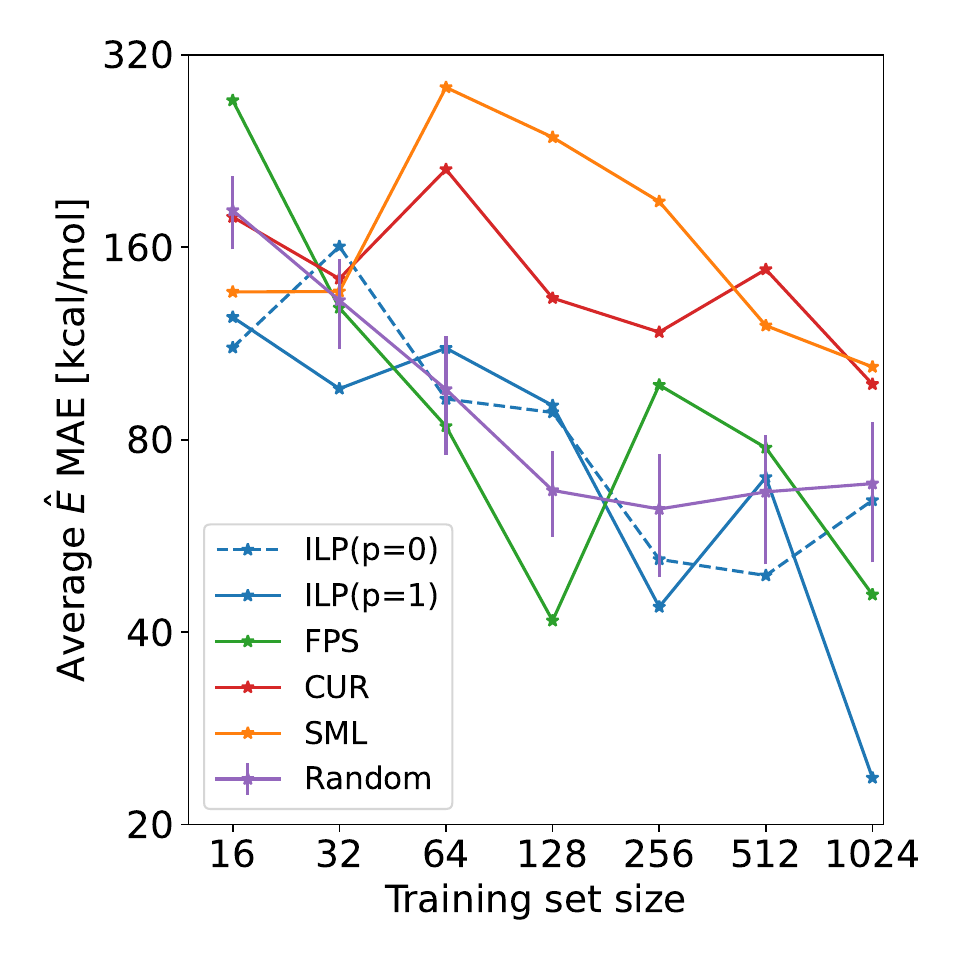}
    \caption{Averaged learning curves for ML models trained on QM7 molecules and testing on drug molecules (shown in Figure~\ref{fig:drug_mols}). Random selection is averaged over 5 random seeds, and error bars show the average standard deviation across predictions.}
    \label{fig:lc_avg_drugs}
\end{figure}

Moving to a more challenging extrapolation example, we predict $\hat{E}$ of ten larger drug molecules containing from 11 to 37 heavy atoms ($27$ on average). Their structures are shown in Figure~\ref{fig:drug_mols} and the significant difference in their energy distribution compared to QM7 is shown in Figure~\ref{fig:data_dists}. Learning curves for all the training set selection approaches are shown in Figure~\ref{fig:lc_avg_drugs}. \\

In this case, ILP($p=1$) gives lower MAEs than ILP($p=0$) and significantly outperforms random selection at both the smallest and the largest training set sizes. ILP($p=1$) also exhibits a more monotonically decreasing learning curve than FPS, which is the only other method that beats random selection at $N=128$ and $1024$ training points -- but has significantly larger error than random at $N=16$ and $256$. Compared to the penalized version, ILP($p=0$) performs quite poorly and remains within the standard deviation of random selection throughout most of the learning curve. For the larger drug molecules, incorporating the penalty term is therefore more beneficial than for the smaller target molecules. \\

SML shows more difficulties to select appropriate training subsets for the larger drug molecules. As discussed, the use of global representations breaks down with a large difference between train and target molecule sizes (for further interpretation of the selection methods, refer to Section~\ref{sec:interpretation}). As for QM9* targets, CUR also results in poor learning curves. Random selection and FPS both result in competitive learning curves, however. The FPS's learning curve is non-monotonic, but for $N=64$, it results in a training set that yields the lowest MAE. Random selection results in a monotonic curve that is competitive with ILP($p=0$) at $N=1024$. As for the QM7/QM9* targets, this suggests that evenly sampling from the QM7 training set is a reasonable strategy for these target molecules. \\

For all methods, the absolute errors on the drug molecular targets are considerably higher than for the QM7 and QM9* targets ($\sim 20$~kcal/mol for the ILP($p=1$) for example). This is because the extrapolation task is fundamentally more challenging than interpolation. Furthermore, $\hat{E}$ is an extensive quantity, with a much larger scale of total energies for drugs (see Figure~\ref{fig:data_dists}). The mean absolute percentage errors (MAPEs), \textit{i.e.}\ $\mathrm{MAE}_i/y_i$ for each target $i$, where $y_i$ is the total energy of the molecule, are comparable to those predicted for QM9* target molecules.

\subsection{Interpretation of the selection methods}\label{sec:interpretation}
\begin{figure}[h!]
    \centering
    \includegraphics[width=\linewidth]{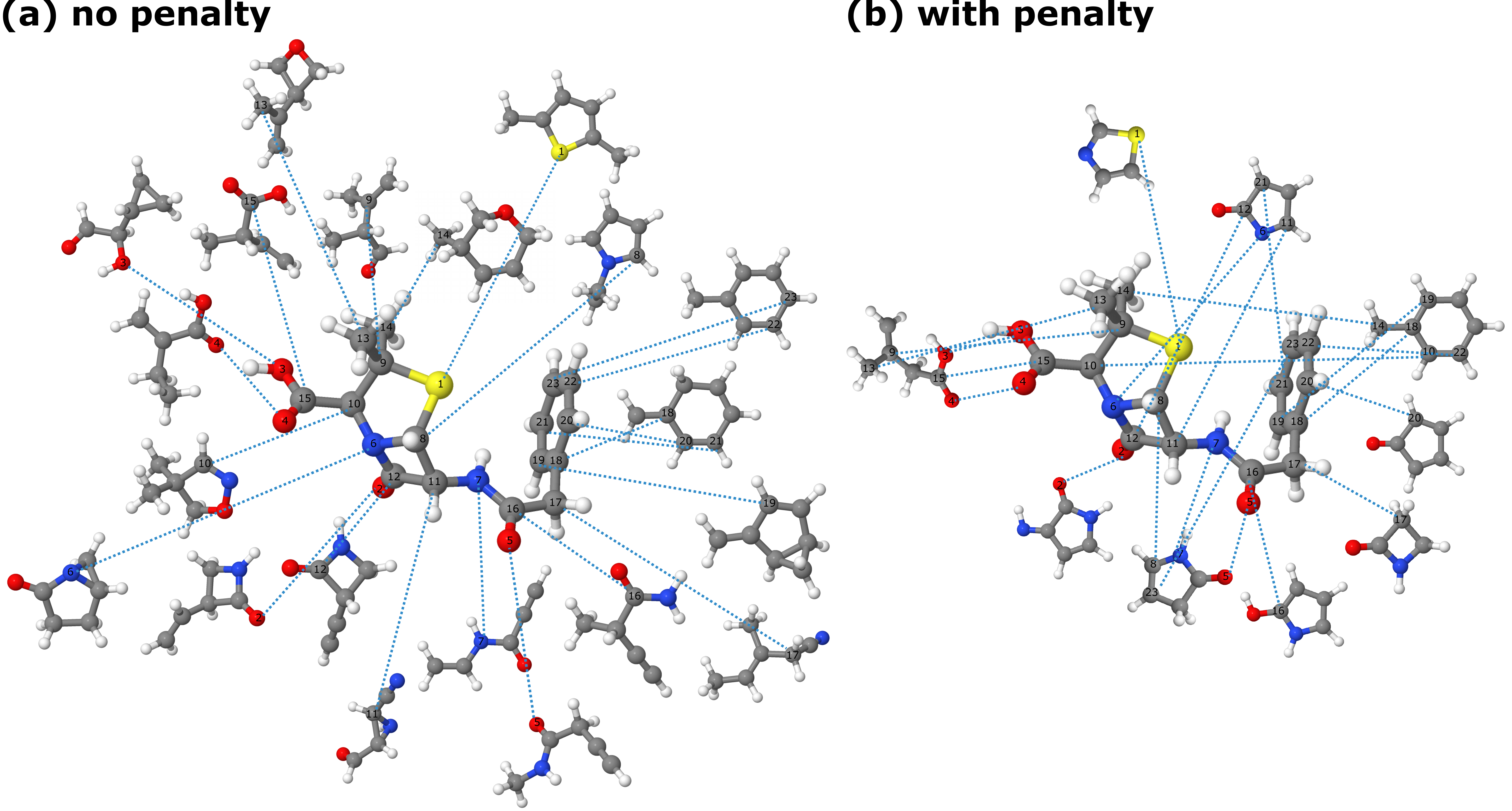}
    \caption{ILP (atom-mapping) solutions with the lowest cost, when not using a penalty term, for the penicillin target molecule.}
    \label{fig:solution_pen_both}
\end{figure}

Figure~\ref{fig:solution_pen_both} illustrates the types of fragments selected by the ILP($p=0$) and ILP($p=1$) variants for the penicillin target, for which the performance of the $p=0/1$ variants (Figure~\ref{S-fig:learning_curves_2}b) differs significantly. The best atom-mapping solution (\textit{i.e.}, with the smallest cost) is shown for ILP($p=0$) in panel (a) and for ILP($p=1$) in (b). Due to the absence of a penalty term in the cost function of ILP($p=0$), the algorithm selects a large number of fragments: in most cases, a molecule is selected to best match a single atom in the target. While these atoms are necessarily the closest in the database according to the Euclidean distance between the fragment and target atom's representations, this process introduces a large amount of molecules for which the remaining atoms may not be relevant for the task at hand. Recall that due to the global nature of the target property ($\hat{E}$), only whole molecules can be included in the training set, even if only some of the atoms are matched to the target. Accordingly, the inclusion of the penalty term in ILP($p=1$) minimizes the amount of molecules containing unnecessary atoms. Furthermore, the flexible ILP formulation allows to easily adapt the penalty strength in case a stronger constraint is required.

\begin{figure}[h!]
    \centering
    \includegraphics[width=0.98\linewidth]{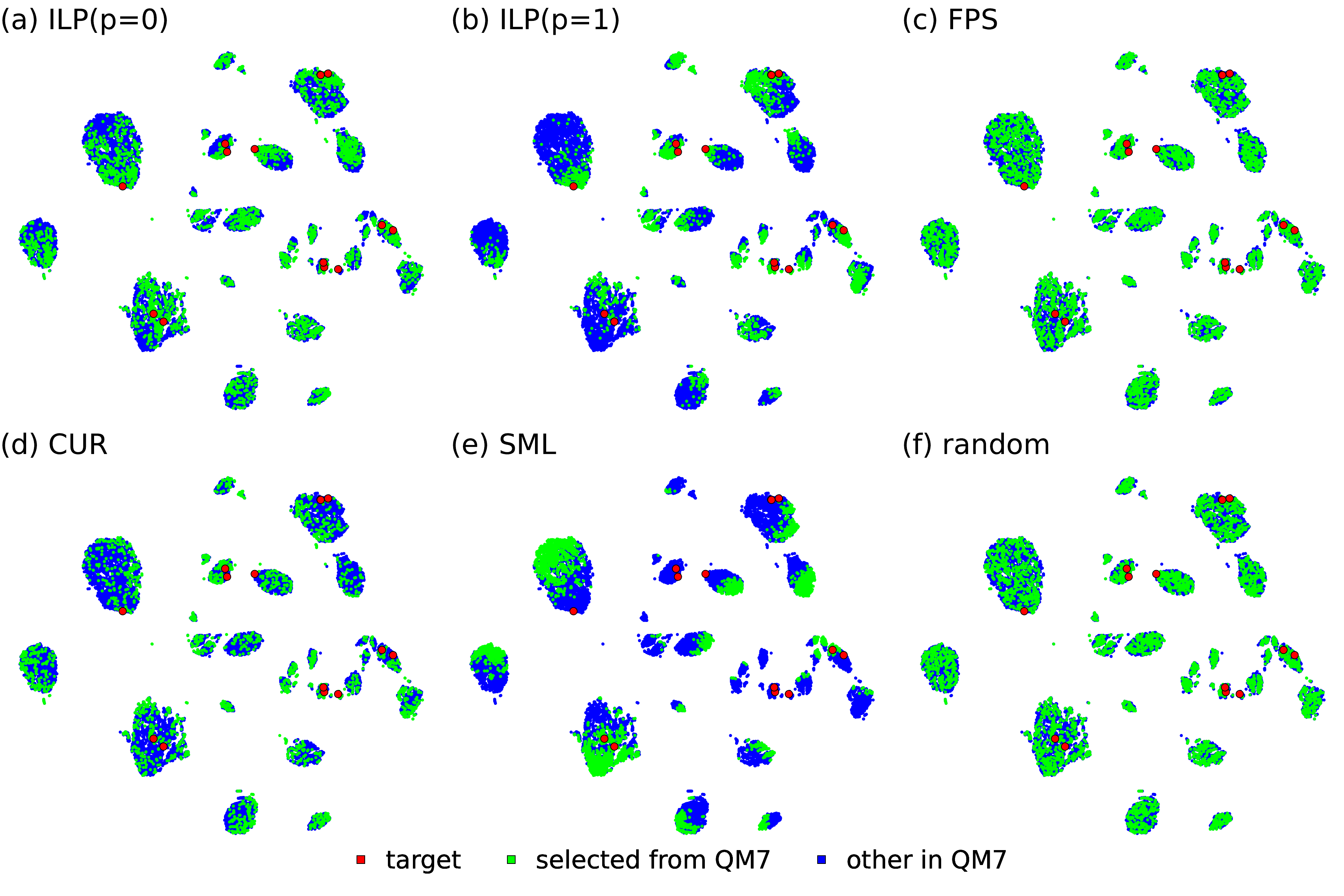}
    \caption{2D t-SNE projection of the carbon atomic environments of the penicillin target molecule (red), those of the selected training molecules for each method (green) and the entire QM7 training set (blue). The projection was performed with the \texttt{openTSNE}\cite{Linderman_2019,Policar2024} package with \emph{perplexity} set to 500.
    }
    \label{fig:tsne}
\end{figure}

We rationalize the observed performance of the different training set selection methods in terms of balance between introducing representative atomic environments from the training set that are most similar to those in the target molecule and minimizing the amount of dissimilar environments. To better visualize how this balance is achieved by each training set selection method, we represent the atomic environments of the target molecules (red), together with those of the selected training molecules (green) and the entire QM7 training set (blue), through a 2D t-SNE projection.\cite{van2008visualizing} The results for the carbon environments of the penicillin molecule (see Section~\ref{S-sec:tsne} of the supplementary information for other examples of elements and target molecules) are presented in Figure~\ref{fig:tsne}. ILP($p=1$) (panel b) has the largest concentration of selected environments in the vicinity of the target atoms and the least amount of environments far away, in agreement with our rationale and its top performance. In contrast, ILP($p=0$) (panel a) includes many environments far away from the target atoms, which leads to a significantly worse result for penicillin (Figure~\ref{S-fig:learning_curves_2}b) and on average (Figure~\ref{fig:lc_avg_drugs}).

Both FPS and random (panels c and f) evenly cover the chemical space of the QM7 database, which leads to a similar middle of the pack result for both methods. Next, CUR decomposition differs from random and FPS by having localized high density in specific regions. However, since those regions are not aligned with the target atomic environments, its performance is worse than random. Finally, SML tends to focus on high-density regions far from the target atoms, almost opposite to ILP ($p=1$), because it relies on global molecular similarity without accounting for individual atomic environments. Consequently, its performance is the worst out of the six methods by far.

Our results demonstrate that a local-similarity based approach to training set selection is a meaningful strategy for predictions of size-extensive molecular properties such as $\hat{E}$.
While the naive ($p=0$) approach performs well for small-molecule interpolation tasks,
for more challenging extrapolation tasks
require additional penalty term ($p=1$) to avoid undesirable solutions.

\subsection{Timings of the ILP approach}
\begin{figure}[h!]
    \centering
    \includegraphics[scale=.38]{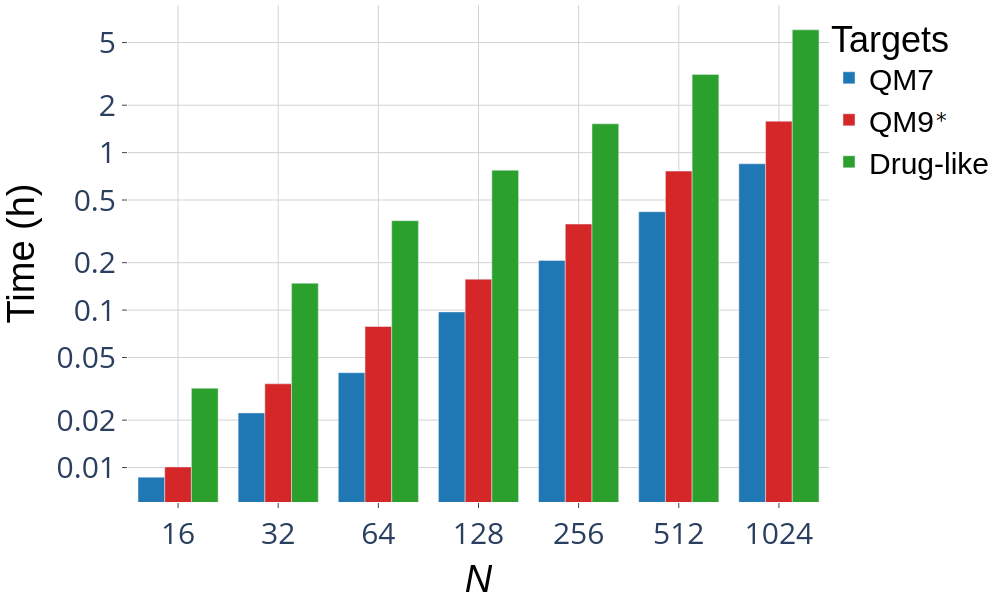}
    \caption{Average time in hours taken to generate $N$ QM7 fragments for each target type. The average is done over $10$ targets of each type. The slopes in the double log scales show a near-linear trend of time as function of the number of fragments to generate.}
    \label{fig:timings}
\end{figure}
Figure~\ref{fig:timings} indicates the time spent on running the ILP algorithm. The timing increases near-linearly with the number of training points $N$. Given the complexity of the optimisation problem (Section~\ref{sec:algo}), such a scaling alludes to efficient performance even if the algorithm adds some cost to the task of training set selection. Section~\ref{sec:limitations} enumerates suggestions for further reducing the complexity of the search problem.
Since it was shown (Figures~\ref{fig:combined}--\ref{fig:lc_avg_drugs}) that the ILP approach outperforms other training set selection methods at low $N$, and that the prediction MAEs will saturate at higher $N$, we suggest the user to opt for a low-to-intermediate value of $N$ for an ideal trade-off between training set selection time and performance on the target. In case the target molecules need to be labelled, a minimal value of $N$ also considerably reduces the computational budget of eventual quantum chemical computations.

\subsection{Open questions}\label{sec:limitations}
Note that there are several alternatives to our approach in incorporating the penalty term. As detailed in Section~\ref{sec:penalty}, a penalty could also be written as a function of the number of non-hydrogen atoms added by the fragments.
Moreover, other representations could be used, or FCHL19 in its current format could be modified (particularly, the cut-off term) to incorporate more distant regions of the molecules. We hypothesized that a penalty is a more effective solution than interfering with the representation for two reasons: first, a penalty can be more generally implemented (independently of the representation). Second, modifying the representations to cover longer-range might be detrimental. For example, most atom-centered representations employ two- and three-body interaction functions that decay rapidly with the distance from the central atom.\cite{langer_representations_2022, musil2021physics, huang2021ab} This is based on the ``locality'' principle, that close-by interactions should be weighted more heavily than distant ones.\cite{gubaev2018machine}
Increasing the cut-off might not be sufficient, and changing the functional forms of the representations would be a more appropriate route. Yet, designing new longer-range representations is outside the scope of this work.\\

The selected solutions (for example, those in Figure~\ref{fig:solution_pen_both}) are often disconnected subgraphs of the target graph. These render the solutions ``non-physical'' which is not necessarily a problem for the eventual regression. Forcing the algorithm to select connected subgraphs could however increase the algorithm's efficiency, \textit{i.e.},\ reducing the number of training points $N$ needed for regression. Substructure-based approaches are common in the classical application of ILP to atom-mapping problems.\cite{bahiense2012maximum, huang2006maximum}
However, we chose to avoid such a strategy, which relies on SMILES strings. By using only 3D molecular coordinates, our training set selection algorithm can handle molecules that are not well-represented by SMILES strings\cite{vela2022cell2mol, krenn2022selfies}, \textit{e.g.}, conformational diversity.\\

The ILP approach, as well as the others tested here (FPS, SML, CUR, random) are ``top-down'' in the sense that the number of training points $N$ is specified in advance. An alternative, ``bottom-up'' option, used by the AMONs\cite{huang2020quantum} defines a variable number of fragments according to how they best represent the target molecule. While the AMONs require the computation of each of the selected molecules, the cost of our ILP algorithm would still remain higher for challenging search problems (large database searches, a complex optimisation landscape, etc.). A hybrid ``top-down'' and ``bottom-up'' approach incorporating sub-structure\cite{bahiense2012maximum, huang2006maximum} or bond counting\cite{first_stereochem_2012} constraints to the algorithm, might be an interesting direction to accelerate the ILP strategy.

Finally, we note that lightweight models trained on a handful of selected data are useful to bootstrap active learning and Bayesian optimization workflows at a low cost. Machine-learned force fields for molecular dynamics simulations are a prime example application, since typically many preliminary simulations are required to expand the training data.\cite{Jurskov2022, Clerse2024}

\section{Conclusion}
Integer linear programming is a powerful tool to solve NP-hard problems, to
date under-utilized in the field of physics-inspired (\textit{i.e.}, atomistic)
machine learning. We present an ILP approach, which searches for optimum
reference atomic environments within an existing molecular dataset to train a
kernel-ridge regression model that predicts size-extensive properties of
out-of-sample molecules. The algorithm is based on atom-to-atom mapping,
constructed according to the similarity of atom-centered physics-based
representations. The ILP approach generally outperforms existing unsupervised
training set selection approaches, especially in the case of
\textit{extrapolation} (with $p=1$), \textit{i.e.}, predicting properties of larger
molecules than those in the training set. We argue that the proposed method has
two distinct advantages: (i) the emphasis on selection based on local
environments and (ii) the intelligent design of an ILP approach with relevant
constraints to the search problem. Further improvements are foreseen through
incorporating additional constraints such as those inspired by the ILP
treatment of atom mapping when describing feasible reaction mechanisms.
Altogether, our training set selection tool provides a means to improve
large-molecule property predictions.

\section{Software Availability}
The optimization algorithms are implemented using the Gurobi Optimization \cite{gurobi} software, using the python interface \texttt{gurobipy}, version 10.0.2.
The algorithms are available at \url{https://github.com/lcmd-epfl/ILPSelect}.

\section*{Acknowledgements}
P.v.G., R.L., J.W. and C.C. acknowledge the National Centre of Competence in Research (NCCR)
``Sustainable chemical process through catalysis (Catalysis)''
of the Swiss National Science Foundation (SNSF, Grant No.~180544) for financial support.
M.H., F.E. and C.C acknowledge the FSB Intrafaculty Seed Funding.
F.E. acknowledges support from the Swiss National Science Foundation (SNSF, Grant No.~185030).
K.R.B. and C.C. acknowledge the European Research Council (ERC, Grant No.~817977)
and the National Centre of Competence in Research (NCCR)
``Materials' Revolution: Computational Design and Discovery of Novel Materials (MARVEL)'',
of the Swiss National Science Foundation (SNSF, Grant No.~205602).

\clearpage

\bibliography{ilp.bib}

@article{bo2018role,
  title   = {The role of computational results databases in accelerating the discovery of catalysts},
  author  = {Bo, Carles and Maseras, Feliu and López, Núria},
  journal = {Nat. Catal.},
  year    = {2018},
  volume  = {1},
  number  = {11},
  pages   = {809--810},
  doi     = {10.1038/s41929-018-0176-4},
}

@article{huang2021ab,
  title   = {Ab initio machine learning in chemical compound space},
  author  = {Huang, Bing and von Lilienfeld, O Anatole},
  journal = {Chem. Rev.},
  year    = {2021},
  volume  = {121},
  number  = {16},
  pages   = {10001--10036},
  doi     = {10.1021/acs.chemrev.0c01303},
}

@article{ramakrishnan2014quantum,
  title   = {Quantum chemistry structures and properties of 134 kilo molecules},
  author  = {Ramakrishnan, Raghunathan and Dral, Pavlo O and Rupp, Matthias and von Lilienfeld, O Anatole},
  journal = {Sci. Data},
  year    = {2014},
  volume  = {1},
  pages   = {140022},
  doi     = {10.1038/sdata.2014.22},
}

@article{smith2017ani,
  title   = {ANI-1, A data set of 20 million calculated off-equilibrium conformations for organic molecules},
  author  = {Smith, Justin S and Isayev, Olexandr and Roitberg, Adrian E},
  journal = {Sci. Data},
  year    = {2017},
  volume  = {4},
  number  = {1},
  pages   = {170193},
  doi     = {10.1038/sdata.2017.193},
}

@article{jain2013commentary,
  title   = {Commentary: The Materials Project: A materials genome approach to accelerating materials innovation},
  author  = {Jain, Anubhav and Ong, Shyue Ping and Hautier, Geoffroy and Chen, Wei and Richards, William Davidson and Dacek, Stephen and Cholia, Shreyas and Gunter, Dan and Skinner, David and Ceder, Gerbrand and others},
  journal = {{APL} Mater.},
  year    = {2013},
  volume  = {1},
  number  = {1},
  pages   = {011002},
  doi     = {10.1063/1.4812323},
}

@article{draxl2018nomad,
  title   = {{NOMAD}: {T}he {FAIR} concept for big data-driven materials science},
  author  = {Draxl, Claudia and Scheffler, Matthias},
  journal = {MRS Bull.},
  year    = {2018},
  volume  = {43},
  number  = {9},
  pages   = {676--682},
  doi     = {10.1557/mrs.2018.208},
}

@article{deringer_gpr,
  title   = {Gaussian Process Regression for Materials and Molecules},
  author  = {Deringer, Volker L. and Bartók, Albert P. and Bernstein, Noam and Wilkins, David M. and Ceriotti, Michele and Csányi, Gábor},
  journal = {Chem. Rev.},
  year    = {2021},
  volume  = {121},
  number  = {16},
  pages   = {10073--10141},
  doi     = {10.1021/acs.chemrev.1c00022},
}

@article{browning2017genetic,
  title   = {Genetic optimization of training sets for improved machine learning models of molecular properties},
  author  = {Browning, Nicholas J and Ramakrishnan, Raghunathan and von Lilienfeld, O Anatole and Roethlisberger, Ursula},
  journal = {J. Chem. Phys. Lett.},
  year    = {2017},
  volume  = {8},
  number  = {7},
  pages   = {1351--1359},
  doi     = {10.1021/acs.jpclett.7b00038},
}

@article{huang2020quantum,
  title   = {Quantum machine learning using atom-in-molecule-based fragments selected on the fly},
  author  = {Huang, Bing and von Lilienfeld, O Anatole},
  journal = {Nat. Chem.},
  year    = {2020},
  volume  = {12},
  number  = {10},
  pages   = {945--951},
  doi     = {10.1038/s41557-020-0527-z},
}

@article{bartok2013representing,
  title   = {On representing chemical environments},
  author  = {Bartók, Albert P and Kondor, Risi and Csányi, Gábor},
  journal = {Phys. Rev. B},
  year    = {2013},
  volume  = {87},
  number  = {18},
  pages   = {184115},
  doi     = {10.1103/PhysRevB.87.184115},
}

@article{christensen2020fchl,
  title   = {FCHL revisited: Faster and more accurate quantum machine learning},
  author  = {Christensen, Anders S and Bratholm, Lars A and Faber, Felix A and von Lilienfeld, O. Anatole},
  journal = {J. Chem. Phys.},
  year    = {2020},
  volume  = {152},
  number  = {4},
  pages   = {044107},
  doi     = {10.1063/1.5126701},
}

@article{musil2021physics,
  title   = {Physics-inspired structural representations for molecules and materials},
  author  = {Musil, Felix and Grisafi, Andrea and Bartók, Albert P and Ortner, Christoph and Csányi, Gábor and Ceriotti, Michele},
  journal = {Chem. Rev.},
  year    = {2021},
  volume  = {121},
  number  = {16},
  pages   = {9759--9815},
  doi     = {10.1021/acs.chemrev.1c00021},
}

@article{rupp2012fast,
  title   = {Fast and accurate modeling of molecular atomization energies with machine learning},
  author  = {Rupp, Matthias and Tkatchenko, Alexandre and Müller, Klaus-Robert and von Lilienfeld, O Anatole},
  journal = {Phys. Rev. Lett.},
  year    = {2012},
  volume  = {108},
  number  = {5},
  pages   = {058301},
  doi     = {10.1103/PhysRevLett.108.058301},
}

@article{nandy2022audacity,
  title   = {Audacity of huge: overcoming challenges of data scarcity and data quality for machine learning in computational materials discovery},
  author  = {Nandy, Aditya and Duan, Chenru and Kulik, Heather J},
  journal = {Curr. Opin. Chem. Eng.},
  year    = {2022},
  volume  = {36},
  pages   = {100778},
  doi     = {10.1016/j.coche.2021.100778},
}

@article{nakata2017pubchemqc,
  title   = {PubChemQC project: a large-scale first-principles electronic structure database for data-driven chemistry},
  author  = {Nakata, Maho and Shimazaki, Tomomi},
  journal = {J. Chem. Inf. Model.},
  year    = {2017},
  volume  = {57},
  number  = {6},
  pages   = {1300--1308},
  doi     = {10.1021/acs.jcim.7b00083},
}

@article{hoja2021qm7,
  title   = {{QM7-X}, a comprehensive dataset of quantum-mechanical properties spanning the chemical space of small organic molecules},
  author  = {Hoja, Johannes and Medrano Sandonas, Leonardo and Ernst, Brian G and Vazquez-Mayagoitia, Alvaro and DiStasio Jr, Robert A and Tkatchenko, Alexandre},
  journal = {Sci. Data},
  year    = {2021},
  volume  = {8},
  number  = {1},
  pages   = {43},
  doi     = {10.1038/s41597-021-00812-2},
}

@article{qm7_b,
  title   = {Machine learning of molecular electronic properties in chemical compound space},
  author  = {Montavon, Grégoire and Rupp, Matthias and Gobre, Vivekanand and Vazquez-Mayagoitia, Alvaro and Hansen, Katja and Tkatchenko, Alexandre and Müller, Klaus-Robert and von Lilienfeld, O. Anatole},
  journal = {New J. Phys.},
  year    = {2013},
  volume  = {15},
  number  = {9},
  pages   = {095003},
  doi     = {10.1088/1367-2630/15/9/095003},
}

@article{tddft_data,
  title   = {{Electronic spectra from TDDFT and machine learning in chemical space}},
  author  = {Ramakrishnan, Raghunathan and Hartmann, Mia and Tapavicza, Enrico and von Lilienfeld, O. Anatole},
  journal = {J. Chem. Phys.},
  year    = {2015},
  volume  = {143},
  number  = {8},
  pages   = {084111},
  doi     = {10.1063/1.4928757},
}

@article{qmugs,
  title   = {QMugs 1.1: Quantum mechanical properties of organic compounds commonly encountered in reactivity datasets},
  author  = {Neeser, Rebecca M. and Isert, Clemens and Stuyver, Thijs and Schneider, Gisbert and Coley, Connor W.},
  journal = {Chem. Data Coll.},
  year    = {2023},
  volume  = {46},
  pages   = {101040},
  doi     = {10.1016/j.cdc.2023.101040},
}

@article{electrolyte_genome_project,
  title   = {The Electrolyte Genome project: A big data approach in battery materials discovery},
  author  = {Qu, Xiaohui and Jain, Anubhav and Rajput, Nav Nidhi and Cheng, Lei and Zhang, Yong and Ong, Shyue Ping and Brafman, Miriam and Maginn, Edward and Curtiss, Larry A. and Persson, Kristin A.},
  journal = {Comput. Mater. Sci.},
  year    = {2015},
  volume  = {103},
  pages   = {56--67},
  doi     = {10.1016/j.commatsci.2015.02.050},
}

@article{sivaraman2020machine,
  title   = {Machine-learned interatomic potentials by active learning: amorphous and liquid hafnium dioxide},
  author  = {Sivaraman, Ganesh and Krishnamoorthy, Anand Narayanan and Baur, Matthias and Holm, Christian and Stan, Marius and Csányi, Gábor and Benmore, Chris and Vázquez-Mayagoitia, {\'A}lvaro},
  journal = {npj Comput. Mater.},
  year    = {2020},
  volume  = {6},
  number  = {1},
  pages   = {104},
  doi     = {10.1038/s41524-020-00367-7},
}

@article{smith2018less,
  title   = {Less is more: Sampling chemical space with active learning},
  author  = {Smith, Justin S and Nebgen, Ben and Lubbers, Nicholas and Isayev, Olexandr and Roitberg, Adrian E},
  journal = {J. Chem. Phys.},
  year    = {2018},
  volume  = {148},
  number  = {24},
  pages   = {241733},
  doi     = {10.1063/1.5023802},
}

@article{cersonsky2021improving,
  title   = {Improving sample and feature selection with principal covariates regression},
  author  = {Cersonsky, Rose K and Helfrecht, Benjamin A and Engel, Edgar A and Kliavinek, Sergei and Ceriotti, Michele},
  journal = {Mach. Learn.: Sci. Technol.},
  year    = {2021},
  volume  = {2},
  number  = {3},
  pages   = {035038},
  doi     = {10.1088/2632-2153/abfe7c},
}

@article{imbalzano2018automatic,
  title   = {Automatic selection of atomic fingerprints and reference configurations for machine-learning potentials},
  author  = {Imbalzano, Giulio and Anelli, Andrea and Giofré, Daniele and Klees, Sinja and Behler, Jörg and Ceriotti, Michele},
  journal = {J. Chem. Phys.},
  year    = {2018},
  volume  = {148},
  number  = {24},
  pages   = {241730},
  doi     = {10.1063/1.5024611},
}

@article{mahoney2009cur,
  title   = {CUR matrix decompositions for improved data analysis},
  author  = {Mahoney, Michael W and Drineas, Petros},
  journal = {Proc. Natl. Acad. Sci. U.S.A.},
  year    = {2009},
  volume  = {106},
  number  = {3},
  pages   = {697--702},
  doi     = {10.1073/pnas.0803205106},
}

@article{fabregat2022local,
  title   = {Local Kernel Regression and Neural Network Approaches to the Conformational Landscapes of Oligopeptides},
  author  = {Fabregat, Raimon and Fabrizio, Alberto and Engel, Edgar A and Meyer, Benjamin and Juraskova, Veronika and Ceriotti, Michele and Corminboeuf, Clemence},
  journal = {J. Chem. Theory Comput.},
  year    = {2022},
  volume  = {18},
  number  = {3},
  pages   = {1467--1479},
  doi     = {10.1021/acs.jctc.1c00813},
}

@article{lemm2023improved,
  title   = {Improved decision making with similarity based machine learning},
  author  = {Lemm, Dominik and von Rudorff, Guido Falk and von Lilienfeld, O. Anatole},
  journal = {Mach. Learn.: Sci. Technol.},
  year    = {2023},
  volume  = {4},
  number  = {4},
  pages   = {045043},
  doi     = {10.1088/2632-2153/ad0fa3},
}

@article{wahab2022compas,
  title   = {The COMPAS Project: A Computational Database of Polycyclic Aromatic Systems. Phase 1: cata-Condensed Polybenzenoid Hydrocarbons},
  author  = {Wahab, Alexandra and Pfuderer, Lara and Paenurk, Eno and Gershoni-Poranne, Renana},
  journal = {J. Chem. Inf. Model.},
  year    = {2022},
  volume  = {62},
  number  = {16},
  pages   = {3704--3713},
  doi     = {10.1021/acs.jcim.2c00503},
}

@article{rossi_solvation_2020,
  title   = {Simulating Solvation and Acidity in Complex Mixtures with First-Principles Accuracy: The Case of CH3SO3H and H2O2 in Phenol},
  author  = {Rossi, Kevin and Jurásková, Veronika and Wischert, Raphael and Garel, Laurent and Corminbœuf, Clémence and Ceriotti, Michele},
  journal = {J. Chem. Theory Comput.},
  year    = {2020},
  volume  = {16},
  number  = {8},
  pages   = {5139--5149},
  doi     = {10.1021/acs.jctc.0c00362},
}

@article{FCHL,
  title   = {Alchemical and structural distribution based representation for universal quantum machine learning},
  author  = {Faber, Felix A. and Christensen, Anders S. and Huang, Bing and von Lilienfeld, O. Anatole},
  journal = {J. Chem. Phys.},
  year    = {2018},
  volume  = {148},
  number  = {24},
  pages   = {241717},
  doi     = {10.1063/1.5020710},
}

@article{lucic_coresets_2018,
  title   = {Training Gaussian Mixture Models at Scale via Coresets},
  author  = {Lucic, Mario and Faulkner, Matthew and Krause, Andreas and Feldman, Dan},
  journal = {J. Mach. Learn. Res.},
  year    = {2018},
  volume  = {18},
  number  = {160},
  pages   = {1--25},
  url     = {https://jmlr.org/papers/v18/15-506.html},
}

@article{podryabinkin2017active,
  title   = {Active learning of linearly parametrized interatomic potentials},
  author  = {Podryabinkin, Evgeny V and Shapeev, Alexander V},
  journal = {Comput. Mater. Sci.},
  year    = {2017},
  volume  = {140},
  pages   = {171--180},
  doi     = {10.1016/j.commatsci.2017.08.031},
}

@article{gubaev2018machine,
  title   = {Machine learning of molecular properties: Locality and active learning},
  author  = {Gubaev, Konstantin and Podryabinkin, Evgeny V and Shapeev, Alexander V},
  journal = {J. Chem. Phys.},
  year    = {2018},
  volume  = {148},
  number  = {24},
  pages   = {241727},
  doi     = {10.1063/1.5005095},
}

@article{rupp2015machine,
  title   = {Machine learning for quantum mechanical properties of atoms in molecules},
  author  = {Rupp, Matthias and Ramakrishnan, Raghunathan and von Lilienfeld, O Anatole},
  journal = {J. Chem. Phys. Lett.},
  year    = {2015},
  volume  = {6},
  number  = {16},
  pages   = {3309--3313},
  doi     = {10.1021/acs.jpclett.5b01456},
}

@article{wilkins2019accurate,
  title   = {Accurate molecular polarizabilities with coupled cluster theory and machine learning},
  author  = {Wilkins, David M and Grisafi, Andrea and Yang, Yang and Lao, Ka Un and DiStasio Jr, Robert A and Ceriotti, Michele},
  journal = {Proc. Natl. Acad. Sci. U.S.A.},
  year    = {2019},
  volume  = {116},
  number  = {9},
  pages   = {3401--3406},
  doi     = {10.1073/pnas.1816132116},
}

@article{jung2020size,
  title   = {Size-Extensive Molecular Machine Learning with Global Representations},
  author  = {Jung, Hyunwook and Stocker, Sina and Kunkel, Christian and Oberhofer, Harald and Han, Byungchan and Reuter, Karsten and Margraf, Johannes T},
  journal = {ChemSystemsChem},
  year    = {2020},
  volume  = {2},
  number  = {4},
  pages   = {e1900052},
  doi     = {10.1002/syst.201900052},
}

@article{grisafi2018transferable,
  title   = {Transferable machine-learning model of the electron density},
  author  = {Grisafi, Andrea and Fabrizio, Alberto and Meyer, Benjamin and Wilkins, David M and Corminboeuf, Clemence and Ceriotti, Michele},
  journal = {ACS Cent. Sci.},
  year    = {2018},
  volume  = {5},
  number  = {1},
  pages   = {57--64},
  doi     = {10.1021/acscentsci.8b00551},
}

@article{Neese2022,
  title   = {Software update: The ORCA program system—Version 5.0},
  author  = {Neese, Frank},
  journal = {Wiley Interdiscip. Rev. Comput. Mol. Sci.},
  year    = {2022},
  volume  = {12},
  number  = {5},
  pages   = {e1606},
  doi     = {10.1002/wcms.1606},
}

@article{Grimme2010,
  title   = {A consistent and accurate ab initio parametrization of density functional dispersion correction (DFT-D) for the 94 elements H-Pu},
  author  = {Grimme, Stefan and Antony, Jens and Ehrlich, Stephan and Krieg, Helge},
  journal = {J. Chem. Phys.},
  year    = {2010},
  volume  = {132},
  number  = {15},
  pages   = {154104},
  doi     = {10.1063/1.3382344},
}

@article{Adamo1999,
  title   = {An accurate density functional method for the study of magnetic properties: the PBE0 model},
  author  = {Adamo, C. and Cossi, M. and Barone, V.},
  journal = {J. Mol. Struct. THEOCHEM},
  year    = {1999},
  volume  = {493},
  number  = {1--3},
  pages   = {145--157},
  doi     = {10.1016/s0166-1280(99)00235-3},
}

@article{Weigend2005,
  title   = {Balanced basis sets of split valence, triple zeta valence and quadruple zeta valence quality for H to Rn: Design and assessment of accuracy},
  author  = {Weigend, Florian and Ahlrichs, Reinhart},
  journal = {Phys. Chem. Chem. Phys.},
  year    = {2005},
  volume  = {7},
  number  = {18},
  pages   = {3297},
  doi     = {10.1039/b508541a},
}

@article{hansen2015machine,
  title   = {Machine learning predictions of molecular properties: Accurate many-body potentials and nonlocality in chemical space},
  author  = {Hansen, Katja and Biegler, Franziska and Ramakrishnan, Raghunathan and Pronobis, Wiktor and von Lilienfeld, O Anatole and Muller, Klaus-Robert and Tkatchenko, Alexandre},
  journal = {J. Chem. Phys. Lett.},
  year    = {2015},
  volume  = {6},
  number  = {12},
  pages   = {2326--2331},
  doi     = {10.1021/acs.jpclett.5b00831},
}

@article{huo2017unified,
  title   = {Unified representation of molecules and crystals for machine learning},
  author  = {Huo, Haoyan and Rupp, Matthias},
  journal = {Mach. Learn.: Sci. Technol.},
  year    = {2022},
  volume  = {3},
  number  = {4},
  pages   = {045017},
  doi     = {10.1088/2632-2153/aca005},
}

@article{nigam2020recursive,
  title   = {Recursive evaluation and iterative contraction of N-body equivariant features},
  author  = {Nigam, Jigyasa and Pozdnyakov, Sergey and Ceriotti, Michele},
  journal = {J. Chem. Phys.},
  year    = {2020},
  volume  = {153},
  number  = {12},
  pages   = {121101},
  doi     = {10.1063/5.0021116},
}

@article{drautz2019atomic,
  title   = {Atomic cluster expansion for accurate and transferable interatomic potentials},
  author  = {Drautz, Ralf},
  journal = {Phys. Rev. B},
  year    = {2019},
  volume  = {99},
  number  = {1},
  pages   = {014104},
  doi     = {10.1103/PhysRevB.99.014104},
}

@article{eq_balancing_ILP,
  title   = {Chemical equation balancing: An integer programming approach},
  author  = {Sen, S.K. and Agarwal, Hans and Sen, Sagar},
  journal = {Math. Comput. Model.},
  year    = {2006},
  volume  = {44},
  number  = {7},
  pages   = {678--691},
  doi     = {10.1016/j.mcm.2006.02.004},
}

@article{bond_order_assignment_ILP,
  title   = {Correct Bond Order Assignment in a Molecular Framework Using Integer Linear Programming with Application to Molecules Where Only Non-Hydrogen Atom Coordinates Are Available},
  author  = {Froeyen, Matheus and Herdewijn, Piet},
  journal = {J. Chem. Inf. Model.},
  year    = {2005},
  volume  = {45},
  number  = {5},
  pages   = {1267--1274},
  doi     = {10.1021/ci049645z},
}

@article{mass_spectra_ILP,
  title   = {Robust Automated Mass Spectra Interpretation and Chemical Formula Calculation Using Mixed Integer Linear Programming},
  author  = {Baran, Richard and Northen, Trent R.},
  journal = {Anal. Chem.},
  year    = {2013},
  volume  = {85},
  number  = {20},
  pages   = {9777--9784},
  doi     = {10.1021/ac402180c},
}

@article{mol_design_ILP_2_2022,
  title   = {A new approach to the design of acyclic chemical compounds using skeleton trees and integer linear programming},
  author  = {Zhang, Fan and Zhu, Jianshen and Chiewvanichakorn, Rachaya and Shurbevski, Aleksandar and Nagamochi, Hiroshi and Akutsu, Tatsuya},
  journal = {Appl. Intell.},
  year    = {2022},
  volume  = {52},
  number  = {15},
  pages   = {17058--17072},
  doi     = {10.1007/s10489-021-03088-6},
}

@article{rxn_networks_ILP_2012,
  title   = {A linear programming approach to weak reversibility and linear conjugacy of chemical reaction networks},
  author  = {Johnston, Matthew D and Siegel, David and Szederkényi, Gábor},
  journal = {J. Math. Chem.},
  year    = {2012},
  volume  = {50},
  pages   = {274--288},
  doi     = {10.1007/s10910-011-9911-7},
}

@article{rxn_networks_ILP_2016,
  title   = {Inference of chemical reaction networks using mixed integer linear programming},
  author  = {Willis, Mark J. and von Stosch, Moritz},
  journal = {Comput. Chem. Eng.},
  year    = {2016},
  volume  = {90},
  pages   = {31--43},
  doi     = {10.1016/j.compchemeng.2016.04.019},
}

@article{mol_design_ILP_2002,
  title   = {On the solution of mixed-integer nonlinear programming models for computer aided molecular design},
  author  = {Ostrovsky, Guennadi M and Achenie, Luke EK and Sinha, Manish},
  journal = {Comput. Chem.},
  year    = {2002},
  volume  = {26},
  number  = {6},
  pages   = {645--660},
  doi     = {10.1016/S0097-8485(02)00049-9},
}

@article{mol_design_ILP_2007,
  title   = {Mixed-integer linear programming algorithm for a computational protein design problem},
  author  = {Zhu, Yushan},
  journal = {Ind. Eng. Chem. Res.},
  year    = {2007},
  volume  = {46},
  number  = {3},
  pages   = {839--845},
  doi     = {10.1021/ie0605985},
}

@article{seq_alignment_ILP_1999,
  title   = {An exact solution for the segment-to-segment multiple sequence alignment problem},
  author  = {Lenhof, H P and Morgenstern, B and Reinert, K},
  journal = {Bioinformatics},
  year    = {1999},
  volume  = {15},
  number  = {3},
  pages   = {203--210},
  doi     = {10.1093/bioinformatics/15.3.203},
}

@article{seq_alignment_ILP_2002,
  title   = {Multiple sequence alignment with arbitrary gap costs: Computing an optimal solution using polyhedral combinatorics},
  author  = {Althaus, Ernst and Caprara, Alberto and Lenhof, Hans-Peter and Reinert, Knut},
  journal = {Bioinformatics},
  year    = {2002},
  volume  = {18},
  number  = {Suppl 2},
  pages   = {S4--S16},
  doi     = {10.1093/bioinformatics/18.suppl_2.s4},
}

@article{seq_alignment_ILP_2006,
  title   = {A branch-and-cut algorithm for multiple sequence alignment},
  author  = {Althaus, Ernst and Caprara, Alberto and Lenhof, Hans-Peter and Reinert, Knut},
  journal = {Math. Program.},
  year    = {2006},
  volume  = {105},
  pages   = {387--425},
  doi     = {10.1007/s10107-005-0659-3},
}

@article{seq_alignment_ILP_2008,
  title   = {A Lagrangian relaxation approach for the multiple sequence alignment problem},
  author  = {Althaus, Ernst and Canzar, Stefan},
  journal = {J. Comb. Optim.},
  year    = {2008},
  volume  = {16},
  pages   = {127--154},
  doi     = {10.1007/s10878-008-9139-z},
}

@article{side_chain_positioning_ILP_2004,
  title   = {{Solving and analyzing side-chain positioning problems using linear and integer programming}},
  author  = {Kingsford, Carleton L. and Chazelle, Bernard and Singh, Mona},
  journal = {Bioinformatics},
  year    = {2004},
  volume  = {21},
  number  = {7},
  pages   = {1028--1039},
  doi     = {10.1093/bioinformatics/bti144},
}

@article{vaccine_design_ILP_2008,
  title   = {A mathematical framework for the selection of an optimal set of peptides for epitope-based vaccines},
  author  = {Toussaint, Nora C and Dönnes, Pierre and Kohlbacher, Oliver},
  journal = {PLoS Comput. Biol.},
  year    = {2008},
  volume  = {4},
  number  = {12},
  pages   = {e1000246},
  doi     = {10.1371/journal.pcbi.1000246},
}

@article{route_retro_2009,
  title   = {Route designer: a retrosynthetic analysis tool utilizing automated retrosynthetic rule generation},
  author  = {Law, James and Zsoldos, Zsolt and Simon, Aniko and Reid, Darryl and Liu, Yang and Khew, Sing Yoong and Johnson, A Peter and Major, Sarah and Wade, Robert A and Ando, Howard Y},
  journal = {J. Chem. Inf. Model.},
  year    = {2009},
  volume  = {49},
  number  = {3},
  pages   = {593--602},
  doi     = {10.1021/ci800228y},
}

@article{route_retro_2015,
  title   = {Route design in the 21st century: the IC SYNTH software tool as an idea generator for synthesis prediction},
  author  = {B{ø}gevig, Anders and Federsel, Hans-Jurgen and Huerta, Fernando and Hutchings, Michael G and Kraut, Hans and Langer, Thomas and Low, Peter and Oppawsky, Christoph and Rein, Tobias and Saller, Heinz},
  journal = {Org. Process. Res. Dev.},
  year    = {2015},
  volume  = {19},
  number  = {2},
  pages   = {357--368},
  doi     = {10.1021/op500373e},
}

@article{segler_retro_2017,
  title   = {Neural-symbolic machine learning for retrosynthesis and reaction prediction},
  author  = {Segler, Marwin HS and Waller, Mark P},
  journal = {Chem. -- Eur. J.},
  year    = {2017},
  volume  = {23},
  number  = {25},
  pages   = {5966--5971},
  doi     = {10.1002/chem.201605499},
}

@article{coley_retro_2020,
  title   = {Data augmentation and pretraining for template-based retrosynthetic prediction in computer-aided synthesis planning},
  author  = {Fortunato, Michael E and Coley, Connor W and Barnes, Brian C and Jensen, Klavs F},
  journal = {J. Chem. Inf. Model.},
  year    = {2020},
  volume  = {60},
  number  = {7},
  pages   = {3398--3407},
  doi     = {10.1021/acs.jcim.0c00403},
}

@article{coley_retro_2018,
  title   = {Machine learning in computer-aided synthesis planning},
  author  = {Coley, Connor W and Green, William H and Jensen, Klavs F},
  journal = {Acc. Chem. Res.},
  year    = {2018},
  volume  = {51},
  number  = {5},
  pages   = {1281--1289},
  doi     = {10.1021/acs.accounts.8b00087},
}

@article{rxn_searching_2016,
  title   = {Automatized Assessment of Protective Group Reactivity: A Step Toward Big Reaction Data Analysis},
  author  = {Lin, Arkadii I. and Madzhidov, Timur I. and Klimchuk, Olga and Nugmanov, Ramil I. and Antipin, Igor S. and Varnek, Alexandre},
  journal = {J. Chem. Inf. Model.},
  year    = {2016},
  volume  = {56},
  number  = {11},
  pages   = {2140--2148},
  doi     = {10.1021/acs.jcim.6b00319},
}

@article{rxn_searching_2002,
  title   = {Over 20 Years of Reaction Access Systems from MDL: A Novel Reaction Substructure Search Algorithm},
  author  = {Chen, Lingran and Nourse, James G. and Christie, Bradley D. and Leland, Burton A. and Grier, David L.},
  journal = {J. Chem. Inf. Comput. Sci.},
  year    = {2002},
  volume  = {42},
  number  = {6},
  pages   = {1296--1310},
  doi     = {10.1021/ci020023s},
}

@article{rxn_classification_2013,
  title   = {Algorithm for Reaction Classification},
  author  = {Kraut, Hans and Eiblmaier, Josef and Grethe, Guenter and Löw, Peter and Matuszczyk, Heinz and Saller, Heinz},
  journal = {J. Chem. Inf. Model.},
  year    = {2013},
  volume  = {53},
  number  = {11},
  pages   = {2884--2895},
  doi     = {10.1021/ci400442f},
}

@article{digdisc,
  title   = {Benchmarking machine-readable vectors of chemical reactions on computed activation barriers},
  author  = {van Gerwen, Puck and Briling, Ksenia R. and Calvino Alonso, Yannick and Franke, Malte and Corminboeuf, Clemence},
  journal = {Dig. Disc.},
  year    = {2024},
  volume  = {3},
  number  = {5},
  pages   = {932--943},
  doi     = {10.1039/D3DD00175J},
}

@article{vangerwen2023equireact,
  title   = {{3DReact}: {Geometric} deep learning for chemical reactions},
  author  = {van Gerwen, Puck and Briling, Ksenia R. and Bunne, Charlotte and Somnath, Vignesh Ram and Laplaza, Ruben and Krause, Andreas and Corminboeuf, Clemence},
  journal = {J. Chem. Inf. Model.},
  year    = {2024},
  volume  = {64},
  number  = {15},
  pages   = {5771--5785},
  doi     = {10.1021/acs.jcim.4c00104},
}

@article{huang2006maximum,
  title   = {Maximum common subgraph: some upper bound and lower bound results},
  author  = {Huang, Xiuzhen and Lai, Jing and Jennings, Steven F},
  journal = {BMC Bioinform.},
  year    = {2006},
  volume  = {7},
  pages   = {S6},
  doi     = {10.1186/1471-2105-7-S4-S6},
}

@article{bahiense2012maximum,
  title   = {The maximum common edge subgraph problem: A polyhedral investigation},
  author  = {Bahiense, Laura and Manić, Gordana and Piva, Breno and {de Souza}, Cid C.},
  journal = {Discrete Appl. Math.},
  year    = {2012},
  volume  = {160},
  number  = {18},
  pages   = {2523--2541},
  doi     = {10.1016/j.dam.2012.01.026},
  issn    = {0166-218X},
}

@article{first_stereochem_2012,
  title   = {Stereochemically Consistent Reaction Mapping and Identification of Multiple Reaction Mechanisms through Integer Linear Optimization},
  author  = {First, Eric L. and Gounaris, Chrysanthos E. and Floudas, Christodoulos A.},
  journal = {J. Chem. Inf. Model.},
  year    = {2012},
  volume  = {52},
  number  = {1},
  pages   = {84--92},
  doi     = {10.1021/ci200351b},
}

@article{mann2014atom,
  title   = {Atom mapping with constraint programming},
  author  = {Mann, Martin and Nahar, Feras and Schnorr, Norah and Backofen, Rolf and Stadler, Peter F and Flamm, Christoph},
  journal = {Algorithms Mol. Bio.},
  year    = {2014},
  volume  = {9},
  pages   = {23},
  doi     = {10.1186/s13015-014-0023-3},
}

@article{msa_2006,
  title   = {Multiple sequence alignment},
  author  = {Edgar, Robert C and Batzoglou, Serafim},
  journal = {Curr. Opin. Struct. Biol.},
  year    = {2006},
  volume  = {16},
  number  = {3},
  pages   = {368--373},
  doi     = {10.1016/j.sbi.2006.04.004},
}

@article{goscinski2023scikit,
  title   = {scikit-matter: A suite of generalisable machine learning methods born out of chemistry and materials science},
  author  = {Goscinski, Alexander and Principe, Victor Paul and Fraux, Guillaume and Kliavinek, Sergei and Helfrecht, Benjamin Aaron and Loche, Philip and Ceriotti, Michele and Cersonsky, Rose Kathleen},
  journal = {Open Res. Eur.},
  year    = {2023},
  volume  = {3},
  pages   = {81},
  doi     = {10.12688/openreseurope.15789.2},
}

@article{janes2013deterministic,
  title   = {Deterministic global optimization in ab-initio quantum chemistry},
  author  = {Janes, Pete P and Rendell, Alistair P},
  journal = {J. Glob. Optim.},
  year    = {2013},
  volume  = {56},
  number  = {2},
  pages   = {537--558},
  doi     = {10.1007/s10898-012-9868-5},
}

@article{Lavor_2007_deterministic,
  title   = {Solving Hartree-Fock systems with global optimization methods},
  author  = {Lavor, C. and Liberti, L. and Maculan, N. and Nascimento, M. A. C.},
  journal = {Europhys. Lett.},
  year    = {2007},
  volume  = {77},
  number  = {5},
  pages   = {50006},
  doi     = {10.1209/0295-5075/77/50006},
}

@article{noe2020machine,
  title   = {Machine learning for molecular simulation},
  author  = {Noé, Frank and Tkatchenko, Alexandre and Müller, Klaus-Robert and Clementi, Cecilia},
  journal = {Annu. Rev. Phys. Chem.},
  year    = {2020},
  volume  = {71},
  pages   = {361--390},
  doi     = {10.1146/annurev-physchem-042018-052331},
}

@article{langer_representations_2022,
  title   = {Representations of molecules and materials for interpolation of quantum-mechanical simulations via machine learning},
  author  = {Langer, Marcel F. and Goessmann, Alex and Rupp, Matthias},
  journal = {npj Comput. Mater.},
  year    = {2022},
  volume  = {8},
  pages   = {41},
  doi     = {10.1038/s41524-022-00721-x},
}

@article{fabregat2022metric,
  title   = {Metric learning for kernel ridge regression: assessment of molecular similarity},
  author  = {Fabregat, Raimon and van Gerwen, Puck and Haeberle, Matthieu and Eisenbrand, Friedrich and Corminboeuf, Clémence},
  journal = {Mach. Learn.: Sci. Technol.},
  year    = {2022},
  volume  = {3},
  number  = {3},
  pages   = {035015},
  doi     = {10.1088/2632-2153/ac8e4f},
}

@article{yamada_transfer_2019,
  title   = {Predicting Materials Properties with Little Data Using Shotgun Transfer Learning},
  author  = {Yamada, Hironao and Liu, Chang and Wu, Stephen and Koyama, Yukinori and Ju, Shenghong and Shiomi, Junichiro and Morikawa, Junko and Yoshida, Ryo},
  journal = {ACS Cent. Sci.},
  year    = {2019},
  volume  = {5},
  number  = {10},
  pages   = {1717--1730},
  doi     = {10.1021/acscentsci.9b00804},
}

@article{grambow_2019_transfer,
  title   = {Accurate Thermochemistry with Small Data Sets: A Bond Additivity Correction and Transfer Learning Approach},
  author  = {Grambow, Colin A. and Li, Yi-Pei and Green, William H.},
  journal = {J. Phys. Chem. A},
  year    = {2019},
  volume  = {123},
  number  = {27},
  pages   = {5826--5835},
  doi     = {10.1021/acs.jpca.9b04195},
}

@article{bai_transfer_2020,
  title   = {Transfer Learning: Making Retrosynthetic Predictions Based on a Small Chemical Reaction Dataset Scale to a New Level},
  author  = {Bai, Renren and Zhang, Chengyun and Wang, Ling and Yao, Chuansheng and Ge, Jiamin and Duan, Hongliang},
  journal = {Molecules},
  year    = {2020},
  volume  = {25},
  number  = {10},
  pages   = {2357},
  doi     = {10.3390/molecules25102357},
}

@article{shim2022transfer,
  title   = {Predicting reaction conditions from limited data through active transfer learning},
  author  = {Shim, Eunjae and Kammeraad, Joshua A and Xu, Ziping and Tewari, Ambuj and Cernak, Tim and Zimmerman, Paul M},
  journal = {Chem. Sci.},
  year    = {2022},
  volume  = {13},
  number  = {22},
  pages   = {6655--6668},
  doi     = {10.1039/D1SC06932B},
}

@article{king2024transfer,
  title   = {Transfer learning for a foundational chemistry model},
  author  = {King-Smith, Emma},
  journal = {Chem. Sci.},
  year    = {2024},
  volume  = {15},
  pages   = {5143--5151},
  doi     = {10.1039/D3SC04928K},
}

@article{jha2019transfer,
  title   = {Enhancing materials property prediction by leveraging computational and experimental data using deep transfer learning},
  author  = {Jha, Dipendra and Choudhary, Kamal and Tavazza, Francesca and Liao, Wei-keng and Choudhary, Alok and Campbell, Carelyn and Agrawal, Ankit},
  journal = {Nat. Commun.},
  year    = {2019},
  volume  = {10},
  number  = {1},
  pages   = {5316},
  doi     = {10.1038/s41467-019-13297-w},
}

@article{jackson2021transfer,
  title   = {TSNet: predicting transition state structures with tensor field networks and transfer learning},
  author  = {Jackson, Riley and Zhang, Wenyuan and Pearson, Jason},
  journal = {Chem. Sci.},
  year    = {2021},
  volume  = {12},
  number  = {29},
  pages   = {10022--10040},
  doi     = {10.1039/D1SC01206A},
}

@article{reker_active_2015,
  title   = {Active-learning strategies in computer-assisted drug discovery},
  author  = {Reker, Daniel and Schneider, Gisbert},
  journal = {Drug Discov. Today},
  year    = {2015},
  volume  = {20},
  number  = {4},
  pages   = {458--465},
  doi     = {10.1016/j.drudis.2014.12.004},
}

@article{wen2023active,
  title   = {Improving molecular machine learning through adaptive subsampling with active learning},
  author  = {Wen, Yujing and Li, Zhixiong and Xiang, Yan and Reker, Daniel},
  journal = {Dig. Disc.},
  year    = {2023},
  volume  = {2},
  number  = {4},
  pages   = {1134--1142},
  doi     = {10.1039/D3DD00037K},
}

@article{dodds2024active,
  title   = {Sample efficient reinforcement learning with active learning for molecular design},
  author  = {Dodds, Michael and Guo, Jeff and Löhr, Thomas and Tibo, Alessandro and Engkvist, Ola and Janet, Jon Paul},
  journal = {Chem. Sci.},
  year    = {2024},
  volume  = {15},
  number  = {11},
  pages   = {4146},
  doi     = {10.1039/d3sc04653b},
}

@article{raffel_transfer_transformer_2020,
  title   = {Exploring the Limits of Transfer Learning with a Unified Text-to-Text Transformer},
  author  = {Raffel, Colin and Shazeer, Noam and Roberts, Adam and Lee, Katherine and Narang, Sharan and Matena, Michael and Zhou, Yanqi and Li, Wei and Liu, Peter J.},
  journal = {J. Mach. Learn. Res.},
  year    = {2020},
  volume  = {21},
  number  = {140},
  pages   = {1--67},
  url     = {http://jmlr.org/papers/v21/20-074.html},
}

@article{taylor2009transfer,
  title   = {Transfer learning for reinforcement learning domains: A survey},
  author  = {Taylor, Matthew E and Stone, Peter},
  journal = {J. Mach. Learn. Res.},
  year    = {2009},
  volume  = {10},
  number  = {7},
  pages   = {1633--1685},
  url     = {http://jmlr.org/papers/v10/taylor09a.html},
}

@article{pesciullesi2020transfer,
  title   = {Transfer learning enables the molecular transformer to predict regio-and stereoselective reactions on carbohydrates},
  author  = {Pesciullesi, Giorgio and Schwaller, Philippe and Laino, Teodoro and Reymond, Jean-Louis},
  journal = {Nat. Commun.},
  year    = {2020},
  volume  = {11},
  number  = {1},
  pages   = {4874},
  doi     = {10.1038/s41467-020-18671-7},
}

@article{behler_generalized_2007,
  title   = {Generalized {Neural}-{Network} {Representation} of {High}-{Dimensional} {Potential}-{Energy} {Surfaces}},
  author  = {Behler, Jörg and Parrinello, Michele},
  journal = {Phys. Rev. Lett.},
  year    = {2007},
  volume  = {98},
  number  = {14},
  pages   = {146401},
  doi     = {10.1103/PhysRevLett.98.146401},
}

@article{nugmanov2019cgrtools,
  title   = {CGRtools: Python library for molecule, reaction, and condensed graph of reaction processing},
  author  = {Nugmanov, Ramil I and Mukhametgaleev, Ravil N and Akhmetshin, Tagir and Gimadiev, Timur R and Afonina, Valentina A and Madzhidov, Timur I and Varnek, Alexandre},
  journal = {J. Chem. Inf. Model.},
  year    = {2019},
  volume  = {59},
  number  = {6},
  pages   = {2516--2521},
  doi     = {10.1021/acs.jcim.9b00102},
}

@article{varnek2005substructural,
  title   = {Substructural fragments: an universal language to encode reactions, molecular and supramolecular structures},
  author  = {Varnek, Alexandre and Fourches, Denis and Hoonakker, Frank and Solov’ev, Vitaly P},
  journal = {J. Comput. Aided Mol. Des.},
  year    = {2005},
  volume  = {19},
  pages   = {693--703},
  doi     = {10.1007/s10822-005-9008-0},
}

@article{hansen2013assessment,
  title   = {Assessment and validation of machine learning methods for predicting molecular atomization energies},
  author  = {Hansen, Katja and Montavon, Grégoire and Biegler, Franziska and Fazli, Siamac and Rupp, Matthias and Scheffler, Matthias and Von Lilienfeld, O Anatole and Tkatchenko, Alexandre and Muller, Klaus-Robert},
  journal = {J. Chem. Theory Comput.},
  year    = {2013},
  volume  = {9},
  number  = {8},
  pages   = {3404--3419},
  doi     = {10.1021/ct400195d},
}

@article{bartok2017machine,
  title   = {Machine learning unifies the modeling of materials and molecules},
  author  = {Bartók, Albert P and De, Sandip and Poelking, Carl and Bernstein, Noam and Kermode, James R and Csányi, Gábor and Ceriotti, Michele},
  journal = {Sci. Adv.},
  year    = {2017},
  volume  = {3},
  number  = {12},
  pages   = {e1701816},
  doi     = {10.1126/sciadv.1701816},
}

@article{unke2017toolkit,
  title   = {Toolkit for the construction of reproducing kernel-based representations of data: Application to multidimensional potential energy surfaces},
  author  = {Unke, Oliver T and Meuwly, Markus},
  journal = {J. Chem. Inf. Model.},
  year    = {2017},
  volume  = {57},
  number  = {8},
  pages   = {1923--1931},
  doi     = {10.1021/acs.jcim.7b00090},
}

@article{gallarati_reaction-based_2021,
  title   = {Reaction-based machine learning representations for predicting the enantioselectivity of organocatalysts},
  author  = {Gallarati, Simone and Fabregat, Raimon and Laplaza, Rubén and Bhattacharjee, Sinjini and Wodrich, Matthew D. and Corminboeuf, Clemence},
  journal = {Chem. Sci.},
  year    = {2021},
  volume  = {12},
  number  = {20},
  pages   = {6879--6889},
  doi     = {10.1039/D1SC00482D},
}

@article{de_comparing_2016,
  title   = {Comparing molecules and solids across structural and alchemical space},
  author  = {De, Sandip and Bartók, Albert P. and Csányi, Gábor and Ceriotti, Michele},
  journal = {Phys. Chem. Chem. Phys.},
  year    = {2016},
  volume  = {18},
  number  = {20},
  pages   = {13754--13769},
  doi     = {10.1039/C6CP00415F},
}

@article{gallarati2024genetic,
  title   = {A genetic optimization strategy with generality in asymmetric organocatalysis as a primary target},
  author  = {Gallarati, Simone and van Gerwen, Puck and Laplaza, Ruben and Brey, Lucien and Makaveev, Alexander and Corminboeuf, Clemence},
  journal = {Chem. Sci.},
  year    = {2024},
  volume  = {15},
  pages   = {3640--3660},
  doi     = {10.1039/D3SC06208B},
}

@article{helfrecht2020structure,
  title   = {Structure-property maps with Kernel principal covariates regression},
  author  = {Helfrecht, Benjamin A and Cersonsky, Rose K and Fraux, Guillaume and Ceriotti, Michele},
  journal = {Mach. Learn.: Sci. Technol.},
  year    = {2020},
  volume  = {1},
  number  = {4},
  pages   = {045021},
  doi     = {10.1088/2632-2153/aba9ef},
}

@article{fabregat2020hamiltonian,
  title   = {Hamiltonian-reservoir replica exchange and machine learning potentials for computational organic chemistry},
  author  = {Fabregat, Raimon and Fabrizio, Alberto and Meyer, Benjamin and Hollas, Daniel and Corminboeuf, Clemence},
  journal = {J. Chem. Theory Comput.},
  year    = {2020},
  volume  = {16},
  number  = {5},
  pages   = {3084--3094},
  doi     = {10.1021/acs.jctc.0c00100},
}

@article{imbalzano2021uncertainty,
  title   = {Uncertainty estimation for molecular dynamics and sampling},
  author  = {Imbalzano, Giulio and Zhuang, Yongbin and Kapil, Venkat and Rossi, Kevin and Engel, Edgar A and Grasselli, Federico and Ceriotti, Michele},
  journal = {J. Chem. Phys.},
  year    = {2021},
  volume  = {154},
  number  = {7},
  pages   = {074102},
  doi     = {10.1063/5.0036522},
}

@article{celerse2024organic,
  title   = {From Organic Fragments to Photoswitchable Catalysts: The OFF--ON Structural Repository for Transferable Kernel-Based Potentials},
  author  = {Célerse, Frédéric and Wodrich, Matthew D and Vela, Sergi and Gallarati, Simone and Fabregat, Raimon and Juraskova, Veronika and Corminboeuf, Clémence},
  journal = {J. Chem. Inf. Model.},
  year    = {2024},
  volume  = {64},
  number  = {4},
  pages   = {1201--1212},
  doi     = {10.1021/acs.jcim.3c01953},
}

@article{mouvet2022recent,
  title   = {Recent advances in first-principles based molecular dynamics},
  author  = {Mouvet, François and Villard, Justin and Bolnykh, Viacheslav and Rothlisberger, Ursula},
  journal = {Acc. Chem. Res.},
  year    = {2022},
  volume  = {55},
  number  = {3},
  pages   = {221--230},
  doi     = {10.1021/acs.accounts.1c00503},
}

@article{raimbault2019using,
  title   = {Using Gaussian process regression to simulate the vibrational Raman spectra of molecular crystals},
  author  = {Raimbault, Nathaniel and Grisafi, Andrea and Ceriotti, Michele and Rossi, Mariana},
  journal = {New J. Phys.},
  year    = {2019},
  volume  = {21},
  number  = {10},
  pages   = {105001},
  doi     = {10.1088/1367-2630/ab4509},
}

@article{fujita1986description,
  title   = {Description of organic reactions based on imaginary transition structures. 1. Introduction of new concepts},
  author  = {Fujita, Shinsaku},
  journal = {J. Chem. Inf. Comput. Sci.},
  year    = {1986},
  volume  = {26},
  number  = {4},
  pages   = {205--212},
  doi     = {10.1021/ci00052a009},
}

@article{vapnik_local,
  title   = {{Local Learning Algorithms}},
  author  = {Bottou, Léon and Vapnik, Vladimir},
  journal = {Neural Comput.},
  year    = {1992},
  volume  = {4},
  number  = {6},
  pages   = {888--900},
  doi     = {10.1162/neco.1992.4.6.888},
}

@article{krenn2022selfies,
  title   = {SELFIES and the future of molecular string representations},
  author  = {Krenn, Mario and Ai, Qianxiang and Barthel, Senja and Carson, Nessa and Frei, Angelo and Frey, Nathan C and Friederich, Pascal and Gaudin, Théophile and Gayle, Alberto Alexander and Jablonka, Kevin Maik and others},
  journal = {Patterns},
  year    = {2022},
  volume  = {3},
  number  = {10},
  pages   = {100588},
  doi     = {10.1016/j.patter.2022.100588},
}

@article{vela2022cell2mol,
  title   = {cell2mol: encoding chemistry to interpret crystallographic data},
  author  = {Vela, Sergi and Laplaza, Ruben and Cho, Yuri and Corminboeuf, Clémence},
  journal = {npj Comput. Mater.},
  year    = {2022},
  volume  = {8},
  pages   = {188},
  doi     = {10.1038/s41524-022-00874-9},
}

@article{van2008visualizing,
  title   = {Visualizing data using t-SNE},
  author  = {Van der Maaten, Laurens and Hinton, Geoffrey},
  journal = {J. Mach. Learn. Res.},
  year    = {2008},
  volume  = {9},
  number  = {11},
  pages   = {2579--2605},
  url     = {https://jmlr.org/papers/v9/vandermaaten08a.html},
}

@article{Linderman_2019,
  title   = {Fast interpolation-based t-{SNE} for improved visualization of single-cell {RNA}-seq data},
  author  = {Linderman, George C. and Rachh, Manas and Hoskins, Jeremy G. and Steinerberger, Stefan and Kluger, Yuval},
  journal = {Nat. Methods},
  year    = {2019},
  volume  = {16},
  number  = {3},
  pages   = {243--245},
  doi     = {10.1038/s41592-018-0308-4},
}

@article{Policar2024,
  title   = {{openTSNE}: {A} modular {P}ython library for t-{SNE} dimensionality reduction and embedding},
  author  = {Poli{\v c}ar, Pavlin G. and Stra{\v z}ar, Martin and Zupan, Bla{\v z}},
  journal = {J. Stat. Softw.},
  year    = {2024},
  volume  = {109},
  number  = {3},
  pages   = {1--30},
  doi     = {10.18637/jss.v109.i03},
}

@article{Dral2015,
  title   = {Machine learning of parameters for accurate semiempirical quantum chemical calculations},
  author  = {Dral, Pavlo O. and von Lilienfeld, O. Anatole and Thiel, Walter},
  journal = {J. Chem. Theory Comput.},
  year    = {2015},
  volume  = {11},
  number  = {5},
  pages   = {2120--2125},
  doi     = {10.1021/acs.jctc.5b00141},
}

@article{Reddy2021,
  title   = {A hybrid quantum regression model for the prediction of molecular atomization energies},
  author  = {Reddy, Pranath and Bhattacherjee, Aranya B},
  journal = {Mach. Learn.: Sci. Technol.},
  year    = {2021},
  volume  = {2},
  number  = {2},
  pages   = {025019},
  doi     = {10.1088/2632-2153/abd486},
}

@article{Jurskov2022,
  title   = {Assessing the persistence of chalcogen bonds in solution with neural network potentials},
  author  = {Jurásková, Veronika and Célerse, Frederic and Laplaza, Ruben and Corminboeuf, Clemence},
  journal = {J. Chem. Phys.},
  year    = {2022},
  volume  = {156},
  number  = {15},
  pages   = {154112},
  doi     = {10.1063/5.0085153},
}

@article{Clerse2024,
  title   = {Capturing dichotomic solvent behavior in solute–solvent reactions with neural network potentials},
  author  = {Célerse, Frédéric and Juraskova, Veronika and Das, Shubhajit and Wodrich, Matthew D. and Corminboeuf, Clemence},
  journal = {J. Chem. Theory Comput.},
  year    = {2024},
  volume  = {20},
  number  = {23},
  pages   = {10350--10361},
  doi     = {10.1021/acs.jctc.4c01201},
}

@article{heinen2023reducing,
  title   = {Reducing Training Data Needs with Minimal Multilevel Machine Learning (M3L)},
  author  = {Heinen, Stefan and Khan, Danish and von Rudorff, Guido Falk and Karandashev, Konstantin and Arrieta, Daniel Jose Arismendi and Price, Alastair J. A. and Nandi, Surajit and Bhowmik, Arghya and Hermansson, Kersti and von Lilienfeld, O. Anatole},
  journal = {arXiv preprint},
  year    = {2023},
  pages   = {2308.11196},
  doi     = {10.48550/arXiv.2308.11196},
}

@article{mirzasoleiman2020coresets,
  title   = {Coresets for Data-efficient Training of Machine Learning Models},
  author  = {Mirzasoleiman, Baharan and Bilmes, Jeff and Leskovec, Jure},
  journal = {arXiv preprint},
  year    = {2020},
  pages   = {1906.01827},
  doi     = {10.48550/arXiv.1906.01827},
}

@article{bachem2017coresets,
  title   = {Practical Coreset Constructions for Machine Learning},
  author  = {Bachem, Olivier and Lucic, Mario and Krause, Andreas},
  journal = {arXiv preprint},
  year    = {2017},
  pages   = {1703.06476},
  doi     = {10.48550/arXiv.1703.06476},
}

@article{mol_design_ILP_2021,
  title   = {A Method for Inferring Polymers Based on Linear Regression and Integer Programming},
  author  = {Ido, Ryota and Cao, Shengjuan and Zhu, Jianshen and Azam, Naveed Ahmed and Haraguchi, Kazuya and Zhao, Liang and Nagamochi, Hiroshi and Akutsu, Tatsuya},
  journal = {arXiv preprint},
  year    = {2021},
  pages   = {2109.02628},
  doi     = {10.48550/arXiv.2109.02628},
}

@article{Dral_2017,
  title   = {Structure-based sampling and self-correcting machine learning for accurate calculations of potential energy surfaces and vibrational levels},
  author  = {Dral, Pavlo O. and Owens, Alec and Yurchenko, Sergei N. and Thiel, Walter},
  journal = {J. Chem. Phys.},
  year    = {2017},
  volume  = {146},
  number  = {24},
  pages   = {244108},
  doi     = {10.1063/1.4989536},
}

@article{Ullah_2024,
  title   = {Molecular quantum chemical data sets and databases for machine learning potentials},
  author  = {Ullah, Arif and Chen, Yuxinxin and Dral, Pavlo O},
  journal = {Mach. Learn.: Sci. Technol.},
  year    = {2024},
  volume  = {5},
  number  = {4},
  pages   = {041001},
  doi     = {10.1088/2632-2153/ad8f13},
}

@book{wolsey2014integer,
  title     = {Integer and combinatorial optimization},
  author    = {Wolsey, Laurence A and Nemhauser, George L},
  year      = {2014},
  publisher = {John Wiley \& Sons},
}

@report{yang_liu_distance_2006,
  title       = {Distance {Metric} {Learning}: {A} {Comprehensive} {Survey}},
  author      = {Yang, Liu and Rong, Jin},
  year        = {2006},
  institution = {Michigan State University},
  type        = {techreport},
  url         = {https://www.cs.cmu.edu/~liuy/frame_survey_v2.pdf},
}

@inproceedings{borsos_coresets_2020,
  title     = {Coresets via Bilevel Optimization for Continual Learning and Streaming},
  author    = {Borsos, Zalán and Mutny, Mojmir and Krause, Andreas},
  year      = {2020},
  volume    = {33},
  pages     = {14879--14890},
  doi       = {10.48550/arXiv.2006.03875},
  booktitle = {Advances in Neural Information Processing Systems},
  editor    = {Larochelle, H. and Ranzato, M. and Hadsell, R. and Balcan, M.F. and Lin, H.},
  publisher = {Curran Associates, Inc.},
}

@inproceedings{mol_design_MILP_2020,
  title     = {A Method for the Inverse QSAR/QSPR Based on Artificial Neural Networks and Mixed Integer Linear Programming},
  author    = {Chiewvanichakorn, Rachaya and Wang, Chenxi and Zhang, Zhe and Shurbevski, Aleksandar and Nagamochi, Hiroshi and Akutsu, Tatsuya},
  year      = {2020},
  pages     = {40--46},
  doi       = {10.1145/3386052.3386054},
  booktitle = {Proceedings of the 2020 10th International Conference on Bioscience, Biochemistry and Bioinformatics},
  location  = {Kyoto, Japan},
  publisher = {Association for Computing Machinery},
  series    = {ICBBB},
}

@inproceedings{mol_design_ILP_2022,
  title     = {A Method for Molecular Design Based on Linear Regression and Integer Programming},
  author    = {Zhu, Jianshen and Azam, Naveed A. and Haraguchi, Kazuya and Zhao, Liang and Nagamochi, Hiroshi and Akutsu, Tatsuya},
  year      = {2022},
  pages     = {21--28},
  doi       = {10.1145/3510427.3510431},
  booktitle = {Proceedings of the 2022 12th International Conference on Bioscience, Biochemistry and Bioinformatics},
  location  = {Tokyo, Japan},
  publisher = {Association for Computing Machinery},
  series    = {ICBBB '22},
}

@inproceedings{rna_alignment_ILP_1998,
  title     = {A polyhedral approach to {RNA} sequence structure alignment},
  author    = {Lenhof, Hans-Peter and Reinert, Knut and Vingron, Martin},
  year      = {1998},
  pages     = {153--162},
  doi       = {10.1145/279069.279109},
  booktitle = {Proceedings of the second annual international conference on Computational molecular biology},
  publisher = {ACM Press},
  series    = {RECOMB ’98},
}

@inproceedings{rna_alignment_ILP_2002,
  title     = {Structural alignment of large—size proteins via lagrangian relaxation},
  author    = {Caprara, Alberto and Lancia, Giuseppe},
  year      = {2002},
  pages     = {100--108},
  doi       = {10.1145/565196.565209},
  booktitle = {Proceedings of the Sixth Annual International Conference on Computational Biology},
  publisher = {ACM},
  series    = {RECOMB '02},
}

@inproceedings{protein_protein_ILP_2000,
  title     = {A combinatorial approach to protein docking with flexible side-chains},
  author    = {Althaus, E. and Kohlbacher, O. and Lenhof, H.-P. and Müller, P.},
  year      = {2000},
  pages     = {15--24},
  doi       = {10.1145/332306.332319},
  booktitle = {Proceedings of the Fourth Annual International Conference on Computational Molecular Biology},
  location  = {Tokyo, Japan},
  publisher = {Association for Computing Machinery},
  series    = {RECOMB '00},
}

@inproceedings{mass_spec_ILP_2008,
  title     = {Computing H/D-exchange speeds of single residues from data of peptic fragments},
  author    = {Althaus, Ernst and Canzar, Stefan and Emmett, Mark R. and Karrenbauer, Andreas and Marshall, Alan G. and Meyer-Baese, Anke and Zhang, Huimin},
  year      = {2008},
  pages     = {1273--1277},
  doi       = {10.1145/1363686.1363981},
  booktitle = {Proceedings of the 2008 ACM Symposium on Applied Computing},
  location  = {Fortaleza, Ceara, Brazil},
  publisher = {Association for Computing Machinery},
  series    = {SAC '08},
}

@inproceedings{weinberger_metric_2007,
  title     = {Metric Learning for Kernel Regression},
  author    = {Weinberger, Kilian Q. and Tesauro, Gerald},
  year      = {2007},
  volume    = {2},
  pages     = {612--619},
  booktitle = {Proceedings of the Eleventh International Conference on Artificial Intelligence and Statistics},
  editor    = {Meila, Marina and Shen, Xiaotong},
  location  = {San Juan, Puerto Rico},
  publisher = {PMLR},
  series    = {Proceedings of Machine Learning Research},
}

@inbook{mol_design_MILP_2_2020,
  title     = {A New Integer Linear Programming Formulation to the Inverse {QSAR}/{QSPR} for Acyclic Chemical Compounds Using Skeleton Trees},
  author    = {Zhang, Fan and Zhu, Jianshen and Chiewvanichakorn, Rachaya and Shurbevski, Aleksandar and Nagamochi, Hiroshi and Akutsu, Tatsuya},
  year      = {2020},
  pages     = {433--444},
  doi       = {10.1007/978-3-030-55789-8_38},
  booktitle = {Trends in Artificial Intelligence Theory and Applications. Artificial Intelligence Practices},
  editor    = {Fujita, Hamido and Fournier-Viger, Philippe and Ali, Moonis and Sasaki, Jun},
  publisher = {Springer International Publishing},
}

@inbook{schrijver,
  title     = {Linear programming methods and the bipartite matching polytope},
  author    = {Schrijver, Alexander},
  year      = {2012},
  booktitle = {Combinatorial Optimisation: Polyhedra and Efficiency},
  chapter   = {17},
  publisher = {Springer},
}

@inbook{book_ILP_bio,
  title     = {Integer Linear Programming in Computational Biology},
  author    = {Althaus, Ernst and Klau, Gunnar W. and Kohlbacher, Oliver and Lenhof, Hans-Peter and Reinert, Knut},
  year      = {2009},
  pages     = {199--218},
  doi       = {10.1007/978-3-642-03456-5_14},
  booktitle = {Efficient Algorithms},
  editor    = {Albers, Susanne and Alt, Helmut and Näher, Stefan},
  publisher = {Springer Berlin Heidelberg},
}

@misc{enamine,
  title        = {Enamine REAL compounds},
  year         = {2020},
  howpublished = {\url{https://enamine.net/library-synthesis/real-compounds}},
}

@misc{gurobi,
  title  = {{Gurobi Optimizer Reference Manual}},
  author = {{Gurobi Optimization, LLC}},
  year   = {2022},
  url    = {https://www.gurobi.com},
}

@misc{qml,
  title        = {{QML}: A Python Toolkit for Quantum Machine Learning},
  author       = {Christensen, Anders S. and Faber, Felix and Huang, Bing and Bratholm, Lars and Tkatchenko, Alexandre and Müller, Klaus-Robert and von Lilienfeld, O. Anatole},
  year         = {2017},
  howpublished = {\url{https://github.com/qmlcode/qml}},
  publisher    = {GitHub},
}

@phdthesis{puck_thesis,
  title   = {Machine-learning quantum-chemical properties of molecules and chemical reactions},
  author  = {van Gerwen, Puck},
  year    = {2024},
  school  = {EPFL},
  address = {Lausanne},
  doi     = {10.5075/epfl-thesis-10980},
  url     = {https://infoscience.epfl.ch/handle/20.500.14299/241305},
}

\clearpage
\end{document}


\title{Supplementary Information for ``Integer Linear Programming for Unsupervised Training Set Selection in Molecular Machine Learning}
\author{
Matthieu Haeberle,\textsuperscript{1,2,\ddag}
Puck van Gerwen,\textsuperscript{1,3,\ddag}\\
Ruben Laplaza,\textsuperscript{1,3}
Ksenia R. Briling,\textsuperscript{1}
Jan Weinreich,\textsuperscript{1,3} \\
Friedrich Eisenbrand\textsuperscript{2} and
Cl\'emence Corminboeuf\textsuperscript{1,3}\footnote{email: clemence.corminboeuf@epfl.ch} \\
\vspace{0.01cm}\\
\small
\textsuperscript{1}Laboratory for Computational Molecular Design,\\
\small Institute of Chemical Sciences and Engineering,\\ \small \'Ecole Polytechnique F\'ed\'erale de Lausanne, \\\small 1015 Lausanne, Switzerland\\
\vspace{0.01cm}\\
\small
\textsuperscript{2}Chair of Discrete Optimisation, \\
\small
\'Ecole Polytechnique F\'ed\'erale de Lausanne, \\
\small
1015 Lausanne, Switzerland\\
\vspace{0.01cm}\\
\small
\textsuperscript{3}National Center for Competence in Research-Catalysis (NCCR-Catalysis),\\\small Zurich, Switzerland \\
\vspace{0.01cm}\\
\small
\textsuperscript{\ddag}These authors contributed equally to this work.
}
\maketitle

\tableofcontents

\section{Hyperparameters}
Table~\ref{tab:hypers_KRR} enumerates the hyperparameters for each target molecule, for each model. In all cases, the hyperparameters are given for the largest training set size. The indices of the selected QM7 and QM9 molecules are given to identify each target.

\begin{sidewaystable}
\centering
\tiny
\begin{tabular}{@{}lllllllllllllllllllll@{}}
\toprule
Target
& \multicolumn{2}{l}{ILP($p=0$)}
& \multicolumn{2}{l}{ILP($p=1$)}
& \multicolumn{2}{l}{FPS}
& \multicolumn{2}{l}{CUR}
& \multicolumn{2}{l}{SML}
& \multicolumn{2}{l}{Random 1}
& \multicolumn{2}{l}{Random 2}
& \multicolumn{2}{l}{Random 3}
& \multicolumn{2}{l}{Random 4}
& \multicolumn{2}{l}{Random 5}
\\
& $\sigma$ & $\gamma$
& $\sigma$ & $\gamma$
& $\sigma$ & $\gamma$
& $\sigma$ & $\gamma$
& $\sigma$ & $\gamma$
& $\sigma$ & $\gamma$
& $\sigma$ & $\gamma$
& $\sigma$ & $\gamma$
& $\sigma$ & $\gamma$
& $\sigma$ & $\gamma$
\\ \midrule
QM7 1246      & 5.62E+00 & 1E-09 & 4.22E+01 & 1E-05 & 5.62E+00 & 1E-04 & 7.50E+00 & 1E-09 & 1.33E+01 & 1E-07 & 7.50E+00 & 1E-06 & 2.37E+00 & 1E-09 & 1.78E+01 & 1E-05 & 5.62E+01 & 1E-06 & 5.62E+01 & 1E-06 \\
QM7 1251      & 1.33E+00 & 1E-09 & 3.16E+00 & 1E-04 & 5.62E+00 & 1E-04 & 7.50E+00 & 1E-09 & 1E+00    & 1E-09 & 7.50E+00 & 1E-09 & 1.33E+01 & 1E-04 & 1E+01    & 1E-04 & 2.37E+00 & 1E-09 & 2.37E+01 & 1E-04 \\
QM7 1513      & 5.62E+00 & 1E-04 & 1.78E+02 & 1E-07 & 5.62E+00 & 1E-04 & 7.50E+00 & 1E-09 & 1.78E+01 & 1E-06 & 1.78E+02 & 1E-08 & 7.50E+02 & 1E-09 & 2.37E+00 & 1E-05 & 5.62E+00 & 1E-04 & 1.78E+00 & 1E-04 \\
QM7 2161      & 7.50E+00 & 1E-09 & 1.78E+00 & 1E-09 & 5.62E+00 & 1E-04 & 7.50E+00 & 1E-09 & 7.50E+00 & 1E-04 & 2.37E+01 & 1E-07 & 1.33E+02 & 1E-09 & 3.16E+02 & 1E-07 & 5.62E+01 & 1E-06 & 3.16E+01 & 1E-09 \\
QM7 3037      & 5.62E+02 & 1E-09 & 5.62E+00 & 1E-04 & 5.62E+02 & 1E-07 & 7.50E+00 & 1E-09 & 2.37E+01 & 1E-07 & 1E+01    & 1E-04 & 4.22E+00 & 1E-04 & 1.33E+01 & 1E-05 & 2.37E+01 & 1E-08 & 5.62E+01 & 1E-09 \\
QM7 3576      & 3.16E+01 & 1E-07 & 1E+02    & 1E-08 & 5.62E+00 & 1E-04 & 7.50E+00 & 1E-09 & 4.22E+00 & 1E-06 & 1E+01    & 1E-09 & 1E+03    & 1E-07 & 1E+00    & 1E-04 & 4.22E+02 & 1E-09 & 5.62E+01 & 1E-09 \\
QM7 5107      & 5.62E+01 & 1E-08 & 7.50E+00 & 1E-06 & 5.62E+00 & 1E-04 & 7.50E+00 & 1E-09 & 1.78E+02 & 1E-09 & 2.37E+01 & 1E-09 & 7.50E+00 & 1E-05 & 7.50E+00 & 1E-05 & 3.16E+00 & 1E-04 & 4.22E+01 & 1E-06 \\
QM7 5245      & 3.16E+01 & 1E-09 & 1E+01    & 1E-05 & 5.62E+00 & 1E-04 & 7.50E+00 & 1E-09 & 7.50E+00 & 1E-07 & 1.78E+02 & 1E-09 & 1E+02    & 1E-05 & 2.37E+01 & 1E-04 & 1.33E+00 & 1E-04 & 5.62E+00 & 1E-09 \\
QM7 6118      & 1.78E+01 & 1E-04 & 4.22E+01 & 1E-09 & 5.62E+00 & 1E-04 & 7.50E+00 & 1E-09 & 1.78E+00 & 1E-09 & 7.50E+01 & 1E-08 & 1E+00    & 1E-04 & 4.22E+02 & 1E-08 & 5.62E+00 & 1E-04 & 7.50E+01 & 1E-07 \\
QM7 6163      & 1.78E+01 & 1E-07 & 5.62E+00 & 1E-05 & 5.62E+00 & 1E-04 & 7.50E+00 & 1E-09 & 3.16E+00 & 1E-09 & 3.16E+01 & 1E-04 & 4.22E+01 & 1E-08 & 1.78E+00 & 1E-09 & 1E+00    & 1E-04 & 1.33E+02 & 1E-09 \\
QM9 5696      & 4.22E+02 & 1E-08 & 7.50E+02 & 1E-07 & 5.62E+00 & 1E-04 & 7.50E+00 & 1E-09 & 5.62E+00 & 1E-04 & 1E+01    & 1E-09 & 1.78E+02 & 1E-04 & 1.78E+01 & 1E-09 & 1E+01    & 1E-05 & 1.33E+01 & 1E-04 \\
QM9 12351     & 7.50E+00 & 1E-06 & 3.16E+00 & 1E-04 & 5.62E+00 & 1E-04 & 7.50E+00 & 1E-09 & 1.78E+00 & 1E-09 & 1E+01    & 1E-09 & 1.78E+02 & 1E-04 & 1.78E+01 & 1E-09 & 1E+01    & 1E-05 & 1.33E+01 & 1E-04 \\
QM9 31476     & 7.50E+00 & 1E-05 & 7.50E+02 & 1E-07 & 5.62E+00 & 1E-04 & 7.50E+00 & 1E-09 & 1.78E+02 & 1E-09 & 1E+01    & 1E-09 & 1.78E+02 & 1E-04 & 1.78E+01 & 1E-09 & 1E+01    & 1E-05 & 1.33E+01 & 1E-04 \\
QM9 35811     & 5.62E+00 & 1E-05 & 3.16E+00 & 1E-05 & 5.62E+00 & 1E-04 & 7.50E+00 & 1E-09 & 2.37E+00 & 1E-04 & 1E+01    & 1E-09 & 1.78E+02 & 1E-04 & 1.78E+01 & 1E-09 & 1E+01    & 1E-05 & 1.33E+01 & 1E-04 \\
QM9 55607     & 1.33E+00 & 1E-09 & 2.37E+01 & 1E-06 & 5.62E+00 & 1E-04 & 7.50E+00 & 1E-09 & 1E+00    & 1E-09 & 1E+01    & 1E-09 & 1.78E+02 & 1E-04 & 1.78E+01 & 1E-09 & 1E+01    & 1E-05 & 1.33E+01 & 1E-04 \\
QM9 68076     & 2.37E+00 & 1E-09 & 5.62E+01 & 1E-09 & 5.62E+00 & 1E-04 & 7.50E+00 & 1E-09 & 7.50E+00 & 1E-05 & 1E+01    & 1E-09 & 1.78E+02 & 1E-04 & 1.78E+01 & 1E-09 & 1E+01    & 1E-05 & 1.33E+01 & 1E-04 \\
QM9 85759     & 3.16E+00 & 1E-05 & 7.50E+01 & 1E-06 & 5.62E+00 & 1E-04 & 7.50E+00 & 1E-09 & 1E+00    & 1E-04 & 1E+01    & 1E-09 & 1.78E+02 & 1E-04 & 1.78E+01 & 1E-09 & 1E+01    & 1E-05 & 1.33E+01 & 1E-04 \\
QM9 96295     & 1.78E+02 & 1E-09 & 1E+02    & 1E-09 & 5.62E+00 & 1E-04 & 7.50E+00 & 1E-09 & 4.22E+00 & 1E-09 & 1E+01    & 1E-09 & 1.78E+02 & 1E-04 & 1.78E+01 & 1E-09 & 1E+01    & 1E-05 & 1.33E+01 & 1E-04 \\
QM9 120425    & 3.16E+00 & 1E-09 & 1E+03    & 1E-07 & 5.62E+00 & 1E-04 & 7.50E+00 & 1E-09 & 3.16E+01 & 1E-09 & 1E+01    & 1E-09 & 1.78E+02 & 1E-04 & 1.78E+01 & 1E-09 & 1E+01    & 1E-05 & 1.33E+01 & 1E-04 \\
QM9 121259    & 2.37E+00 & 1E-06 & 5.62E+02 & 1E-08 & 5.62E+00 & 1E-04 & 7.50E+00 & 1E-09 & 2.37E+00 & 1E-09 & 1E+01    & 1E-09 & 1.78E+02 & 1E-04 & 1.78E+01 & 1E-09 & 1E+01    & 1E-05 & 1.33E+01 & 1E-04 \\
apixaban      & 1E+02    & 1E-04 & 2.37E+00 & 1E-04 & 5.62E+00 & 1E-04 & 7.50E+00 & 1E-09 & 2.37E+01 & 1E-04 & 1E+01    & 1E-09 & 1.78E+02 & 1E-04 & 1.78E+01 & 1E-09 & 1E+01    & 1E-05 & 1.33E+01 & 1E-04 \\
imatinib      & 3.16E+01 & 1E-06 & 2.37E+02 & 1E-07 & 5.62E+00 & 1E-04 & 7.50E+00 & 1E-09 & 1.78E+00 & 1E-09 & 1E+01    & 1E-09 & 1.78E+02 & 1E-04 & 1.78E+01 & 1E-09 & 1E+01    & 1E-05 & 1.33E+01 & 1E-04 \\
oseltamivir   & 4.22E+00 & 1E-04 & 1E+02    & 1E-07 & 5.62E+00 & 1E-04 & 7.50E+00 & 1E-09 & 4.22E+01 & 1E-09 & 1E+01    & 1E-09 & 1.78E+02 & 1E-04 & 1.78E+01 & 1E-09 & 1E+01    & 1E-05 & 1.33E+01 & 1E-04 \\
oxycodone     & 2.37E+01 & 1E-08 & 1.33E+00 & 1E-09 & 5.62E+00 & 1E-04 & 7.50E+00 & 1E-09 & 2.37E+00 & 1E-04 & 1E+01    & 1E-09 & 1.78E+02 & 1E-04 & 1.78E+01 & 1E-09 & 1E+01    & 1E-05 & 1.33E+01 & 1E-04 \\
pemetrexed    & 3.16E+01 & 1E-08 & 1.33E+00 & 1E-09 & 5.62E+00 & 1E-04 & 7.50E+00 & 1E-09 & 1E+02    & 1E-09 & 1E+01    & 1E-09 & 1.78E+02 & 1E-04 & 1.78E+01 & 1E-09 & 1E+01    & 1E-05 & 1.33E+01 & 1E-04 \\
penicillin    & 1E+02    & 1E-08 & 7.50E+01 & 1E-04 & 5.62E+00 & 1E-04 & 7.50E+00 & 1E-09 & 1.78E+01 & 1E-06 & 1E+01    & 1E-09 & 1.78E+02 & 1E-04 & 1.78E+01 & 1E-09 & 1E+01    & 1E-05 & 1.33E+01 & 1E-04 \\
pregabalin    & 5.62E+00 & 1E-09 & 3.16E+01 & 1E-06 & 5.62E+00 & 1E-04 & 7.50E+00 & 1E-09 & 1E+01    & 1E-05 & 1E+01    & 1E-09 & 1.78E+02 & 1E-04 & 1.78E+01 & 1E-09 & 1E+01    & 1E-05 & 1.33E+01 & 1E-04 \\
salbutamol    & 1.78E+00 & 1E-04 & 1.78E+01 & 1E-09 & 5.62E+00 & 1E-04 & 7.50E+00 & 1E-09 & 1.78E+01 & 1E-07 & 1E+01    & 1E-09 & 1.78E+02 & 1E-04 & 1.78E+01 & 1E-09 & 1E+01    & 1E-05 & 1.33E+01 & 1E-04 \\
sildenafil    & 2.37E+00 & 1E-05 & 3.16E+00 & 1E-09 & 5.62E+00 & 1E-04 & 7.50E+00 & 1E-09 & 5.62E+00 & 1E-04 & 1E+01    & 1E-09 & 1.78E+02 & 1E-04 & 1.78E+01 & 1E-09 & 1E+01    & 1E-05 & 1.33E+01 & 1E-04 \\
troglitazone  & 7.50E+00 & 1E-04 & 5.62E+00 & 1E-09 & 5.62E+00 & 1E-04 & 7.50E+00 & 1E-09 & 1.33E+00 & 1E-05 & 1E+01    & 1E-09 & 1.78E+02 & 1E-04 & 1.78E+01 & 1E-09 & 1E+01    & 1E-05 & 1.33E+01 & 1E-04 \\
 \bottomrule
\end{tabular}
\caption{Hyperparameters of kernel models for the maximum training set size (1024). $\sigma$ is the kernel width and $\lambda$ the regularization parameter.}
\label{tab:hypers_KRR}
\end{sidewaystable}

\clearpage
\section{Distance and size plots}
\begin{figure}[h!]
    \centering
    \includegraphics[width=0.8\linewidth]{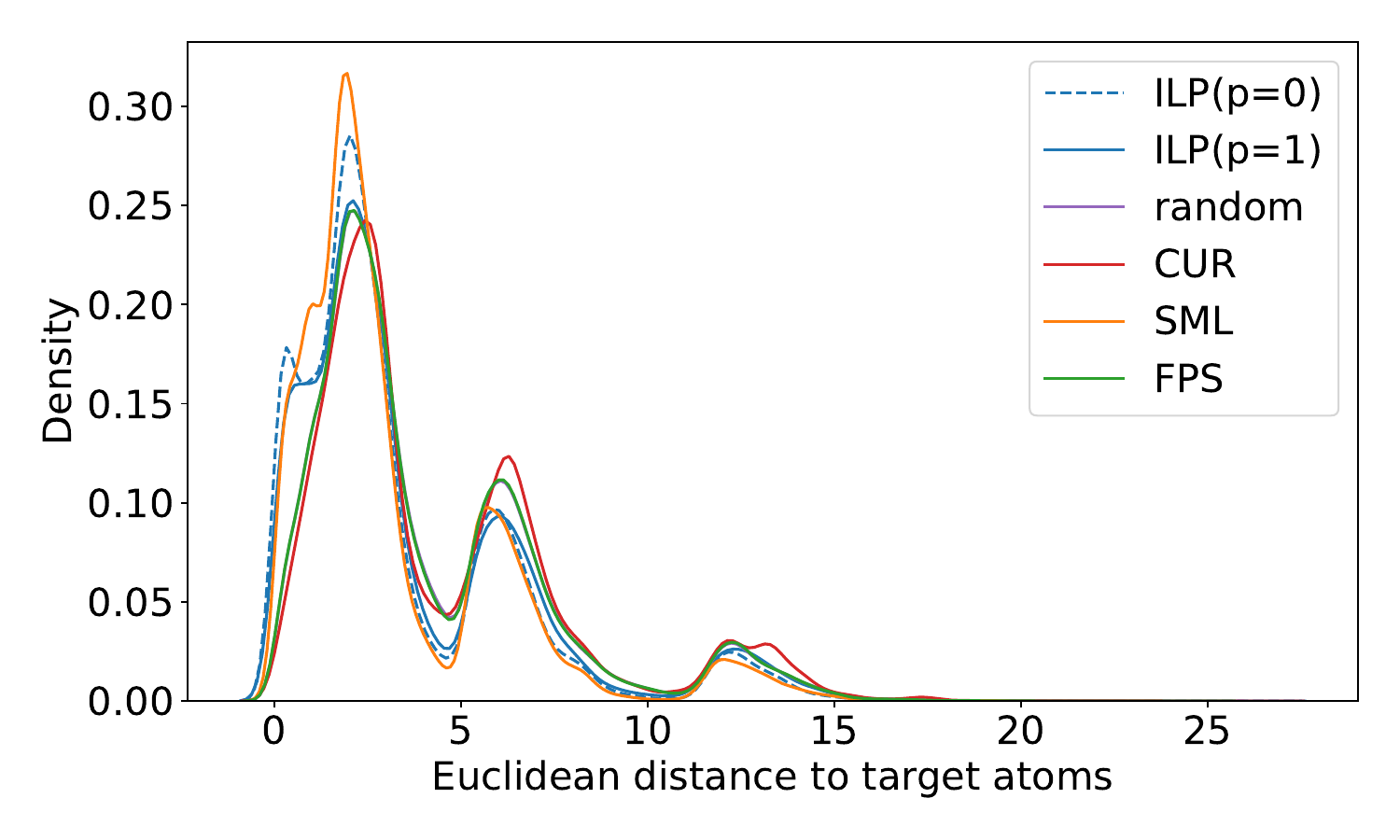}
    \caption{Density (kernel density estimate) of the subsets selected by each method according to their Euclidean distance to the target molecule atoms, for the QM7 target molecules.}
    \label{fig:distances_qm7}
\end{figure}

\begin{figure}[h!]
    \centering
    \includegraphics[width=0.8\linewidth]{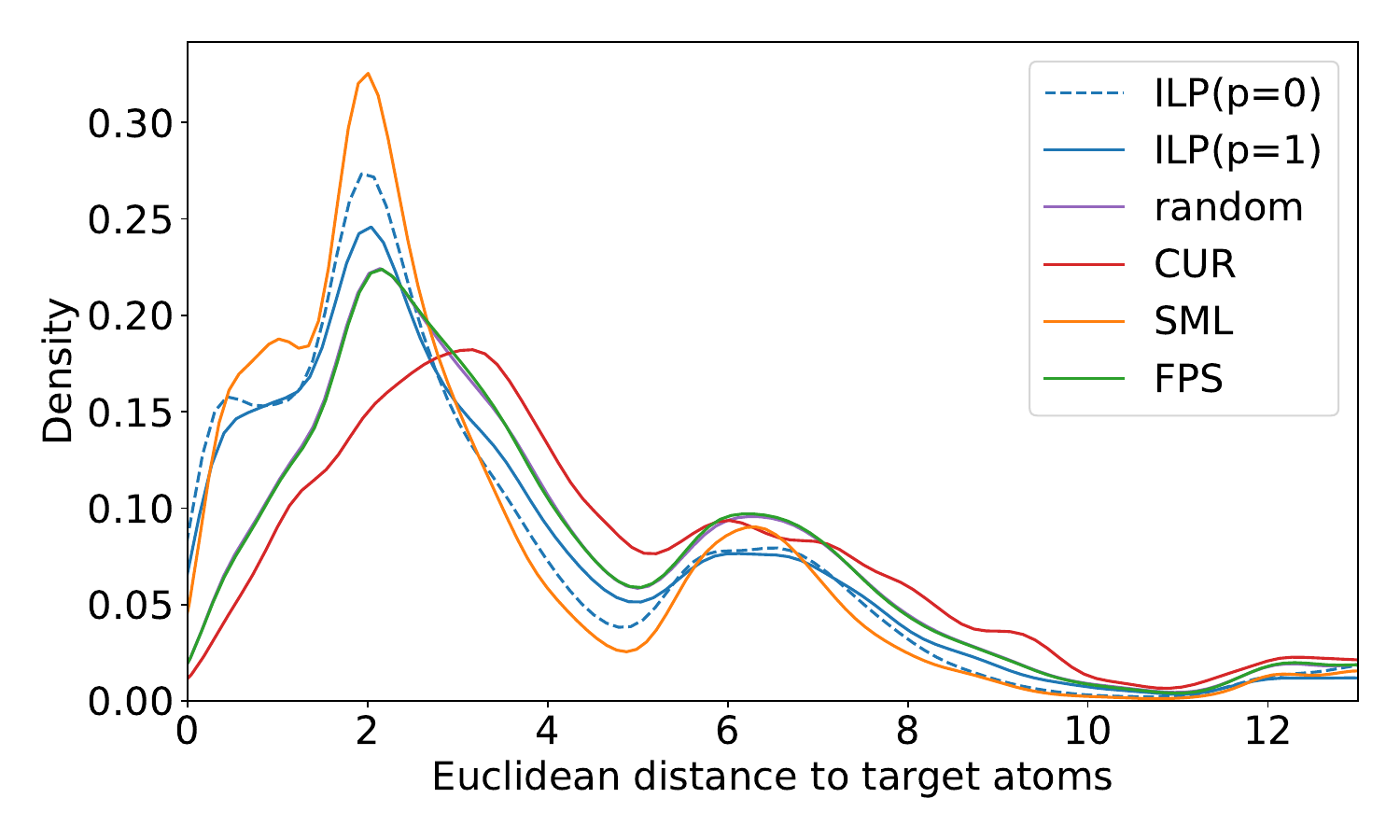}
    \caption{Density (kernel density estimate) of the subsets selected by each method according to their Euclidean distance to the target molecule atoms, for the QM9 target molecules (QM9*).}
    \label{fig:distances_qm9}
\end{figure}

Figures~\ref{fig:distances_qm7} and \ref{fig:distances_qm9} show the distance plots shown in the main text for the drug target molecules, now for QM7 and QM9 target molecules, respectively.
While the hierarchy of the density of close environments to the target is not directly correlated to their behavior in the learning curves, there is still a close relationship: for example, here, SML results in the most similar environments to the target, which results in good predictions (\textit{vs.} the equivalent plots for drug-like target molecules). The ILP result in a lower density around a distance of \SI{2}{\AA} but ILP($p=0$) shows an additional peak at an even closer distance to the target ($< \SI{1}{\AA}$), resulting in the lowest error for this model. Nevertheless, the ILP($p=1$) results in improved performance compared to SML, while demonstrating a lower density of close target atoms. In addition, CUR results in the least similar environments of the algorithms tested, but predictions on par with FPS. This illustrates that selecting similar environments alone is not enough to result in good predictions (in an interpolation regime), but in general explains the trend in the different models' behavior.

\begin{figure}[h!]
    \centering
    \includegraphics[width=0.8\linewidth]{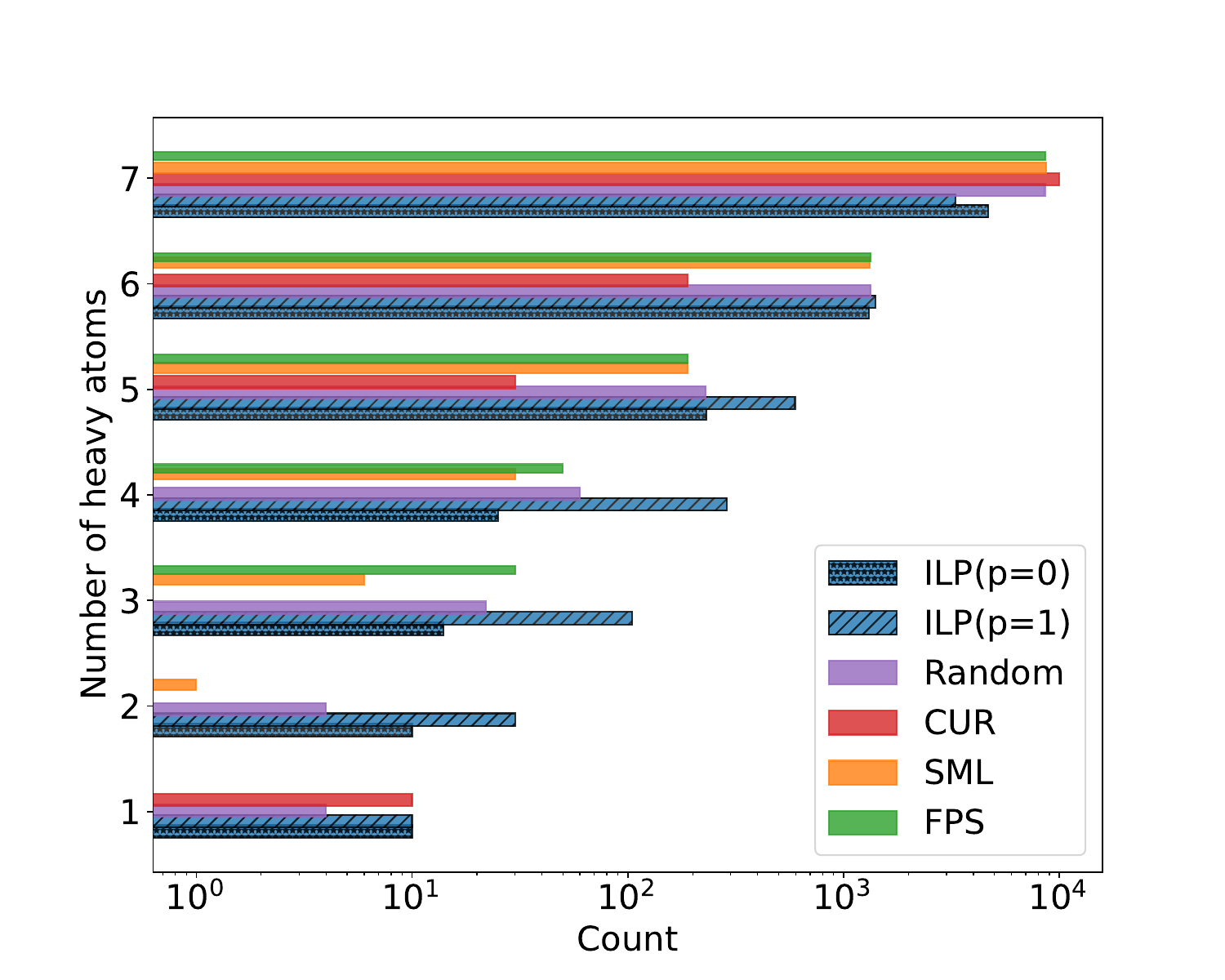}
    \caption{Counts of sizes of fragments selected by each search algorithm, for all QM7 target molecules combined. The count ($x$-axis) is on a log scale.}
    \label{fig:size_plots_qm7}
\end{figure}

\begin{figure}[h!]
    \centering
    \includegraphics[width=0.8\linewidth]{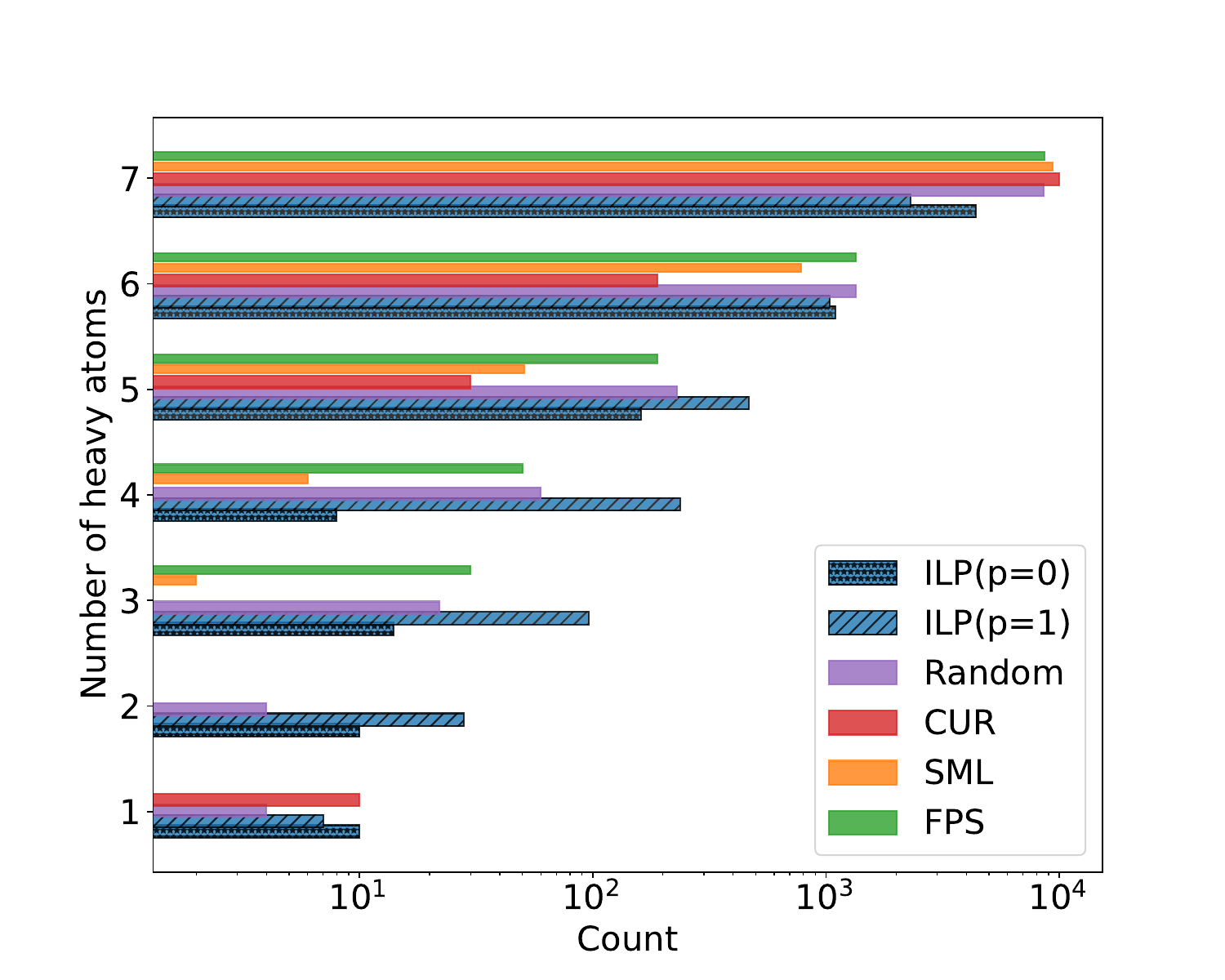}
    \caption{Counts of sizes of fragments selected by each search algorithm, for all QM9 target molecules (QM9*) combined. The count ($x$-axis) is on a log scale.}
    \label{fig:size_plots_qm9}
\end{figure}

\begin{figure}[h!]
    \centering
    \includegraphics[width=0.8\linewidth]{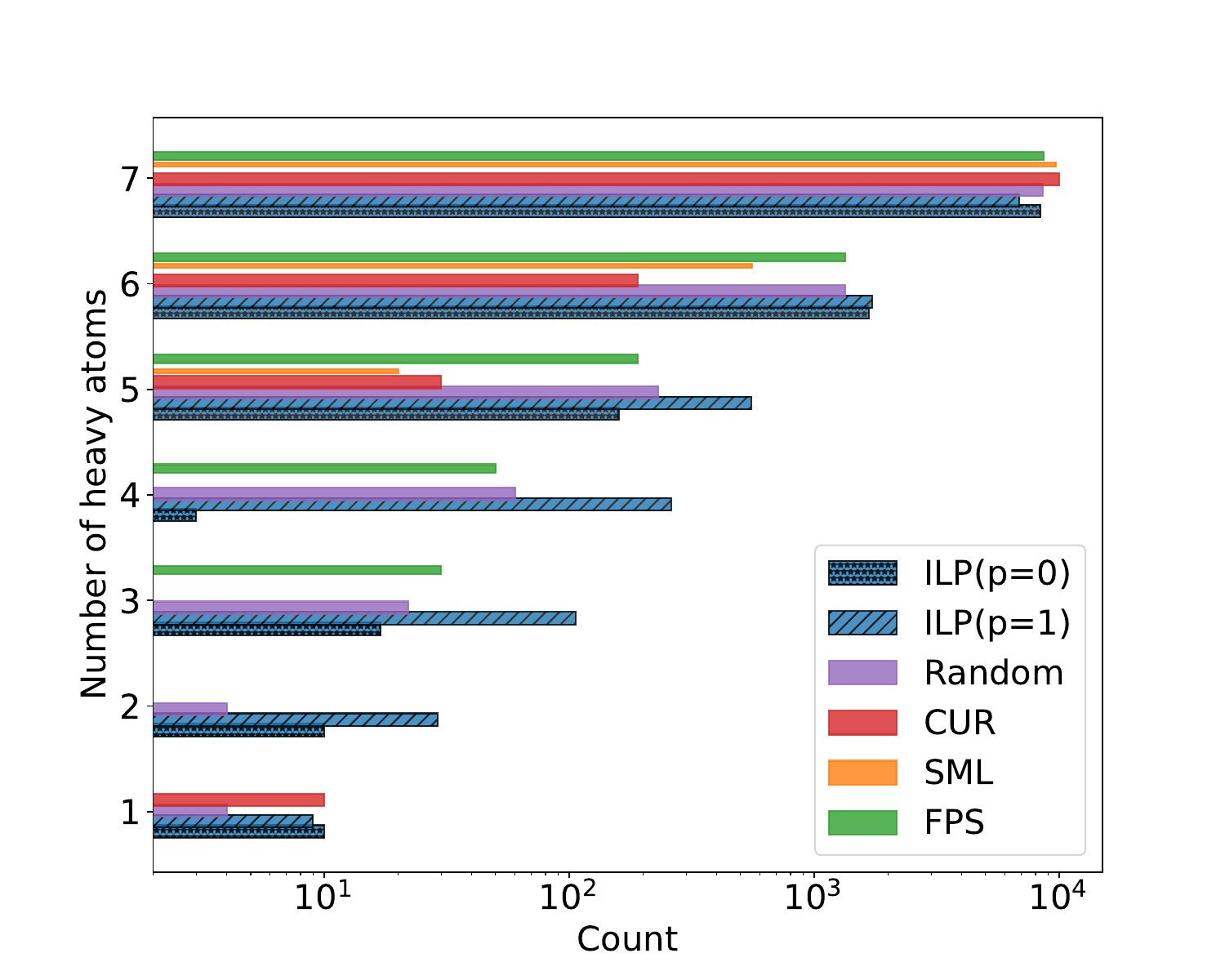}
    \caption{Counts of sizes of fragments selected by each search algorithm, for all drug target molecules combined. The count ($x$-axis) is on a log scale.}
    \label{fig:size_plots_drugs}
\end{figure}

Figures~\ref{fig:size_plots_qm7}--\ref{fig:size_plots_drugs} illustrate the sizes of the fragments selected by each search algorithm, for the QM7, QM9* and drug targets respectively. This illustrates the inherent behavior of the different algorithms --- keeping in mind that FPS, CUR and random will result in the same pool of fragments regardless of the target molecule (and are therefore the same for the drug, QM7 or QM9* targets).

\clearpage
\section{Individual learning curves}
\label{sec:learning_curves}

\begin{figure*}[h!]
    \centering
    \begin{subfigure}[t]{0.5\textwidth}
        \centering
        \includegraphics[width=0.9\linewidth]{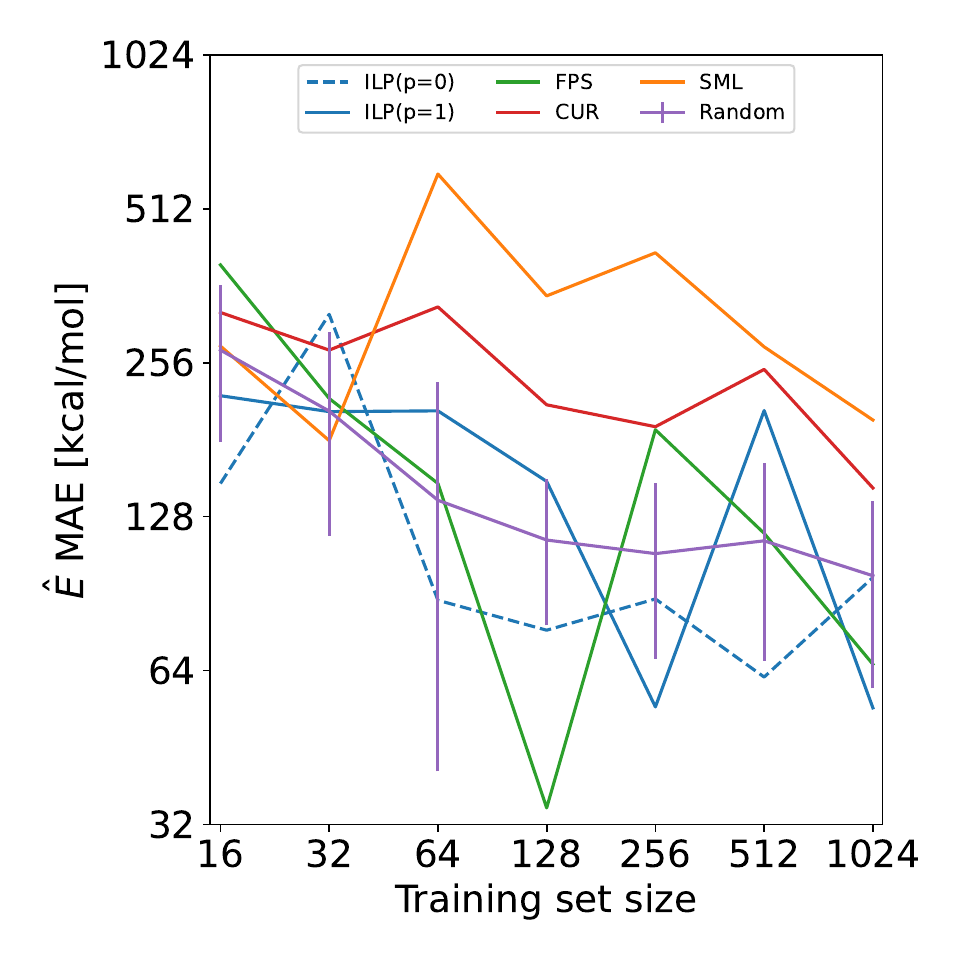}
        \caption{}
    \end{subfigure}%
    ~
    \begin{subfigure}[t]{0.5\textwidth}
        \centering
        \includegraphics[width=0.9\linewidth]{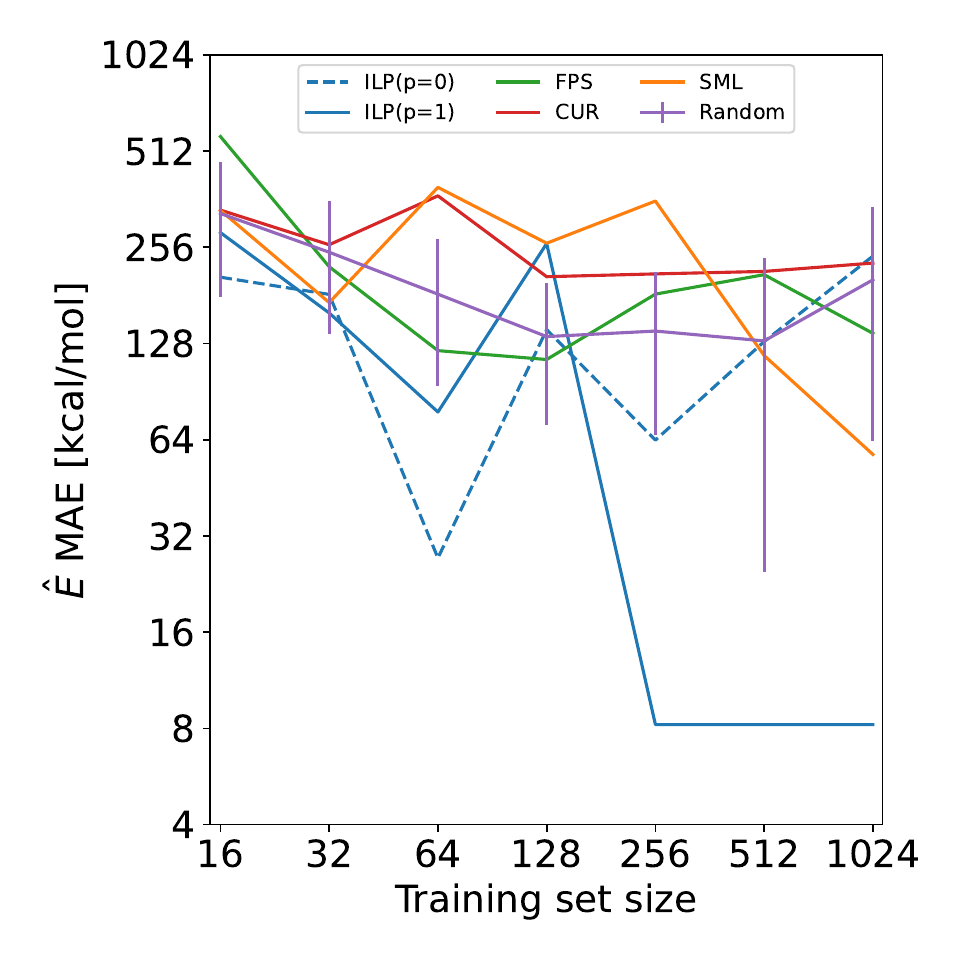}
        \caption{}
    \end{subfigure}
    \\
        \begin{subfigure}[t]{0.5\textwidth}
        \centering
        \includegraphics[width=0.9\linewidth]{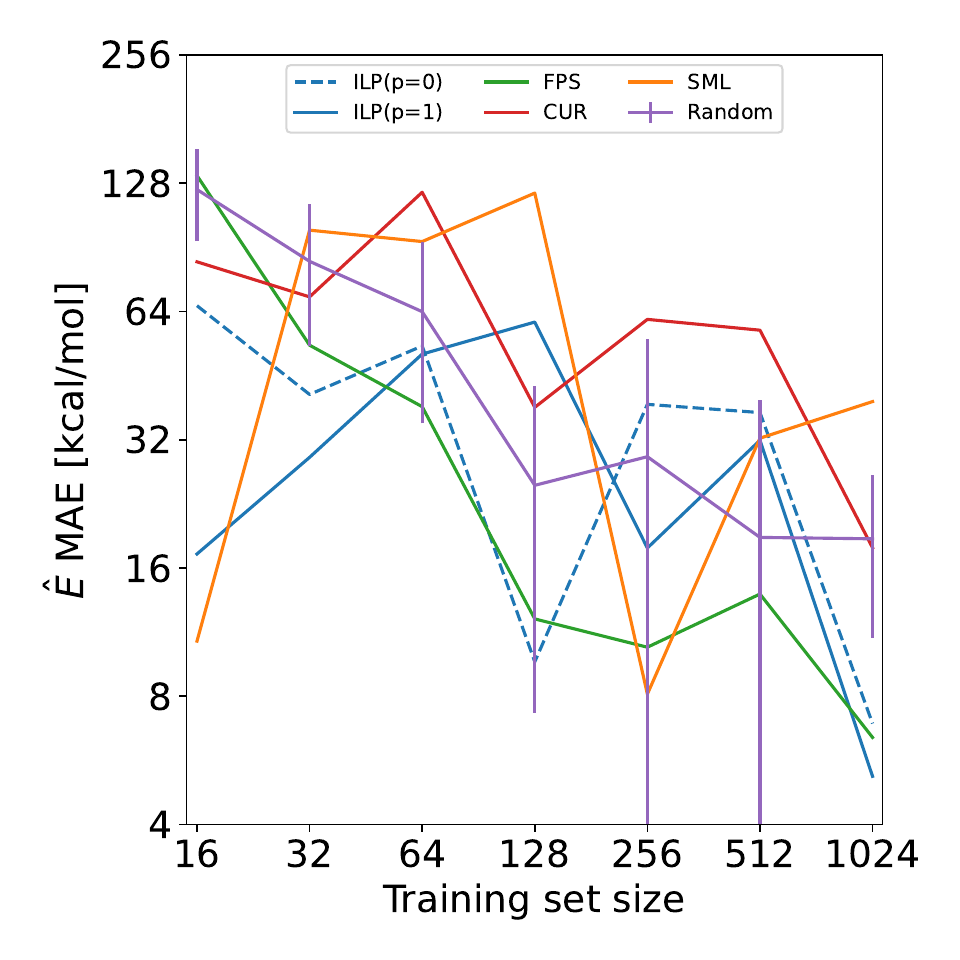}
        \caption{}
    \end{subfigure}%
    ~
    \begin{subfigure}[t]{0.5\textwidth}
        \centering
        \includegraphics[width=0.9\linewidth]{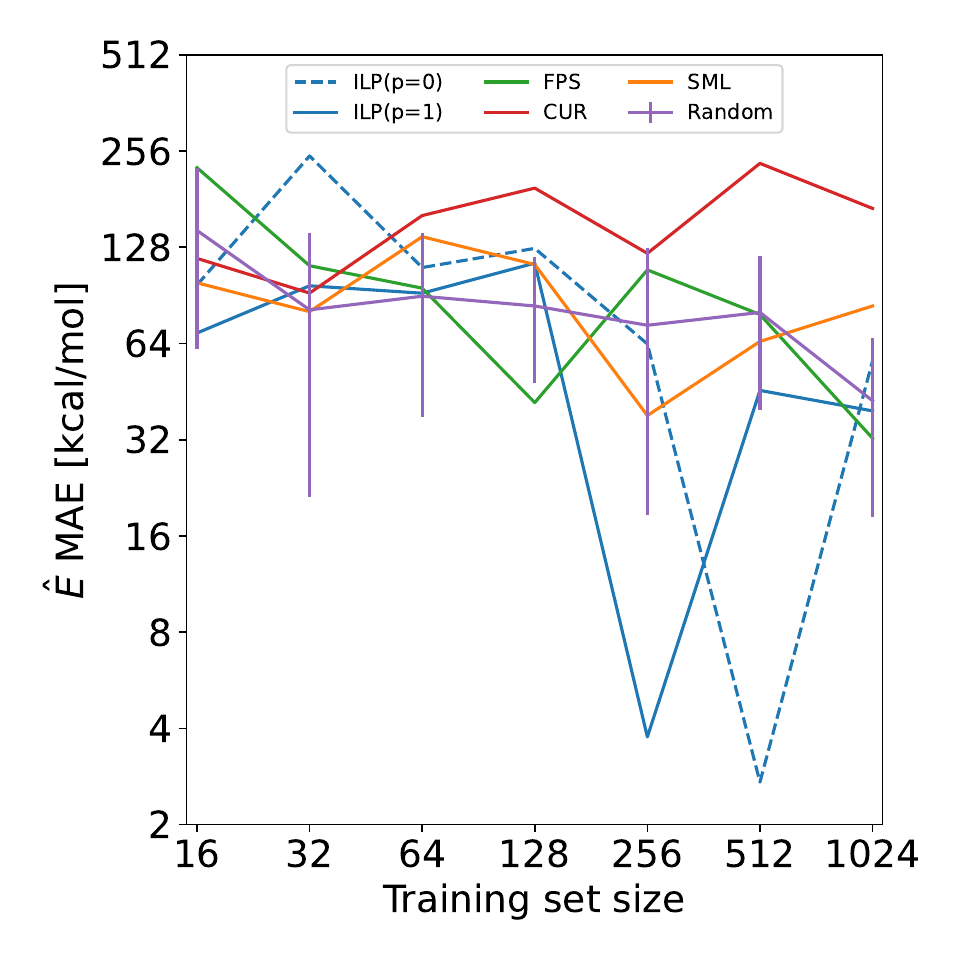}
        \caption{}
    \end{subfigure}
        \caption{Individual learning curves of drugs apixaban, imatinib, oseltamivir and oxycodone (a--d).}
        \label{fig:learning_curves_1}
\end{figure*}

\begin{figure*}[h!]
    \centering
        \begin{subfigure}[t]{0.5\textwidth}
        \centering
        \includegraphics[width=0.9\linewidth]{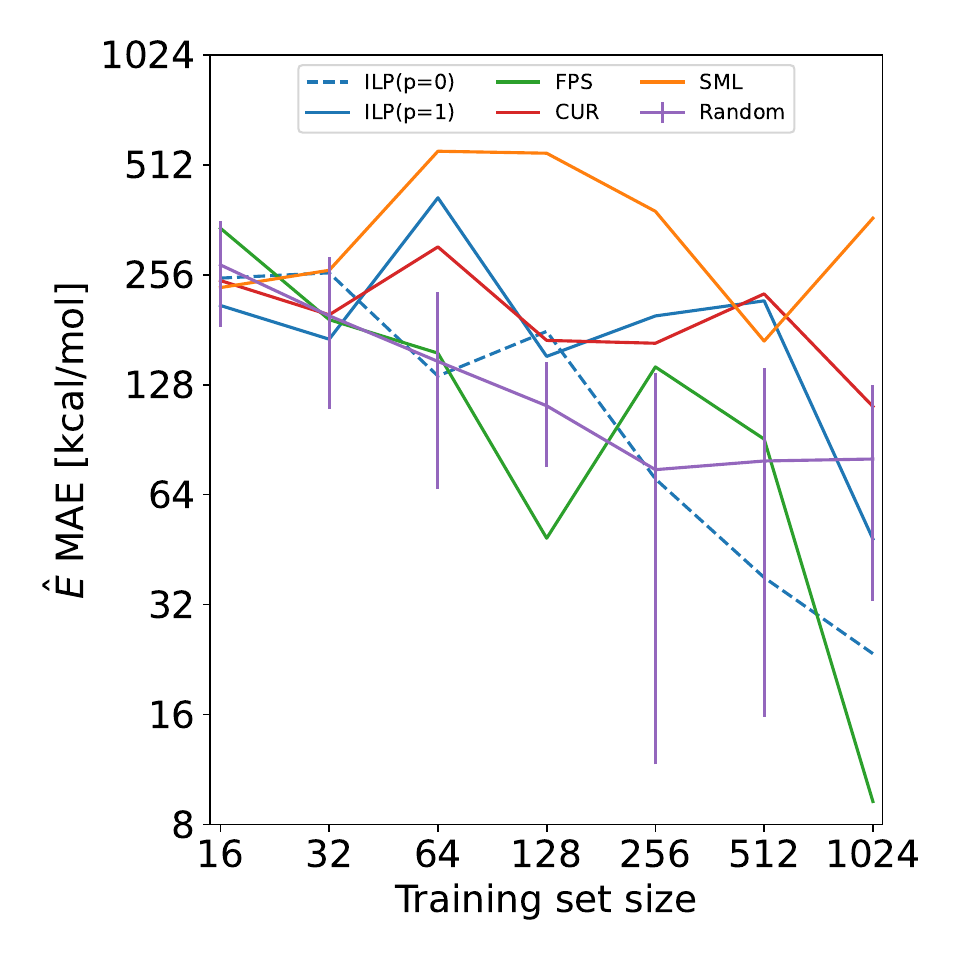}
        \caption{}
    \end{subfigure}%
    ~
    \begin{subfigure}[t]{0.5\textwidth}
        \centering
        \includegraphics[width=0.9\linewidth]{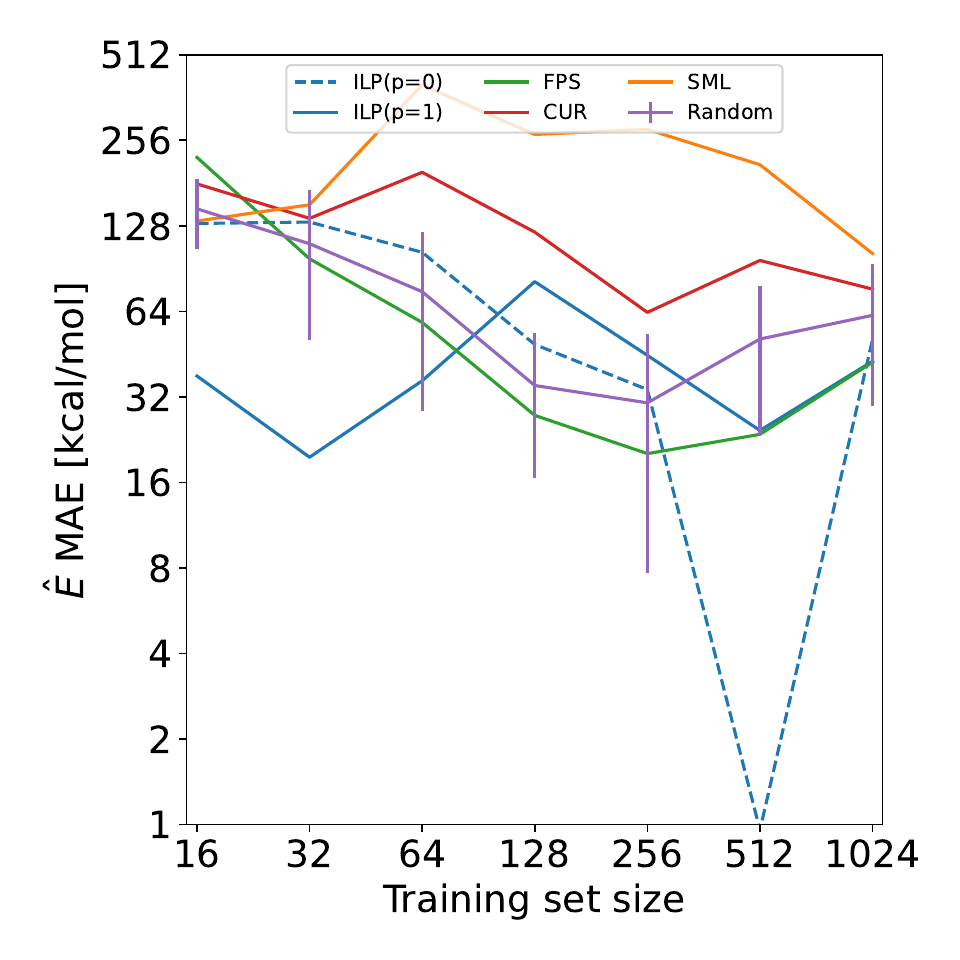}
        \caption{}
    \end{subfigure}
    \\
        \begin{subfigure}[t]{0.5\textwidth}
        \centering
        \includegraphics[width=0.9\linewidth]{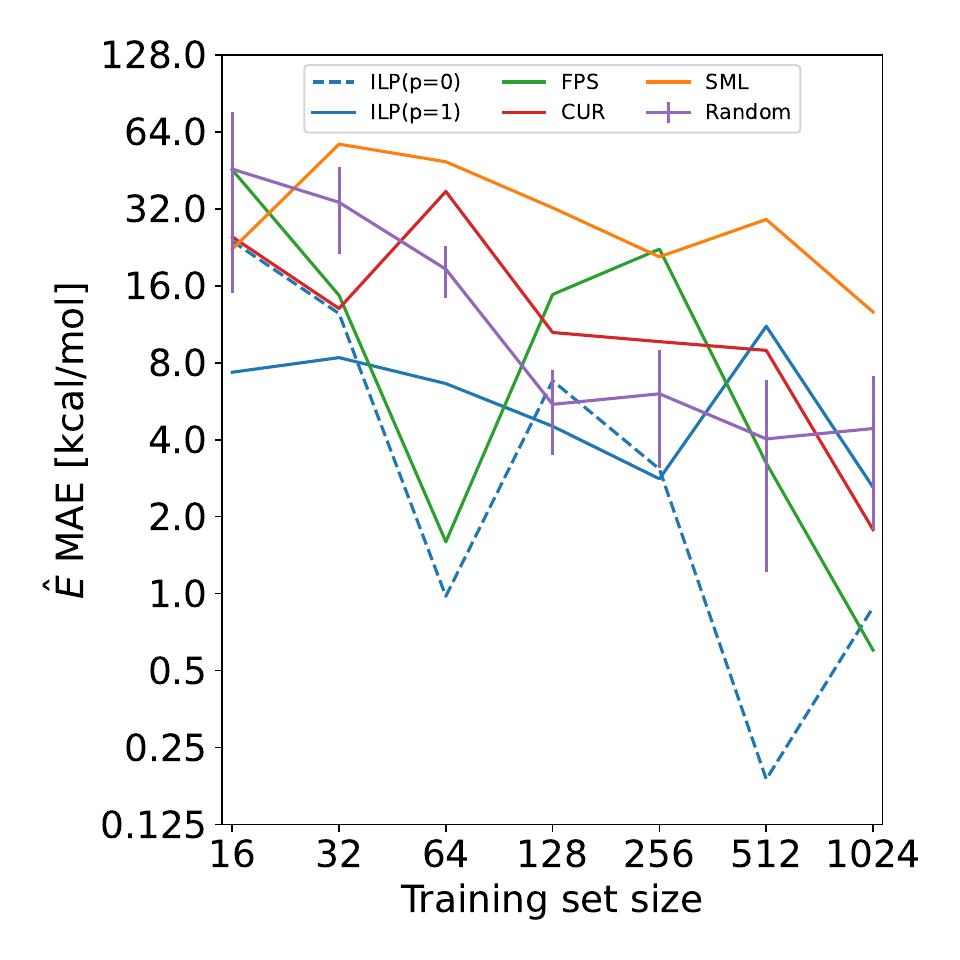}
        \caption{}
    \end{subfigure}%
    ~
    \begin{subfigure}[t]{0.5\textwidth}
        \centering
        \includegraphics[width=0.9\linewidth]{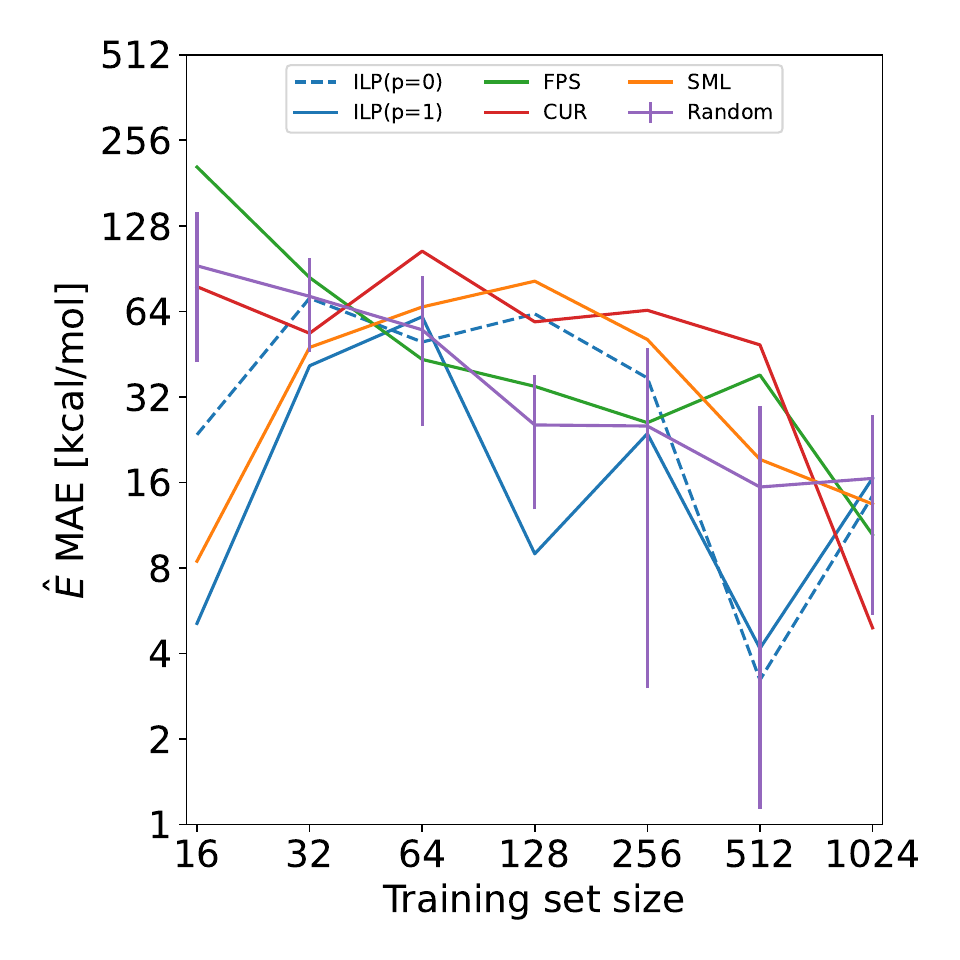}
        \caption{}
    \end{subfigure}
    \\
        \begin{subfigure}[t]{0.5\textwidth}
        \centering
        \includegraphics[width=0.9\linewidth]{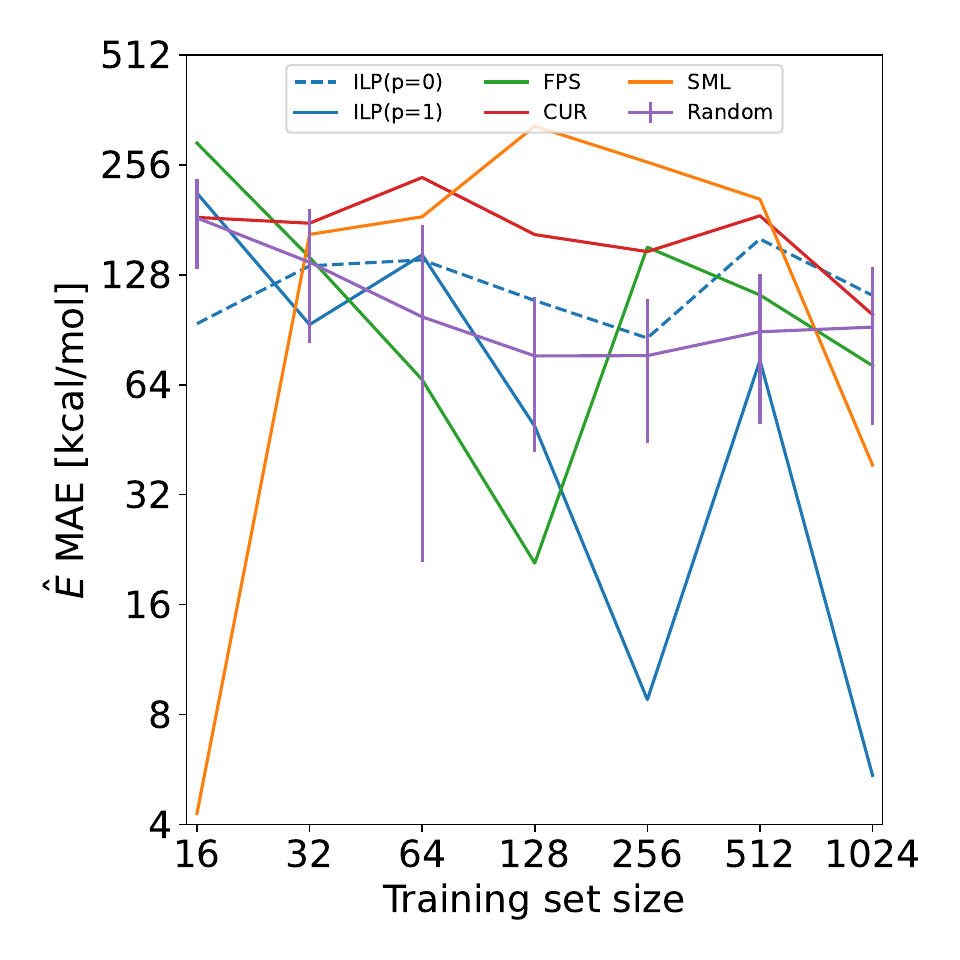}
        \caption{}
    \end{subfigure}%
    ~
    \begin{subfigure}[t]{0.5\textwidth}
        \centering
        \includegraphics[width=0.9\linewidth]{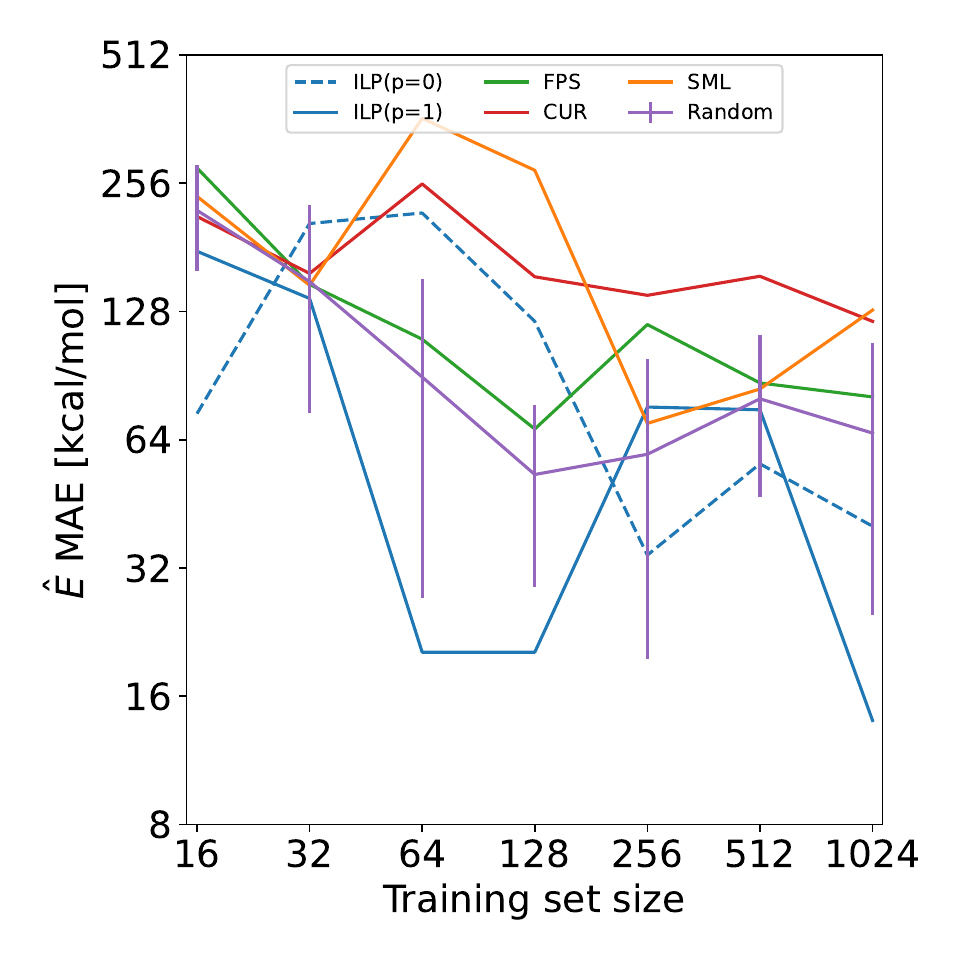}
        \caption{}
    \end{subfigure}
        \caption{Individual learning curves of drugs pemetrexed, penicillin, pregabalin, salbutamol, sildenafil, troglitazone (a--f).}
        \label{fig:learning_curves_2}
\end{figure*}

Individual learning curves for each of the 10 drug target molecules are given in Figures~\ref{fig:learning_curves_1} and \ref{fig:learning_curves_2}.

\clearpage
\section{t-SNE maps} 
\label{sec:tsne}

\begin{figure}[h!]
    \centering
    \includegraphics[width=0.85\linewidth]{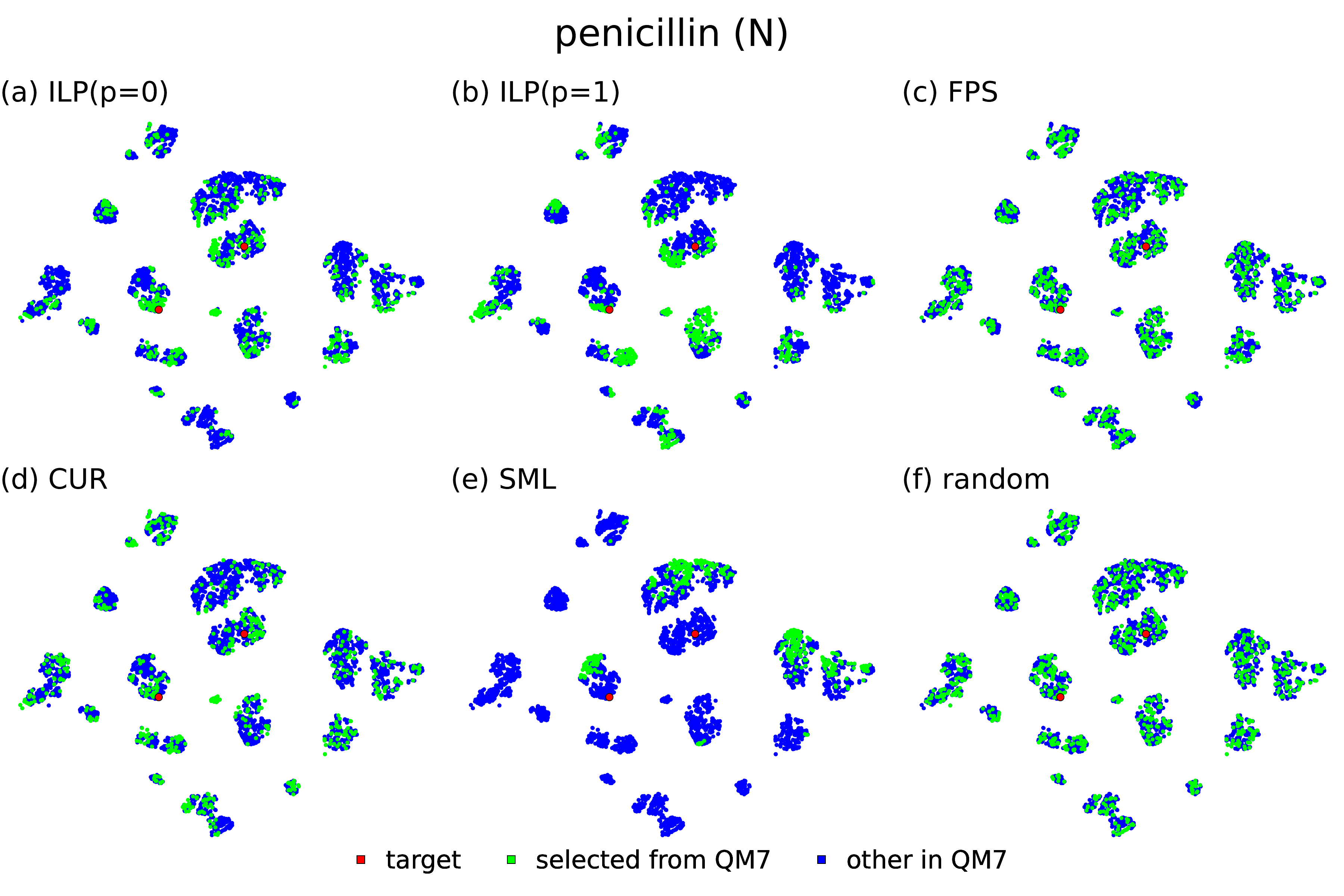}

    \vspace{2.5ex}

    \includegraphics[width=0.85\linewidth]{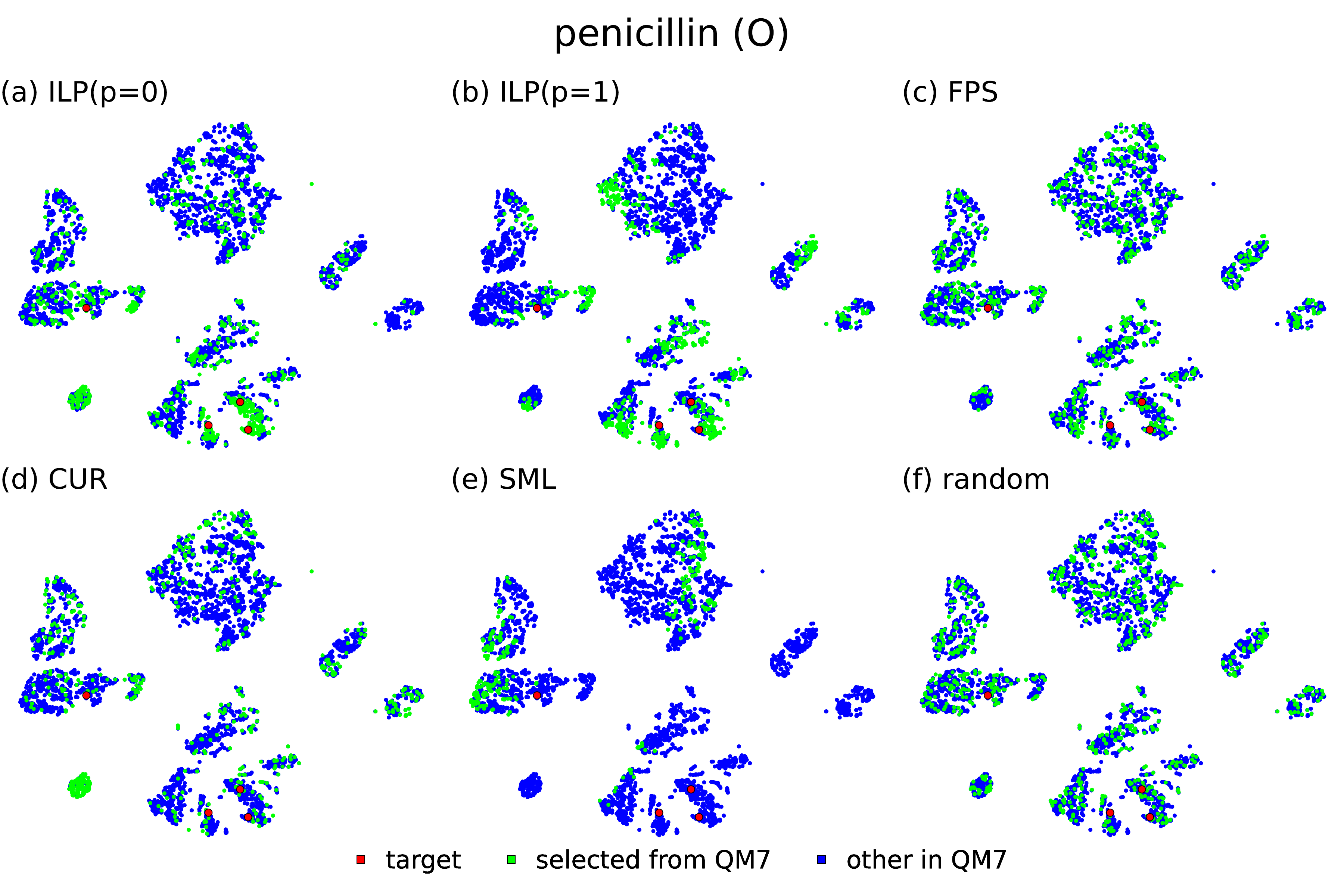}
    \caption{2D t-SNE projection of the nitrogen (perplexity $=90$) and oxygen (perplexity $=80$) atomic environments of the penicillin target molecule (red), those of the selected training molecules for each method (green) and the entire QM7 training set (blue).}
    \label{fig:tsne_penicillin1}
\end{figure}

\begin{figure}[h!]
    \centering
    \includegraphics[width=0.85\linewidth]{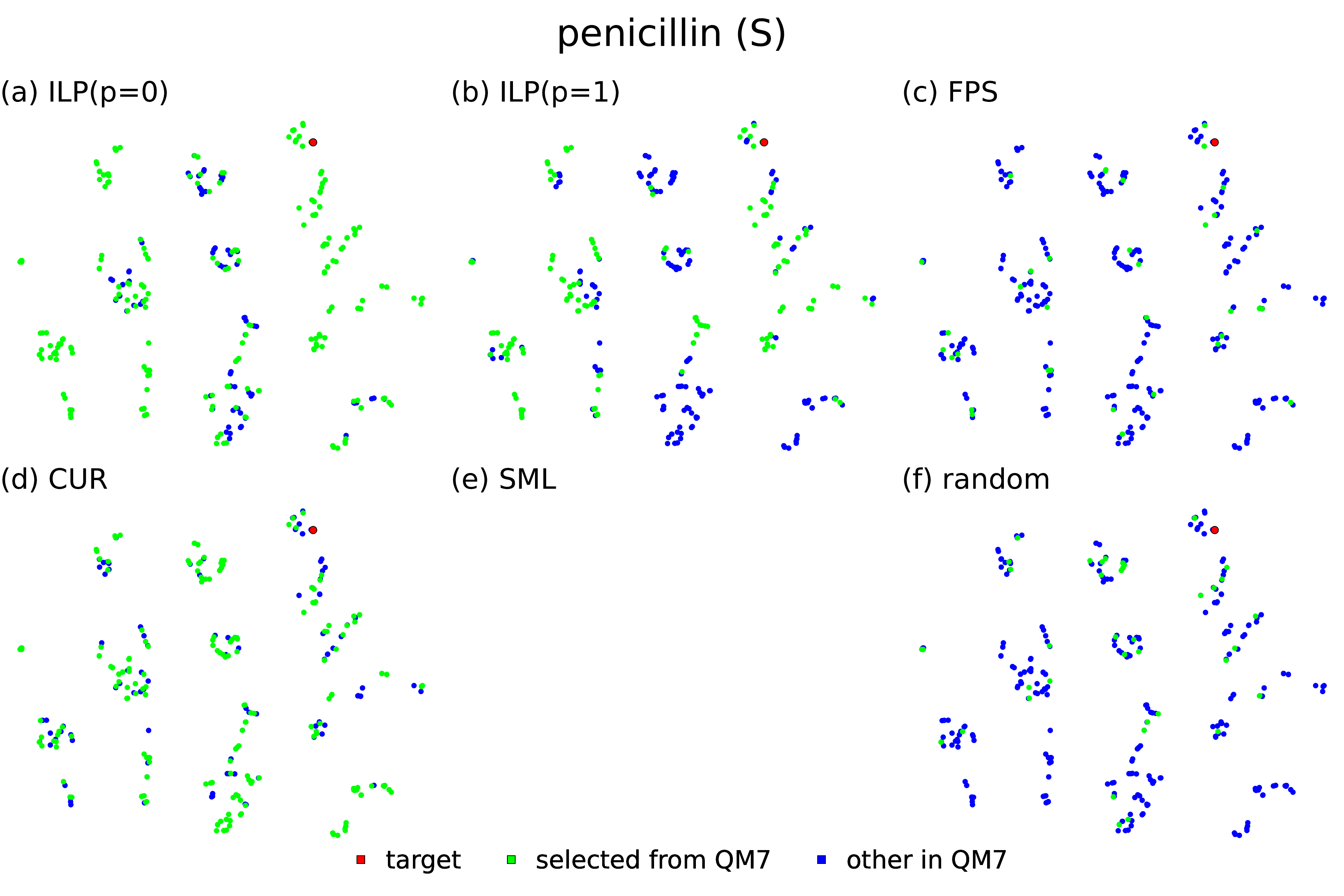}
    \caption{2D t-SNE projection of the sulphur (perplexity $=4$) atomic environments of the penicillin target molecule (red), those of the selected training molecules for each method (green) and the entire QM7 training set (blue). SML selects no sulfur atoms.}
    \label{fig:tsne_penicillin2}
\end{figure}

\begin{figure}[h!]
    \centering
    \includegraphics[width=0.85\linewidth]{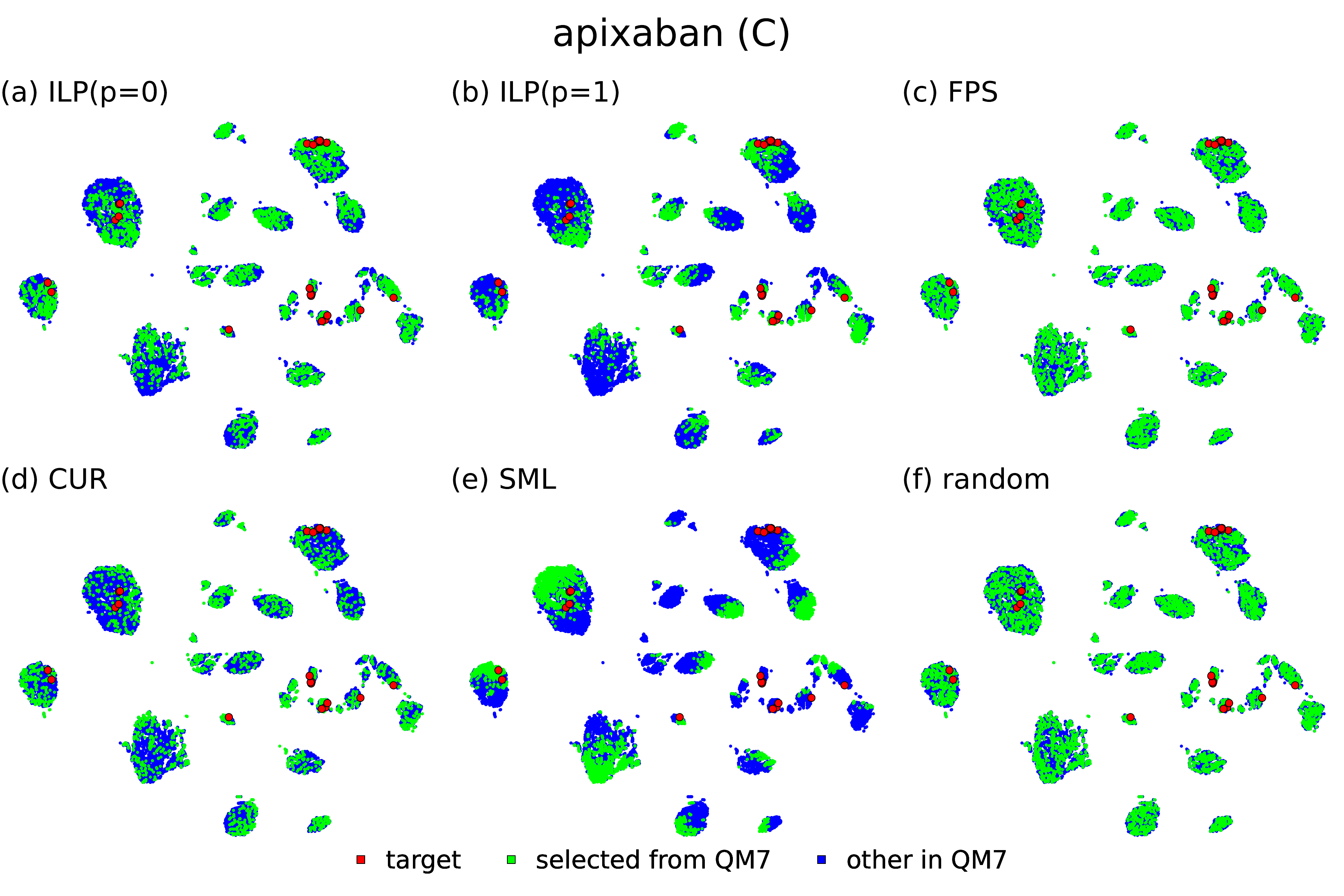}

    \vspace{2.5ex}

    \includegraphics[width=0.85\linewidth]{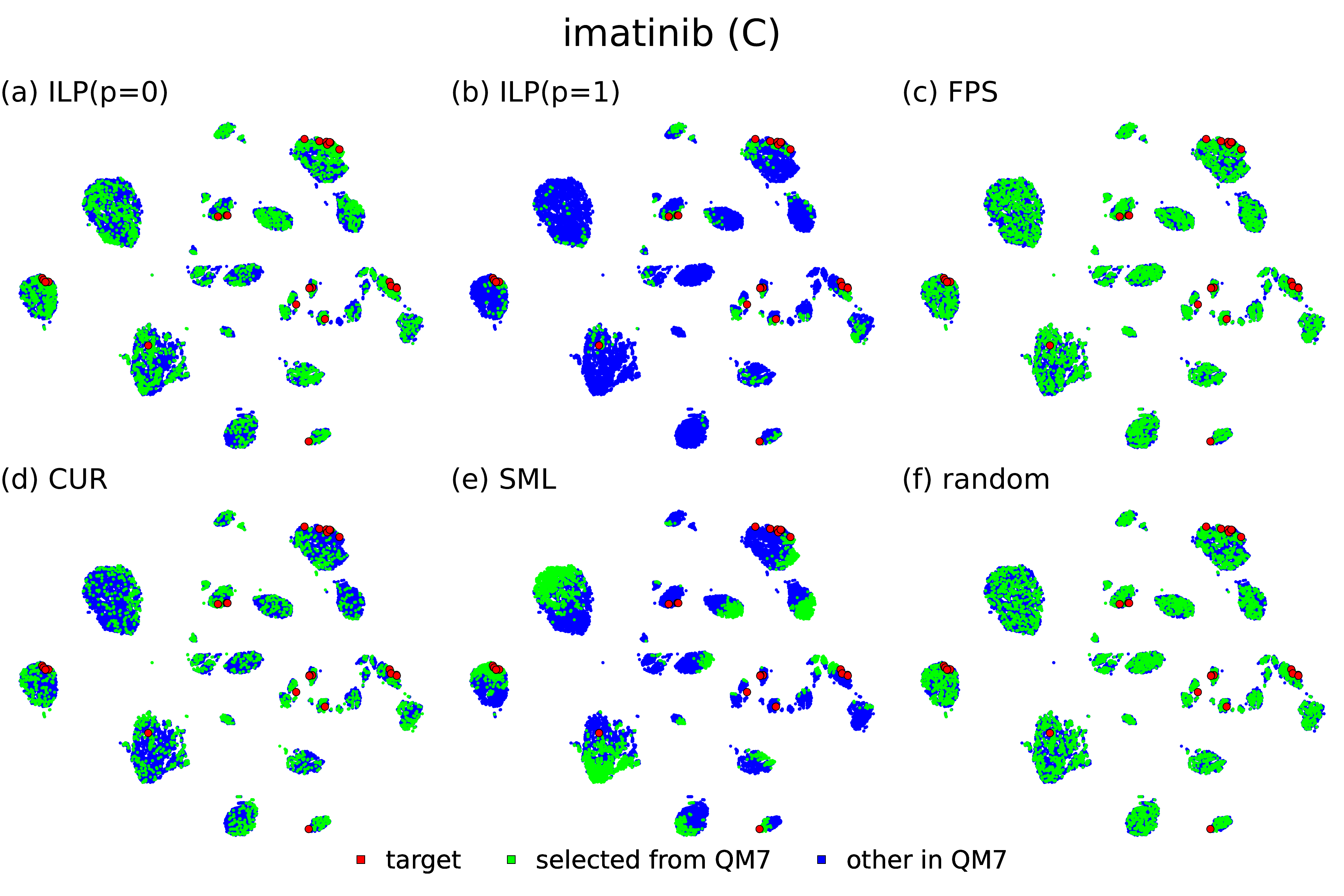}
    \caption{2D t-SNE projection of the carbon (perplexity $=500$) atomic environments of the indicated target molecule (red), those of the selected training molecules for each method (green) and the entire QM7 training set (blue).}
    \label{fig:tsne_C1}
\end{figure}

\begin{figure}[h!]
    \centering
    \includegraphics[width=0.85\linewidth]{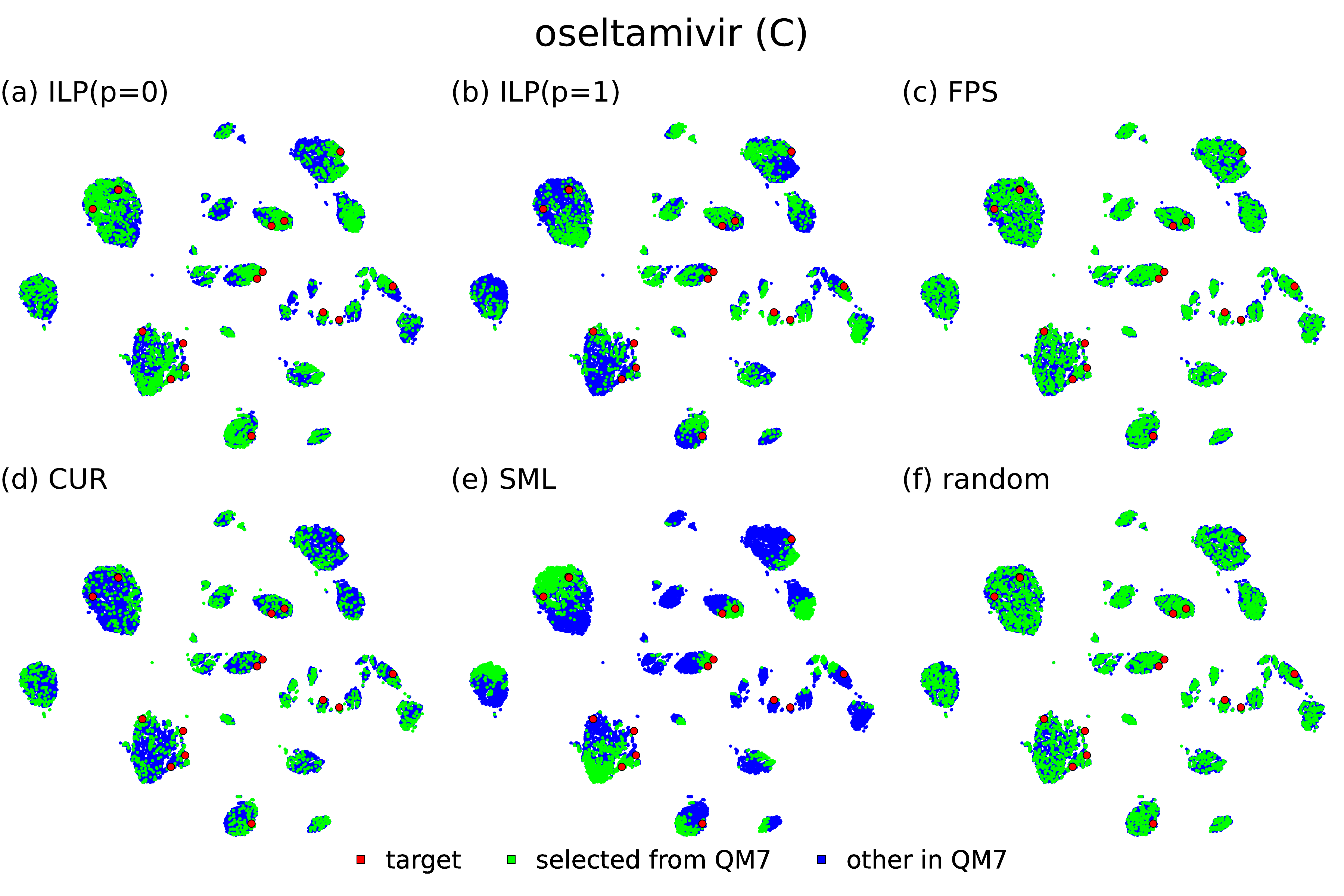}

    \vspace{2.5ex}

    \includegraphics[width=0.85\linewidth]{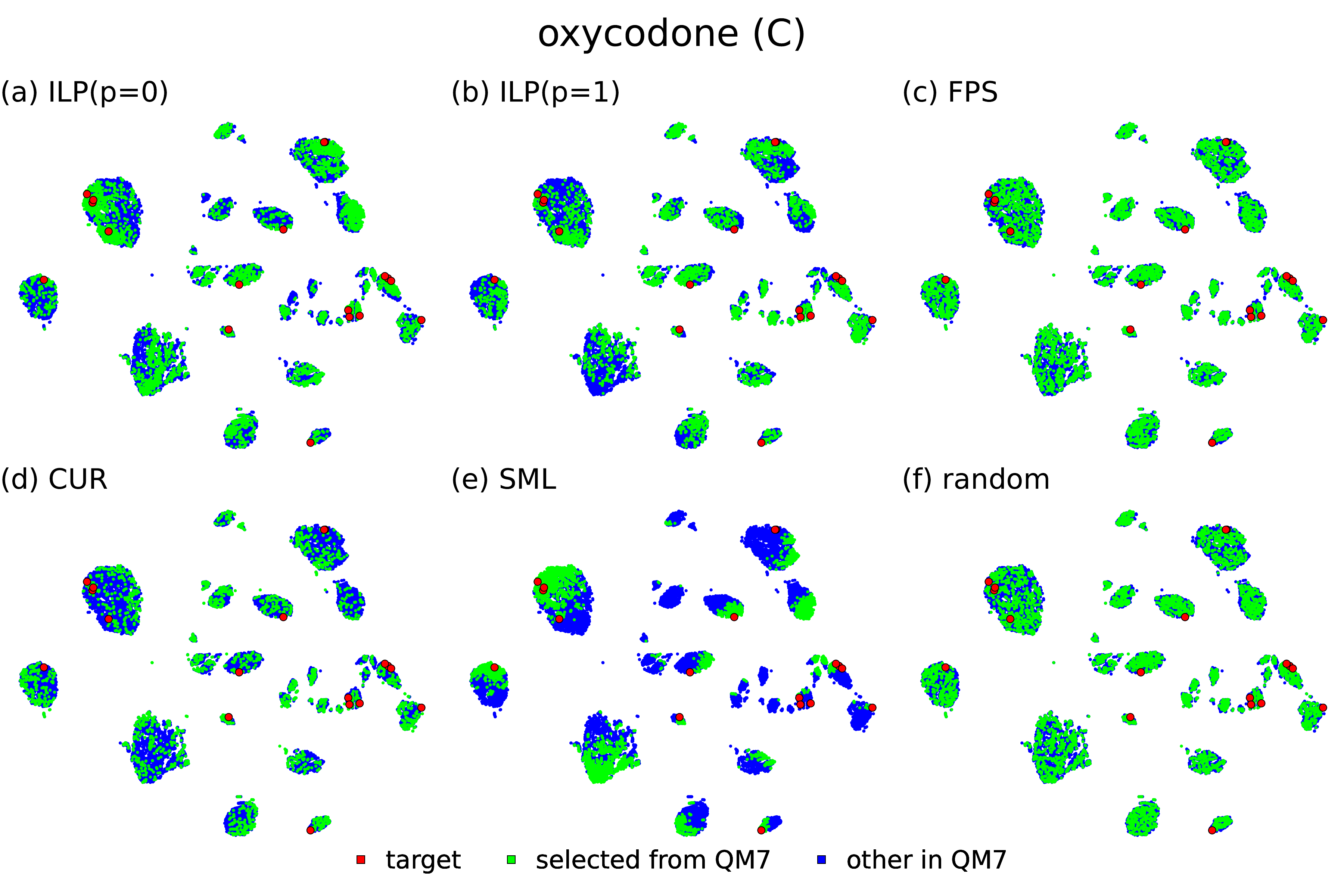}
    \caption{2D t-SNE projection of the carbon (perplexity $=500$) atomic environments of the indicated target molecule (red), those of the selected training molecules for each method (green) and the entire QM7 training set (blue).}
    \label{fig:tsne_C2}
\end{figure}

\begin{figure}[h!]
    \centering
    \includegraphics[width=0.85\linewidth]{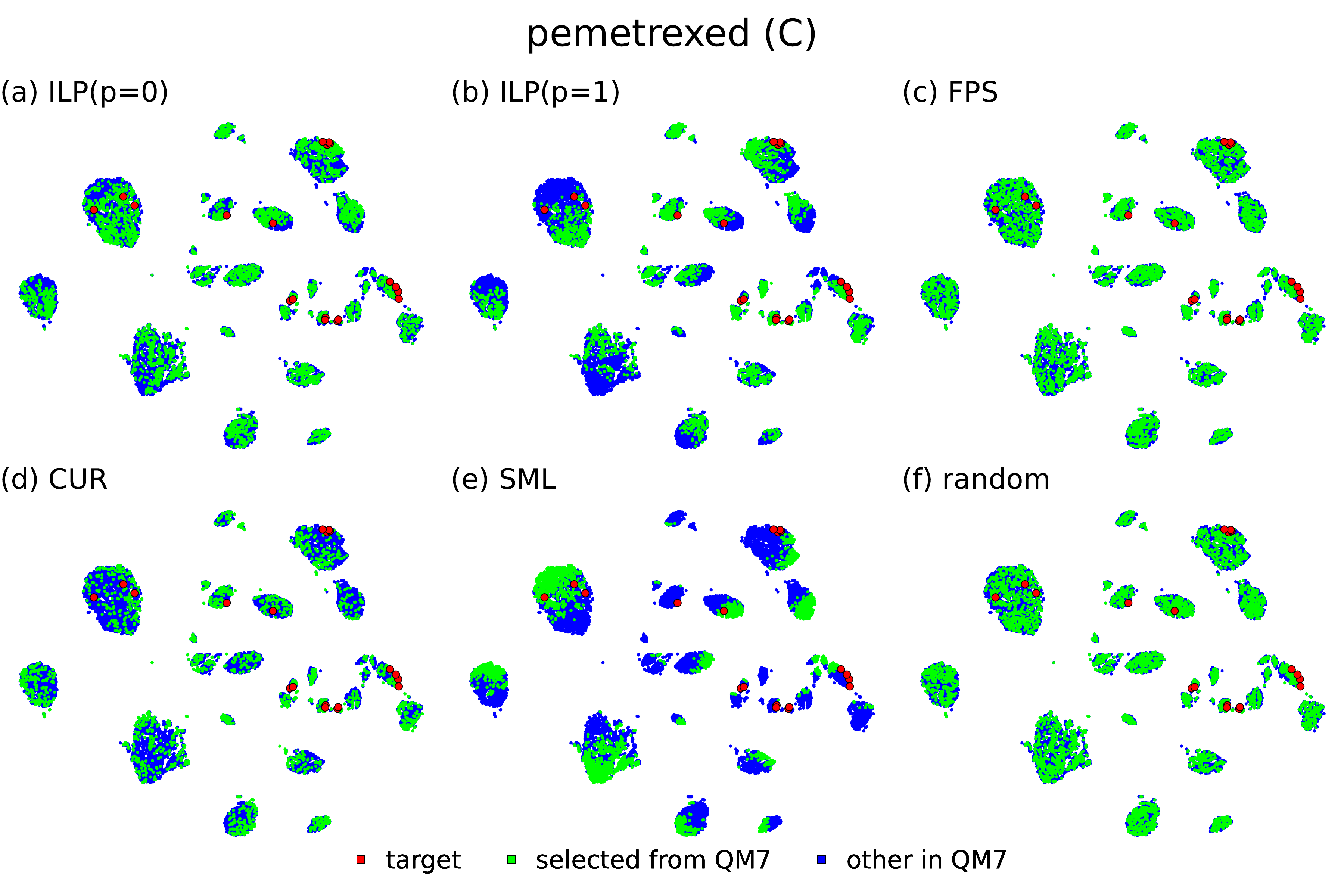}

    \vspace{2.5ex}

    \includegraphics[width=0.85\linewidth]{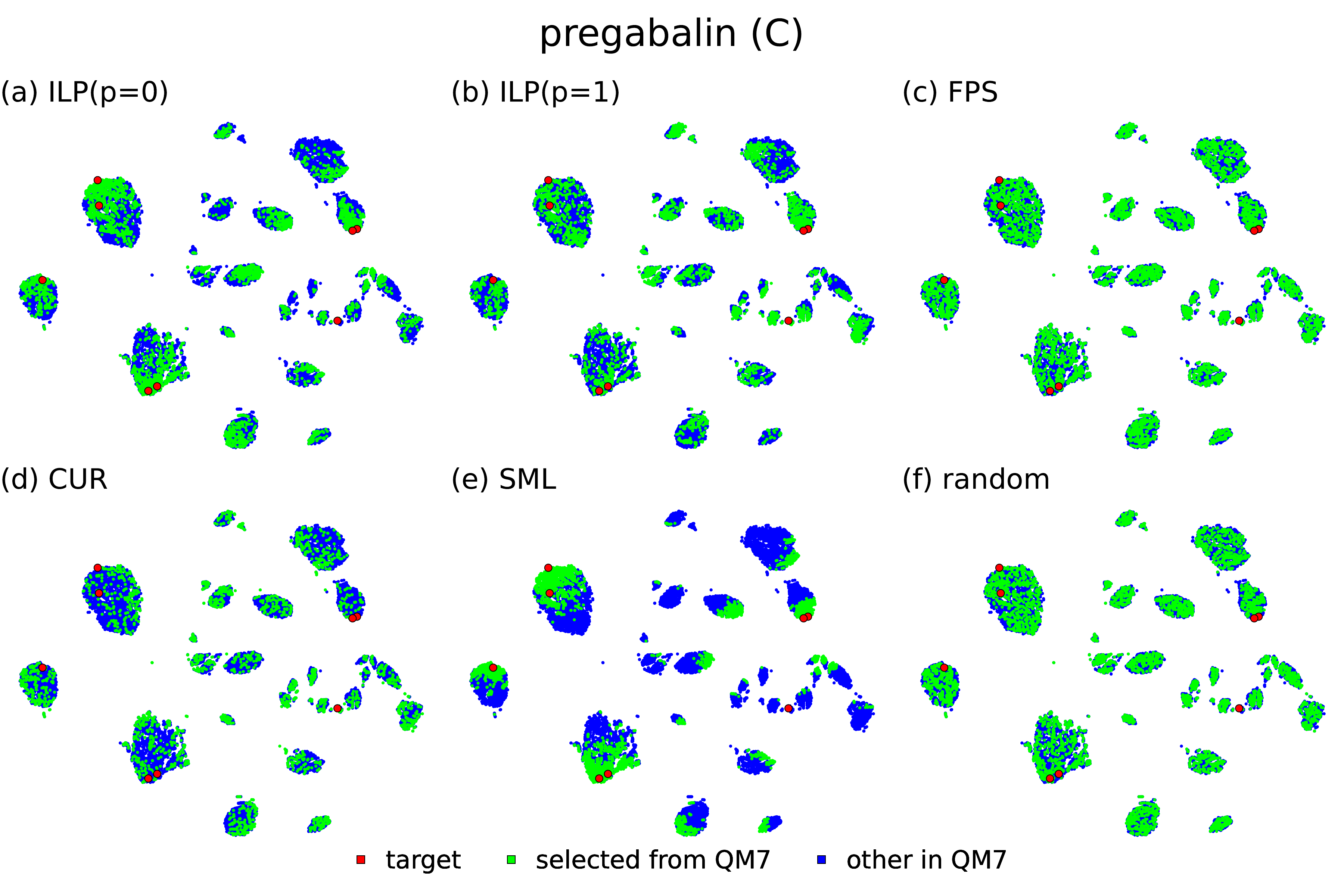}
    \caption{2D t-SNE projection of the carbon (perplexity $=500$) atomic environments of the indicated target molecule (red), those of the selected training molecules for each method (green) and the entire QM7 training set (blue).}
    \label{fig:tsne_C3}
\end{figure}

\begin{figure}[h!]
    \centering
    \includegraphics[width=0.85\linewidth]{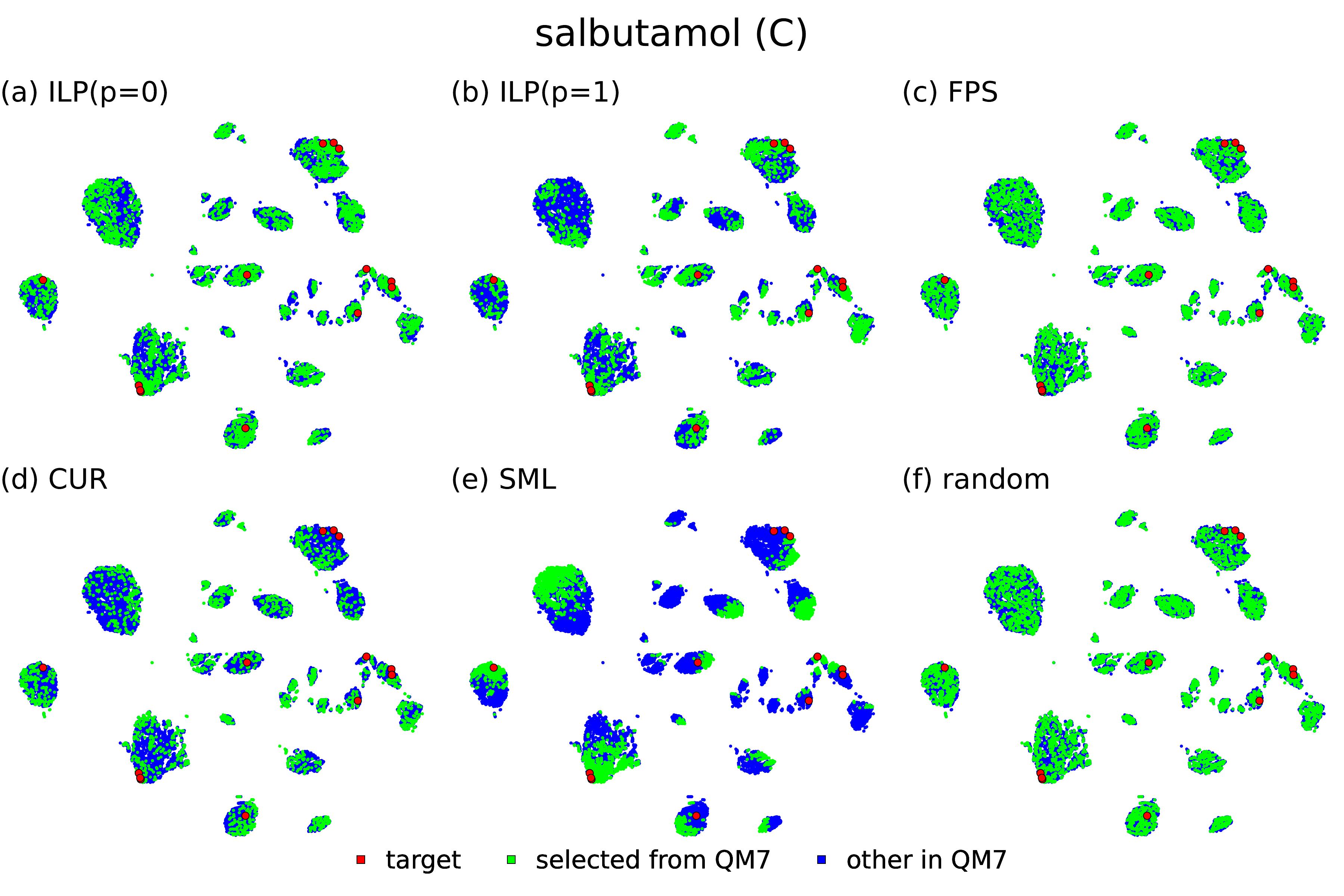}

    \vspace{2.5ex}

    \includegraphics[width=0.85\linewidth]{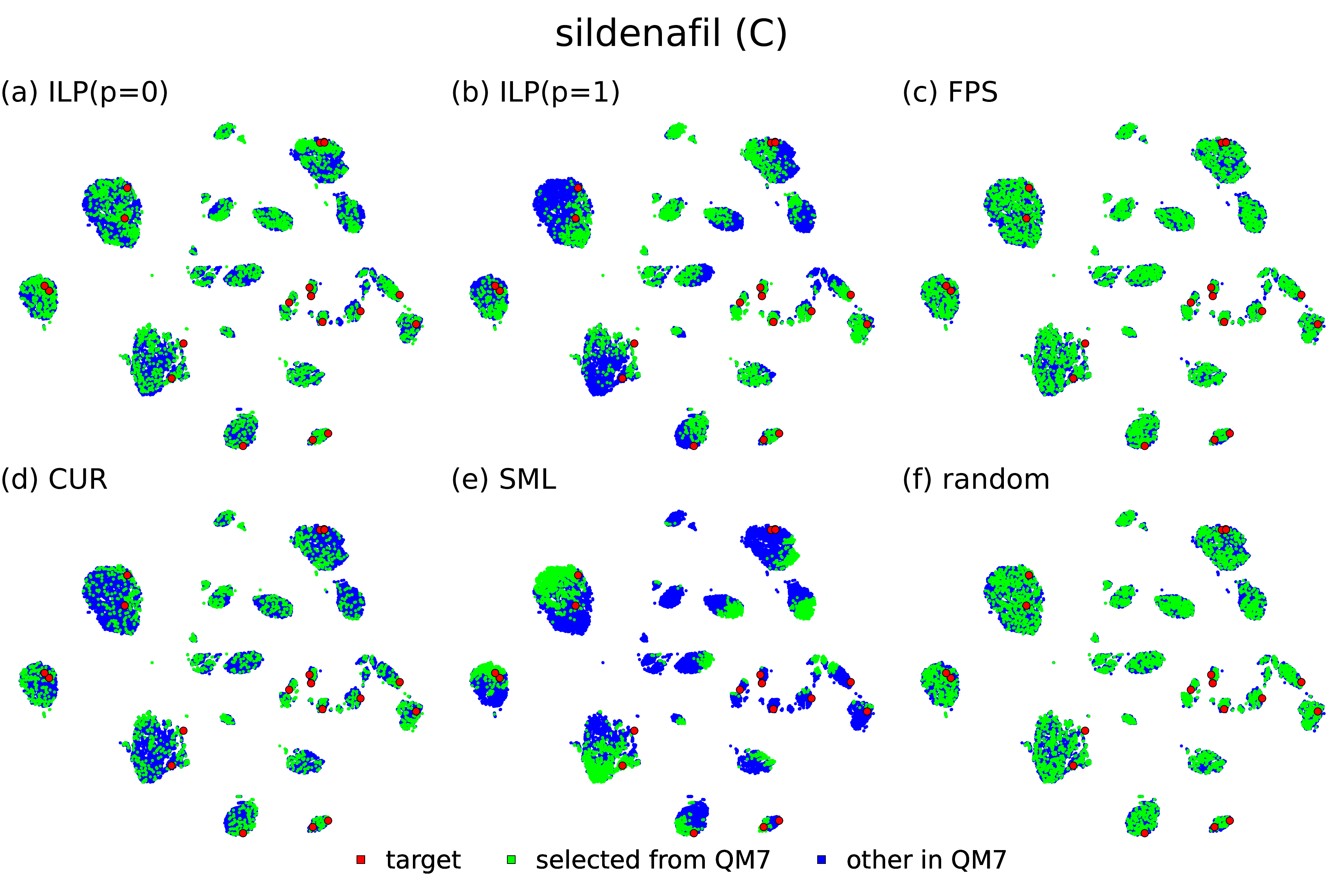}
    \caption{2D t-SNE projection of the carbon (perplexity $=500$) atomic environments of the indicated target molecule (red), those of the selected training molecules for each method (green) and the entire QM7 training set (blue).}
    \label{fig:tsne_C4}
\end{figure}

\begin{figure}[h!]
    \centering
    \includegraphics[width=0.85\linewidth]{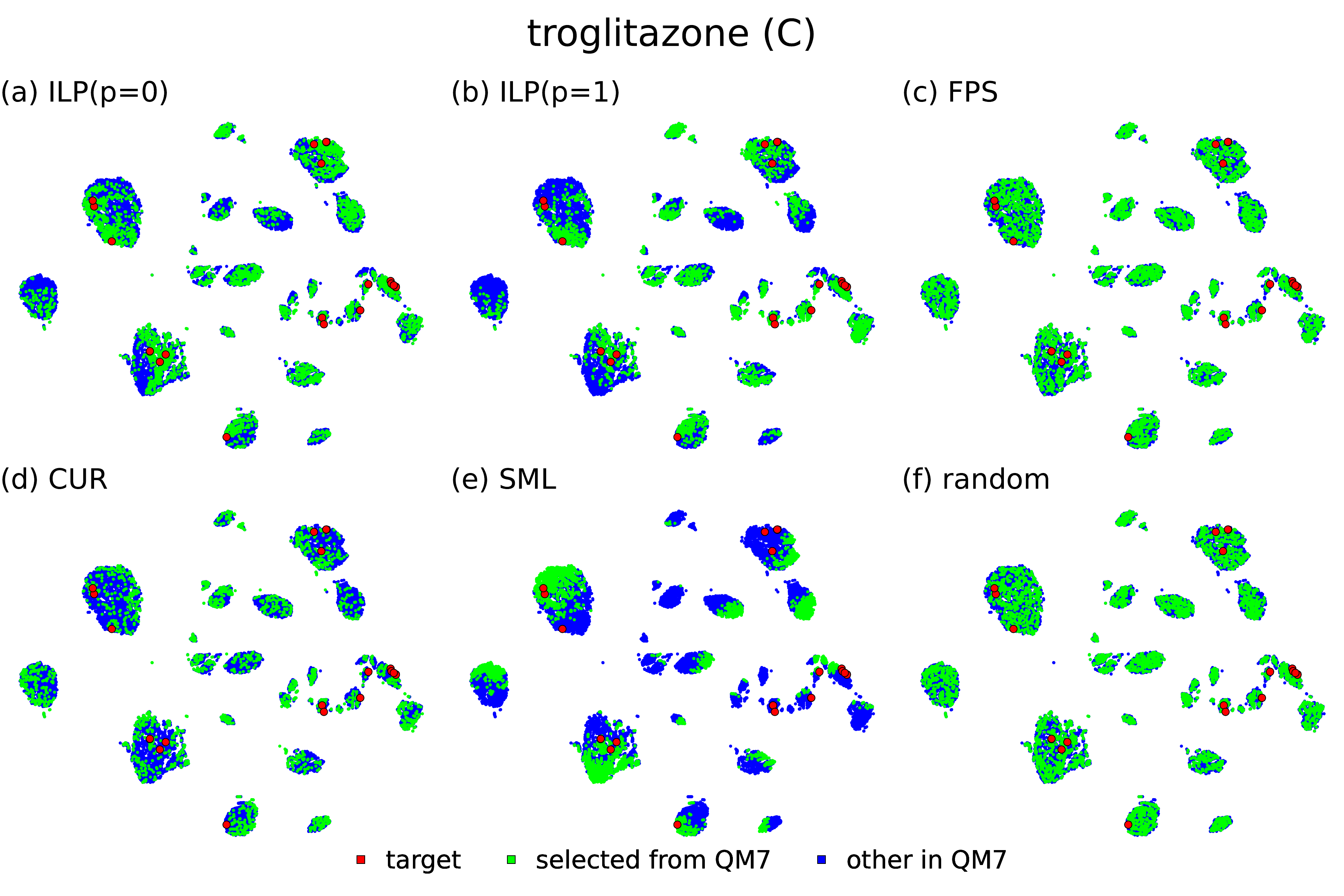}
    \caption{2D t-SNE projection of the carbon (perplexity $=500$) atomic environments of the indicated target molecule (red), those of the selected training molecules for each method (green) and the entire QM7 training set (blue).}
    \label{fig:tsne_C5}
\end{figure}

2D t-SNE plots\cite{van2008visualizing} were generated using the \texttt{openTSNE}\cite{Linderman_2019,Policar2024} package. We fit the embedding using the atomic representations of the specified element (\textit{i.e.,} \ce{C}, \ce{N}, \ce{O}, \ce{S}) of the entire QM7 database, and then project the atomic representations of the same element in the target molecules to this embedding. Figures~\ref{fig:tsne_penicillin1} and \ref{fig:tsne_penicillin2} show the t-SNE plots for \ce{N}, \ce{O} and \ce{S} in penicillin, whereas Figures~\ref{fig:tsne_C1}--\ref{fig:tsne_C5} present the results for the carbon environments in other drugs.

\clearpage
\newrefcontext[labelprefix=S]
\printbibliography